\newcommand\ARXIV

\ifdefined\ARXIV
\documentclass[a4paper,twocolumn,9pt]{extarticle}
\else
\documentclass[final,authoryear,5p,times,twocolumn]{elsarticle}
\fi

\usepackage{graphicx}
\usepackage{color}

\usepackage{layouts} 

\usepackage{lcbmaths}
\usepackage{downsample}
\usepackage{enumitem}
\usepackage{array}

\ifdefined\ARXIV

\usepackage{geometry}
\usepackage{natbib}

\newcommand\apxref[1] {Appendix~\ref{#1}}

\else

\newcommand\apxref[1] {\ref{#1}}

\fi

\usepackage[colorlinks,citecolor=blue,breaklinks=true]{hyperref}

\begin{document}

\ifdefined\ARXIV

\title{Detectability of Granger causality for subsampled continuous-time neurophysiological processes}

\author{Lionel Barnett and Anil K. Seth\\Sackler Centre for Consciousness Science\\School of Engineering and Informatics\\University of Sussex, BN1 9QJ, UK}

\date{\today}

\maketitle

\begin{abstract}
\noindent Granger causality is well established within the neurosciences for inference of directed functional connectivity from neurophysiological data. These data usually consist of time series which subsample a continuous-time biophysiological process. While it is well-known that subsampling can lead to imputation of spurious causal connections where none exist, here we address the equally important issue of the effects of subsampling on the ability to reliably \emph{detect} causal connections which \emph{do} exist.

Neurophysiological processes typically feature signal propagation delays on multiple time scales; accordingly, we base our analysis on a distributed-lag, continuous-time stochastic model, and consider Granger causality in continuous time at finite prediction horizons. Via exact analytical solutions, we identify relationships among sampling frequency, underlying causal time scales and detectability of causalities. Our analysis reveals complex interactions between the time scale(s) of neural signal propagation and sampling frequency: we demonstrate that Granger causality decays exponentially as the sample time interval increases beyond causal delay times, identify detectability ``black spots'' and ``sweet spots'', and show that subsampling may sometimes improve detectability. We also demonstrate that the invariance of Granger causality under causal, invertible filtering fails at finite prediction horizons. We discuss the implications of our results for inference of Granger causality at the neural level from various neurophysiological recording modes, and emphasise that sampling rates for causal analysis of neurophysiological time series should be informed by domain-specific time scales.
\end{abstract}

\else

\journal{Journal of Neuroscience Methods}

\begin{frontmatter}

\title{Detectability of Granger causality for subsampled continuous-time neurophysiological processes}

\author{Lionel Barnett\corref{cor1}}
\ead{l.c.barnett@sussex.acf.uk}

\author{Anil K. Seth\corref{cor2}}
\ead{a.k.seth@sussex.ac.uk}

\cortext[cor1]{Corresponding author}

\address{
Sackler Centre for Consciousness Science and School of Engineering and Informatics,\\
University of Sussex, Brighton BN1 9QJ, UK
}

\begin{abstract}
\textit{Background}

Granger causality is well established within the neurosciences for inference of directed functional connectivity from neurophysiological data. These data usually consist of time series which subsample a continuous-time biophysiological process. While it is well known that subsampling can lead to imputation of spurious causal connections where none exist, less is known about the effects of subsampling on the ability to reliably \emph{detect} causal connections which \emph{do} exist.

\ \\\noindent\textit{New Method}

\noindent We present a theoretical analysis of the effects of subsampling on Granger-causal inference. Neurophysiological processes typically feature signal propagation delays on multiple time scales; accordingly, we base our analysis on a distributed-lag, continuous-time stochastic model, and consider Granger causality in continuous time at finite prediction horizons. Via exact analytical solutions, we identify relationships among sampling frequency, underlying causal time scales and detectability of causalities.

\ \\\noindent\textit{Results}

\noindent We reveal complex interactions between the time scale(s) of neural signal propagation and sampling frequency. We demonstrate that detectability decays exponentially as the sample time interval increases beyond causal delay times, identify detectability ``black spots'' and ``sweet spots'', and show that downsampling may potentially improve detectability. We also demonstrate that the invariance of Granger causality under causal, invertible filtering fails at finite prediction horizons, with particular implications for inference of Granger causality from fMRI data.

\ \\\noindent\textit{Comparison with Existing Method(s)}

\noindent Our analysis emphasises that sampling rates for causal analysis of neurophysiological time series should be informed by domain-specific time scales, and that state-space modelling should be preferred to purely autoregressive modelling.

\ \\\noindent\textit{Conclusions}

\noindent On the basis of a very general model that captures the structure of neurophysiological processes, we are able to help identify confounds, and offer practical insights, for successful detection of causal connectivity from neurophysiological recordings.
\end{abstract}


\end{frontmatter}

\fi

\section{Introduction} \label{sec:intro}

Neurophysiological recordings are generally obtained by sampling, at regular discrete time intervals, a continuous-time analogue signal associated with some underlying biophysiological processes. Thus, for example, electroencephalography (EEG) records electrical activity arising from ionic current flows in the brain, magnetoencephalography (MEG) records the weak magnetic fields produced by neuronal currents, while functional magnetic resonance imaging (fMRI) measures changes in blood oxygenation level associated with neural activity \citep{LogothetisEtal:2001}. Even spike train recordings are typically derived from a continuous analogue measurement of cellular membrane potentials.

Wiener-Granger causality \citep{Wiener:1956,Granger:1963,Granger:1969,Granger:1981,Geweke:1982}---henceforth just Granger causality, or GC---is a popular technique for inferring directed functional connectivity of the underlying process in the neurosciences \citep{Seth:gcneuro:2015}, from (discrete-time) subsampled\footnote{The term ``subsample'' refers throughout to sampling of a discrete- or continuous-time process at \emph{regular intervals}. We reserve the term ``downsample'' for the further subsampling of an already-sampled discrete-time process.} process. Granger causality is premised on a notion of causality whereby cause (a) precedes effect, and (b) contains unique information about effect. This idea is commonly (but not exclusively) operationalised within a vector autoregressive (VAR) modelling framework. At this point, we recognise that the ascription of a ``causal'' interpretation to GC is clearly problematic to some. Our view is that Granger causality represents \emph{a} rather than \emph{the} notion of causality, an avowedly \emph{statistical}, as opposed, \eg, to ``interventionist'' notions \citep{Pearl:2009}. As such, its strengths and limitations have been widely discussed [see \eg~\cite{ValdesSosaEtal:2011} for a review of the issues involved with regard to biophysical modelling; also \cite{ChicharroPanzeri:2014}], and we do not enter that debate here. We remark, however, that Granger causality also has a principled interpretation---through its intimate relationship \citep{Barnett:tegc:2009,Barnett:teml:2012} with the information-theoretic \emph{transfer entropy} \citep{Schreiber:2000,Palus:2001}---as a measure of \emph{information transfer}, and we generally prefer this interpretation  \citep{LizierProkopenko:2010}, particularly with regard to functional connectivity analysis.

Problems associated with Granger-causal inference from subsampled (or otherwise aggregated) time series have long been noted \citep{Granger:1969,Sims:1971,Wei:1981,Marcellino:1999,BreitungSwanson:2002}. Specifically, it has been observed that subsampling may distort GC values. This may be considered especially problematic in two distinct aspects:
\begin{enumerate}
\item[i] \emph{Spurious causality}, where GC is absent at the finer time scale, but non-zero for the subsampled process \citep{ComteRenault:1996,RenaultEtal:1998,BreitungSwanson:2002,McCrorieChambers:2006,Solo:2007,Solo:2016}, and
\item[ii] \emph{Undetectable causality}, where GC is present at the finer time scale, but zero (or too small to detect reliably) for the subsampled process \citep{Barnett:gcfilt:2011,Seth:gcfmri:2013,Zhou:2014}.
\end{enumerate}
Subsampling may, in addition, distort the \emph{relative strengths} of causalities \citep{Solo:2016}.

\cite{Solo:2007,Solo:2016}, drawing on previous work by \cite{Caines:1976}, distinguishes between the conventional ``weak'' causality and ``strong'' causality (see \secref{sec:gc}), and concludes that only strong causality remains undistorted by subsampling. \cite{Seth:gcfmri:2013} demonstrate that GC inference from fMRI data may be severely degraded by the sample rates, slow in comparison to underlying neural time scales, of fMRI recording technologies. More recently, \cite{Zhou:2014} report oscillations in estimated causalities with varying sampling frequency, with causal estimates almost vanishing at some frequencies, as well as inference of spurious causalities.

Although Granger himself was clearly concerned about the detectability problem---in \cite{Granger:1969} he notes that ``[...] a simple causal mechanism can appear to be a feedback mechanism\footnote{Here, by ``feedback mechanism'', Granger refers to \emph{contemporaneous} feedback between time series [\citet{Geweke:1982} terms this ``instantaneous feedback''], as opposed to \emph{time-delayed} feedback, which in his theory underpins ``causal mechanism''; see \secref{sec:gc} for details.} if the sampling period for the data is so long that details of causality cannot be picked out''---subsequent studies have concentrated mostly on spurious causality. Here we investigate detectability: specifically, we examine how the relationship between the underlying time scale of causal mechanisms and the sampling time scale mediates the distortion of (non-zero) Granger causalities, and how this distortion impacts on statistical inference of Granger causality from empirical data. We discuss the implications of our results with regard to the successful inference of Granger causalities at the structural (neural) level, from neurophysiological recordings.

\subsection{Contributions of this study} \label{sec:contrib}

A significant feature of the neuronal systems underlying such measurements is the potential range of signal propagation delays due to variation in biophysical parameters such as axonal length, diameter, conduction velocity and myelination \citep{Miller:1994,BuddKisvarday:2012,CaminitiEtal:2013}. Here we model the underlying analogue signal as a stochastic linear autoregression in continuous time. Unlike prevailing continuous-time stochastic process models in the neurosciences, our model accommodates distributed lags on arbitrary time scales, and is thus able to reflect variability of signal propagation delays. This leads, via consideration of prediction at \emph{finite} time horizons, to a novel and intuitive definition of Granger causality at multiple time scales for continuous-time processes. In contrast to previous work on continuous-time Granger causality, in which various statistical (non)causality test criteria have been proposed, our definition is \emph{quantitative}, furnishing a Granger-Geweke measure with an information-theoretic interpretation.

Using discrete-time VAR modelling, we then analyse the properties of processes obtained by subsampling the temporally multiscale continuous-time process, and relate the spectral and causal properties of the subsampled process to those of the underlying continuous-time model. Having defined continuous-time, finite-horizon GC---which represents a target for statistical analysis---we investigate the extent to which it may be inferred, and in particular \emph{detected}, by discrete-time VAR analysis of the subsampled processes.

We focus on the practical questions of the feasibility and reliability of causal inference on sampling frequency and the (dominant) time scale of causal feedback in the generative process. We investigate in detail the relationship between sampling frequency and the quality of causal inference via a fully analytic solution of a minimal, but non-trivial, bivariate model in continuous time, with finite causal delay.

On the basis of our theoretical and empirical analysis, we identify critical relationships between causal delay, sampling interval and detectability of Granger causality. These include exponential decay of subsampled Granger causalities with increasing sampling interval, resonance between sampling frequency and causal delay frequency, potential detectability ``black spots'', and the existence of a non-zero optimal sampling interval (\ie, detectability may sometimes be improved by downsampling). We also discover a hitherto unremarked non-invariance of finite-horizon/multistep GC under causal, invertible filtering (in contrast with the known invariance of single-step discrete-time GC).

Finally, we discuss the implications of our findings for Granger-causal inference of neural functional relationships from neurophysiological recordings under various technologies - including fMRI, which continues to generate controversy.

\subsection{Organisation} \label{sec:organise}

The paper is organised as follows: in \secref{sec:varGC} we review essential aspects of the theory of VAR processes and Granger causality in discrete time. In \secref{sec:voudl} we introduce CTVAR (continuous-time vector autoregressive) processes as continuous-time, distributed-lag analogues of discrete-time VAR processes, and derive a principled extension of Granger causality to such processes, based on finite-temporal horizon prediction. We analyse discrete-time processes derived by subsampling a CTVAR process, and demonstrate the consistency of GC in the limit as the subsampling interval shrinks to zero. In \secref{sec:ouminlag} we present a detailed analytic solution of the subsampling problem for a non-trivial minimal bivariate CTVAR process with finite causal delay, and address the issue of statistical inference (detectability) for GC. Lastly, in \secref{sec:disc} we discuss the implications of our results presented in the setting of analysis of neurophysiological data. Technical details, where they would detract from the narrative flow, are presented in Appendices.

\subsection{Notation and conventions} \label{sec:noco}

The principal objects of study in this paper are random vectors in a real Euclidean space $\reals^n$ and vector stochastic processes; \ie\ sequences of random vectors in discrete or continuous time. Time sequences are generally written as $x = \{x_k \,|\, k \in \integers\}$ ($\integers$ denotes the set of integers) and $x = \{x(t) \,|\, t \in \reals\}$ in discrete and continuous time respectively, where the $x_k$ or $x(t)$ could be real or complex, random or deterministic scalars, vectors, matrices, \etc; note that when we refer to a sequence as a whole, we shall frequently drop the time index/variable. Vectors in $\reals^n$ are generally written in bold type and random variables in upper case; thus, \eg, a vector stochastic process in discrete time is generally represented as $\{\bX_k \,|\, k \in \integers\}$, and the entire process referred to simply as $\bX$. For avoidance of consideration of initial conditions, process time (discrete or continuous) is assumed to extend into the infinite past.

Time is assumed measured in a standard unit, which we take to be milliseconds (ms). For discrete-time sequences, we require that a sample interval (time step) $\Dt$ be specified [equivalently, a sampling frequency $f_s \equiv 1/\Dt$, measured in kilohertz (kHz)]. To emphasize the dependence of a quantity on sample interval, $\Dt$ is included as a function argument. In particular, this study is concerned with the \emph{regular subsampling} of continuous-time processes. If $x = \{x(t) \,|\, t \in \reals\}$ is a (random or deterministic, scalar, vector, \etc) continuous-time sequence, we write $x(\Dt) \equiv \{x(k\Dt) \,|\, k \in \integers\}$ for the \emph{discrete}-time sequence obtained by sampling $x$ at regular intervals $\Dt$, which we refer to as a \emph{$\Dt$-subsampling} of $x$.

Much of the analysis presented here takes place naturally in the spectral domain. Ordinary frequencies are generally written as $-\infty < \lambda < \infty$, measured in kHz. In discrete time with sampling interval $\Dt$, spectral quantities are periodic in $\lambda$ with period $f_s = 1/\Dt$, the sampling frequency; we shall sometime restrict such quantities to the interval $-1/(2\Dt) \le \lambda < 1/(2\Dt)$, where $1/(2\Dt) = f_s/2$ is the \emph{Nyqvist frequency}. For continuous-time sequences, spectral quantities are not generally periodic. For discrete-time spectral quantities, it is sometimes convenient to work instead with the \emph{angular frequency} $\omega \equiv 2\pi\Dt \lambda$. We may then consider spectral quantities as defined on the unit circle $|z| = 1$ in the complex plane $\complexes$, with $z = e^{-i\omega}$, $-\pi \le \omega < \pi$. In continuous time we occasionally use a normalised frequency $\omega \equiv 2\pi\lambda$, and spectral quantities may be considered defined on the imaginary line $\re(\zeta) = 0$ in the complex plane, with $\zeta = i\omega$, $-\infty < \omega < \infty$. In the time domain, $z$ and $\zeta$ may be interpreted as \emph{lag (backshift) operators} in discrete and continuous time respectively.

The link between time and frequency domains is the \emph{Fourier transform}. Here, Fourier transforms are always defined in terms of ordinary frequency $\lambda$ and indicated with a hat symbol over over the corresponding sequence specifier; \eg\ $\hat{x}(\lambda)$, or just $\hat{x}$ when the entire transform is referenced. In light of the proliferation of conventions, our definitions for Fourier transforms are set out in \apxref{sec:ftran}; discrete-time transforms are scaled by the sample time step $\Dt$ in order to ensure that dimensions are always the same in discrete and continuous time, and that, in particular, limiting values for $\Dt$-subsampled continuous-time sequences tend to corresponding continuous-time values as $\Dt \to 0$; see \apxref{sec:ftran} for details. Generally, we take care to scale by sampling interval so that (almost) all measurable quantities (\tabref{tab:quants}) have the same dimensions in discrete and continuous time, and comparisons of magnitudes are thus meaningful, in particular in the limit $\Dt \to 0$.
\begin{table}[t]
\centering
\begin{tabular} {|l|l|l|}
\hline
quantity & description & unit \\
\hline\hline
$k,\ell,m,\ldots$ & discrete time index & $\mathsf 1$ \\
$t,u,h,\ldots$ & continuous time parameter & $\mathsf t$ \\
$\Dt$ & sample interval & $\mathsf t$ \\
$\lambda$ & ordinary frequency & $\mathsf t^{-1}$ \\
$\omega$ & angular frequency & $\mathsf{rad.}$ \\
$\bX_k$, $\bX(t)$ & stochastic process & $\mathsf x$ \\
$\Sigma$ & residual noise intensity & $\mathsf x^2 \mathsf t^{-1}$ \\
$\MSE_m$, $\MSE(h)$ & mean-square prediction error (MSPE) & $\mathsf x^2$ \\
$\Gamma_k$, $\Gamma(t)$ & autocovariance sequence/funtion & $\mathsf x^2$ \\
$H(\lambda)$ & transfer function & $\mathsf t$ \\
$S(\lambda)$ & cross-power spectral density (CPSD) & $\mathsf x^2 \mathsf t$ \\
$\gc\bY\bX$ & Granger causality (time domain) & $\mathsf b$ \\
$\sgc\bY\bX(\lambda)$ & Granger causality (frequency domain) & $\mathsf b$ \\
\hline
\end{tabular}
\caption{Notation and dimensions: $\mathsf t$ denotes units of time (\eg, ms, so that $\mathsf t^{-1}$ is measured in kHz), $\mathsf x$ denotes the units of the neural signal under consideration (\eg, volts, tesla, \etc) and $\mathsf b$ denotes the unit of information (bits or nats, depending on whether base 2 or natural logarithms are used).} \label{tab:quants}
\end{table}
Although this convention may appear cumbersome, particularly in our analysis of discrete-time processes (\secref{sec:varGC}), the payoff is a more harmonious and intuitive tie-in with the continuous-time case and subsampling analysis (\secref{sec:voudl}).

Throughout, superscript ``$\transop$'' denotes matrix transpose, superscript ``$^*$'' matrix conjugate transpose and $|\cdot|$ the determinant of a square matrix. A dot over a symbol denotes differentiation with respect to a continuous time parameter.

\section{Discrete VAR processes and Granger Causality} \label{sec:varGC}

We briefly outline the VAR theory that we shall require; the reader is referred to standard texts \citep{Hamilton:1994,Lutkepohl:2005} for details. Let $\bX \equiv \{\bX_k \,|\, k \in \integers\}$ be a discrete-time, purely nondeterministic, zero-mean, wide-sense stationary vector process. With a view to Granger-causal analysis, we assume the same conditions on the process $\bX$ as in \cite{Geweke:1982}. By Wold's Theorem \citep{Wold:1938,Lutkepohl:2005} $\bX$ has a \emph{vector moving-average} (VMA) representation
\begin{equation}
	\bX_k = \sum_{\ell=0}^\infty B_\ell \beps_{k-\ell} \label{eq:vma}
\end{equation}
with square-summable coefficient matrices $B_\ell$ ($B_0 = I$), and $\beps$ a white-noise process. Considering $z$ as the \emph{lag operator} $z \cdot \bX_k = \bX_{k-1}$, we can write \eqref{eq:vma} in compact form as
\begin{equation}
	\bX_k = \Psi(z) \cdot \beps_k \label{eq:vmaconv}
\end{equation}
where the MA operator
\begin{equation}
	\Psi(z) \equiv \sum_{\ell = 0}^\infty B_\ell z^\ell \label{eq:vmagen}
\end{equation}
has the minimum-phase property that $|\Psi(z)| \ne 0$ for complex $z$ on the unit disc $|z| \le 1$. We shall require that the VMA representation \eqref{eq:vma} may be inverted to yield a \emph{vector autoregressive} (VAR) representation
\begin{equation}
	\bX_k = \sum_{\ell=1}^\infty A_\ell \bX_{k-\ell} + \beps_k \label{eq:var}
\end{equation}
with square-summable coefficient matrices $A_\ell$. \cite{Geweke:1982} supplies a condition on the spectrum of $\bX$ (see below)---that it is bounded away from zero almost everywhere---which suffices for invertibility of the VMA representation \citep{Rozanov:1967}. The condition \citep[see][eq.~2.4]{Geweke:1982}, which we assume here, also guarantees that any vector \emph{sub}process of $\bX$ has a VAR representation. We may write the VAR \eqref{eq:var} as
\begin{equation}
	\Phi(z) \cdot \bX_k = \beps_k \label{eq:varconv}
\end{equation}
with
\begin{equation}
	\Phi(z) \equiv I - \sum_{\ell = 1}^\infty A_\ell z^\ell \label{eq:vargen}
\end{equation}
Stability of the process requires that  $|\Phi(z)| \ne 0$ on $|z| \le 1$, and we have $\Phi(z) = \Psi(z)^{-1}$ on $|z| \le 1$. Henceforth, by ``VAR process'' we mean a process satisfying all of the above conditions.

In accordance with the conventions described in the Introduction, we assume a sample time step $\Dt$ is always given, and parametrise the magnitude of the residual noise $\beps$ by its \emph{intensity} $\Sigma \equiv \Dt^{-1} \covs{\beps_k}$. This is consistent with the additivity of variance, and ensures that the dimensions of $\Sigma$ are consistent with the corresponding quantity in the continuous-time processes we shall encounter later (\secref{sec:voudl}).

The \emph{autocovariance sequence} $\Gamma \equiv \{\Gamma_k \,|\, k \in \integers\}$ is given by
\begin{equation}
	\Gamma_k \equiv \Cov{\bX_{k+\ell}}{\bX_\ell} \label{eq:acov}
\end{equation}
(by stationarity, this does not depend on $\ell$) and satisfies $\Gamma_{-k} = \trans{\Gamma_k}$, and the \emph{Yule-Walker equations}
\begin{equation}
	\Gamma_k = \sum_{\ell = 1}^\infty A_\ell\Gamma_{k-\ell} + \delta_{k0} \Dt\Sigma \qquad k \ge 0 \label{eq:yw}
\end{equation}
In terms of the VMA coefficients, it is straightforward to show that
\begin{equation}
	\Gamma_k = \Dt \sum_{\ell = 0}^\infty B_{k+\ell} \Sigma \trans{B_\ell} \qquad k \ge 0 \label{eq:ywma}
\end{equation}

The \emph{cross-power spectral density} (CPSD) $S$ for $\bX$ is defined for $-\infty < \lambda < \infty$ by
\begin{equation}
	S(\lambda) \equiv \lim_{K \to \infty} \frac1{2K\Dt} \Covs{\hbX_K(\lambda)} \label{eq:WKT}
\end{equation}
where $\hbX_K(\lambda) \equiv \Dt \sum_{k = -K}^K \bX_k e^{-2\pi i \Dt\lambda k}$ is the truncated Fourier transform of the process $\bX$. The $S(\lambda)$ are Hermitian matrices and the \emph{Wiener-Khintchine Theorem} \citep{Wiener:1930,Khintchine:1934} states that:
\begin{equation}
	S(\lambda) = \widehat\Gamma(\lambda) \label{eq:cpsd}
\end{equation}
at all frequencies; \ie\ the CPSD is the Fourier transform of the autocovariance sequence.

The \emph{transfer function} for the VAR \eqref{eq:var} is defined to be the Fourier transform of the MA coefficients:
\begin{equation}
	 H(\lambda) \equiv \widehat B(\lambda) = \Dt\Psi\bracr{e^{-2\pi i \Dt\lambda}} = \Dt \sum_{\ell = 0}^\infty B_\ell e^{-2\pi i \Dt\lambda \ell} \label{eq:trfun}
\end{equation}
which may also be written as
\begin{equation}
	 H(\lambda) = \Dt\Phi\bracr{e^{-2\pi i \Dt\lambda \ell}}^{-1} = \Dt \bracr{I - \sum_{\ell = 1}^\infty A_\ell e^{-2\pi i \Dt\lambda \ell}}^{-1} \label{eq:trfun1}
\end{equation}
We then have the  \emph{spectral factorisation} formula \citep{Masani:1966}
\begin{equation}
	S(\lambda) = H(\lambda) \Sigma \ctran{H(\lambda)} \label{eq:specfac}
\end{equation}
which holds for all $\lambda$. A classical result states that, given a CPSD $S(\lambda)$ satisfying certain regularity conditions \citep{Masani:1966,Wilson:1972}, there exists a unique $\Psi(z)$ holomorphic on the disc $|z| \le 1$ with $\Psi(0) = I$, and a unique positive-definite symmetric matrix $\Sigma$, such that setting  $H(\lambda) = \Dt\Psi\big(e^{-2i\pi\Dt\lambda}\big)$, \eqref{eq:specfac} is satisfied. In other words, for a class of CPSDs the spectral factorisation \eqref{eq:specfac} is \emph{uniquely solvable} for $H(\lambda)$ and $\Sigma$, and hence parameters for a VAR model with the given CPSD may be obtained. Although this result is not constructive---there is no known algorithm for analytic factorisation of an arbitrary CPSD---in specific cases, in particular for \emph{rational} spectral densities \citep{Kucera:1991}], it is frequently feasible; see \secref{sec:ouminlag:subsamp} and \apxref{sec:minoulagdsgc} for a concrete, nontrivial example.

A VAR of the form \eqref{eq:var} is equivalently specified by the VAR parameters $(A,\Sigma)$, the autocovariance sequence $\Gamma$ or the CPSD $S$. The Yule-Walker equations \eqref{eq:yw}, Wiener-Khintchine Theorem \eqref{eq:cpsd} and spectral factorisation formula \eqref{eq:specfac} establish reciprocal relationships between the respective representations. \cite{Barnett:mvgc:2014} exploit these relationships to design efficient computational pathways for the numerical computation of Granger causalities (\secref{sec:gc} below). Analytically, we are also free to choose the representation appropriate to the task at hand.

\subsection{Granger Causality} \label{sec:gc}

Granger causality is most commonly framed in terms of \emph{prediction}. Usually, only $1$-step-ahead prediction is considered. Here, for reasons that will become clear (\secref{sec:gcouc}), we consider Granger causality at arbitrary prediction horizons \citep{Lutkepohl:1993,DufourRenault:1998}. The optimal (in the least-squares sense) $m$-step-ahead prediction ($m$ = 1,2,\ldots) of the stable VAR \eqref{eq:var} based on all information contained in its own (infinite) past---\ie, the optimal prediction of $\bX_{k+m}$ given the history $\bX^-_k \equiv \{\ldots,\bX_{k-2},\bX_{k-1}, \bX_k\}$ of the process up to and including the $k$th step---is given by the orthogonal projection $\cexpect{\bX_{k+m}}{\bX^-_k}$ of $\bX_{k+m}$ onto $\bX^-_k$. A standard result \citep{Hamilton:1994} states that
\begin{equation}
	\cexpect{\bX_{k+m}}{\bX^-_k} = \sum_{\ell=m}^\infty B_\ell \beps_{k+m-\ell} \label{eq:varpred}
\end{equation}
It follows that the \emph{mean-square prediction error} (MSPE) at a prediction horizon of time $m\Dt$ into the future, is given by
\begin{equation}
	\MSE_m \equiv \covs{\cexpect{\bX_{k+m}}{\bX^-_k} - \bX_{k+m}}  = \Dt \sum_{\ell=0}^{m-1} B_\ell \Sigma \trans{B_\ell} \label{eq:varpredmse}
\end{equation}
In particular, $\MSE_1$ is just the residual noise covariance $\Dt\Sigma$, while from \eqref{eq:ywma}, $\MSE_m \to \Gamma_0$ as $m \to \infty$; it also makes sense to define $\MSE_0 \equiv 0$, as prediction at zero horizon is exact.

Our exposition of Granger causality follows in spirit the standard formulation of \cite{Geweke:1982}. Suppose that we have two jointly distributed discrete-time vector stochastic processes (``variables'') $\bX,\bY$ so that the joint process $\trans{[\trans\bX \trans\bY]}$ is a VAR of the form \eqref{eq:var}. By assumption, the subprocesses $\bX$ and $\bY$ also have VAR representations. We may then, for a given prediction horizon $m\Dt$ ($m = 1,2,\ldots$), compare the MSPE $\MSE_{m,xx}$ of the prediction $\cexpect{\bX_{k+m}}{\bX^-_k,\bY^-_k}$ of $X_{k+m}$ based on the \emph{joint} history of $\bX$ and $\bY$ (the ``full regression''), with the MSPE $\MSE'_{m,xx}$ of the prediction $\cexpect{\bX_{k+m}}{\bX^-_k}$  of $X_{k+m}$ based only on the \emph{self}-history of the subprocess $\bX$ (the ``reduced regression''\footnote{We generally indicate quantities associated with the reduced, as opposed to full regression, with a prime.}). If inclusion of the history $\bY^-_k$ improves the prediction of $\bX_{k+m}$, then we say that $\bY$ (the ``source'' variable) \emph{Granger-causes} $\bX$ (the ``target'' variable) at prediction horizon $m\Dt$. \cite{Geweke:1982} proposed that prediction be quantified by \emph{generalised variance}\footnote{For a discussion on the preferability of the generalised variance $\dett\Sigma$ over the \emph{total variance} $\trace\Sigma$, see \cite{Barrett:2010}; see also the maximum likelihood interpretation outlined in \apxref{sec:statinf}.} \citep{Wilks:1932}---that is, the determinant of the MSPE---leading to the definition
\begin{equation}
	\gc\bY{\bX,m} \equiv \log\frac{\dett{\MSE'_{m,xx}}}{\dett{\MSE_{m,xx}}} \label{eq:hgc}
\end{equation}
$\gc\bX{\bY,m}$ is defined symmetrically.

If $m = 1$ ($1$-step prediction), we drop the $m$ subscript. Note that $\MSE_{1,xx} = \Dt \Sigma_{xx}$, where $\Sigma_{xx}$ is the $xx$ component of the noise intensity $\Sigma$ of the joint process, while $\MSE'_{1,xx} = \Dt\Sigma'_{xx}$ where $\Sigma'_{xx}$ is the noise intensity of the subprocess $\bX$, considered as a VAR. Thus we obtain the standard $1$-step Geweke measure
\begin{equation}
	\gc\bY\bX \equiv \gc\bY{\bX,1} = \log\frac{\dett{\Sigma'_{xx}}}{\dett{\Sigma_{xx}}} \label{eq:gc}
\end{equation}

$\gc\bY{\bX,m} \ge 0$ always, since inclusion of the history $\bY^-_k$ in the full regression can only decrease the prediction error. It may be shown, furthermore \citep{Sims:1972,Caines:1976}, that $\gc\bY\bX = 0$ iff $\Psi_{xy}(z) \equiv 0$. But from \eqref{eq:vmaconv} it follows that $\Psi_{xy}(z) \equiv 0\implies \Psi'_{xx}(z) = \Psi_{xx}(z)$, and, since all the $B_\ell$ are lower block-triangular, from \eqref{eq:varpredmse} we have $\MSE'_{m,xx} = \MSE_{m,xx}$ for any $m$. Thus we may state:
\begin{equation}
	\gc\bY\bX = 0 \iff \Psi_{xy}(z) \equiv 0 \iff \gc\bY{\bX,m} = 0 \; \forall m > 0 \label{eq:gcvanish}
\end{equation}
That is, \emph{vanishing $1$-step GC implies vanishing GC at \emph{any} prediction horizon}. The converse does not hold, though: $\gc\bY{\bX,m}$ may vanish for $m > 1$ even if $\gc\bY\bX > 0$ (\cf~\apxref{sec:nofinv}). In general, $\gc\bY{\bX,m}$ will depend on the prediction horizon. Since both $\MSE_{m,xx}$ and $\MSE'_{m,xx} \to \Gamma_{0,xx}$ as $m \to \infty$, $\gc\bY{\bX,m} \to 0$ as $m \to \infty$, so that $\gc\bY{\bX,m}$ attains a maximum at some finite value(s) of $m$.

At this point we note, as alluded to in the Introduction (\secref{sec:intro}), that Granger causality has a clear information-theoretic interpretation: \cite{Barnett:tegc:2009} show that for Gaussian processes, Granger causality is entirely equivalent to the non-parametric information-theoretic \emph{transfer entropy} measure \citep{Schreiber:2000,Palus:2001}, and for general Markovian processes (under a mild ergodicity assumption) the log-likelihood ratio statistic for the Markov model [\cf\ \eqref{eq:egc}] converges in the large-sample limit to the corresponding transfer entropy \citep{Barnett:teml:2012}. TE, as a conditional mutual information---and by extension GC---is naturally measured in units of bits (or nats, if natural logarithms are used).

Regarding Granger's other requirement for causal effect, that the information that $\bY$ contains about (the future of) $\bX$ be \emph{unique}, here we note just that the effect on Granger causalities of other (accessible) variables jointly distributed with $\bX$ and $\bY$ may be discounted by including them in both the full and reduced predictor sets\footnote{Of course this is only possible for \emph{accessible} variables - inaccessible (hidden, latent) influences are in general problematic for causal analysis in a broader sense \citep{ValdesSosaEtal:2011}.}. This leads to the definition of \emph{conditional} Granger causality \citep{Geweke:1984}. While all Granger causalities discussed in this paper have conditional counterparts, here we restrict our attention to the unconditional case.

\cite{Geweke:1982} refers to the Granger causality $\gc\bY\bX$ as the ``linear feedback'' from $\bY$ to $\bX$, and goes on to define the ``instantaneous feedback'' or \emph{instantaneous causality}:
\begin{equation}
	\igc\bX\bY \equiv \log\frac{\dett{\Sigma_{xx}} \dett{\Sigma_{yy}}}{\dett\Sigma} \label{eq:igc}
\end{equation}
which vanishes iff the residuals $\beps_{x,k},\beps_{y,k}$ are contemporaneously uncorrelated. \cite{Solo:2007} distinguishes ``strong'' Granger causality from the conventional (``weak'') variety, noting that only (the existence of) strong causality is strictly preserved under subsampling. In a VAR framework, strong causality from $\bY \to \bX$ replaces the full ($1$-step) predictor set $X^-_k, \bY^-_k$ with the predictor set $X^-_k, \bY^-_{k+1}$ [\cf\ \citet[eq.~2.9]{Geweke:1982}]; that is, the contemporaneous source term $\bY_{k+1}$ is included in the full predictor set (the reduced predictor set remains unaltered). The residual errors of the strong least-squares prediction are $\beps_{x,k} - \Sigma_{xy} \Sigma^{-1}_{yy} \beps_{y,k}$ \citep{Geweke:1982}, so that the MSPE is $\Dt$ $\times$ the \emph{partial} residual noise intensity matrix
\begin{equation}
	\Sigma_{xx|y} \equiv \Sigma_{xx} - \Sigma_{xy} \Sigma_{yy}^{-1} \Sigma_{yx} \label{eq:rescovpar}
\end{equation}
This leads to the statistic
\begin{equation}
	\ggc\bY\bX \equiv \log\frac{\dett{\,\Sigma'_{xx}\;}}{\dett{\Sigma_{xx|y}}} = \gc\bY\bX + \igc\bX\bY \label{eq:ggc}
\end{equation}
(the last equality follows from block-decomposition of the determinant $|\Sigma|$). Strong GC, while invariant under subsampling is, however, unsatisfactory as a \emph{directional} measure, since it is not generally possible to disentangle the directional and instantaneous contributions.

Although not the focus of this paper, a significant feature of Granger-causal analysis is that (time domain) GC may be decomposed in a natural way by frequency. The resulting frequency-domain, or \emph{spectral} Granger causality integrates to the time-domain GC \eqref{eq:gc}. For a full derivation and discussion we refer to \cite{Geweke:1982}; here we just present the definition of the spectral GC from $\bY$ to $\bX$:
\begin{equation}
    \sgc\bY\bX(\lambda) \equiv \log{\frac{\dett{S_{xx}(\lambda)}}{\dett{S_{xx}(\lambda) - H_{xy}(\lambda) \Sigma_{yy|x} H_{xy}(\lambda)^*}}} \label{eq:sgc}
\end{equation}
where $\Sigma_{yy|x} \equiv \Sigma_{yy} - \Sigma_{yx} \Sigma_{xx}^{-1} \Sigma_{xy}$ [\cf\ \eqref{eq:rescovpar}]. $\sgc\bY\bX(\lambda)$ is always nonnegative, and Geweke's fundamental spectral decomposition of Granger causality applies\footnote{Strictly speaking, equality in \eqref{eq:gcint} holds provided the condition $\dett{A_{yy}(z) - \Sigma_{yx} \Sigma_{xx}^{-1} A_{xy}(z)} \ne 0$ is satisfied for all $z$ on the unit disc $|z| \le 1$; otherwise it should be replaced by $\le$. In practice, according to \cite{Geweke:1982}, the equality condition is ``almost always'' satisfied.}
\begin{equation}
    \gc\bY\bX = \Dt \int_{-\frac1{2\Dt}}^{\frac1{2\Dt}} \sgc\bY\bX(\lambda) \,d\lambda \label{eq:gcint}
\end{equation}
The spectral GC \eqref{eq:sgc} is, at any specific frequency $\lambda$, also a quantity of information measured in bits or nats, and \eqref{eq:gcint} presents time-domain GC as an average over all frequencies of spectral GC. We remark that a spectral counterpart for $\igc\bX\bY$ has been defined \citep{DingEtal:2006}, but is somewhat unsatisfactory insofar as it may become negative at some frequencies and lacks a compelling physical interpretation.

It is known \citep{Geweke:1982,Barnett:gcfilt:2011,Solo:2016} that 1-step Granger causality, in both time and frequency domains, is invariant under (almost) arbitrary causal, invertible (stable, minimum-phase\footnote{Minimum phase requires that the inverse filter also be stable; note that in \cite{Barnett:gcfilt:2011} this requirement is erroneously overlooked [thanks to Victor Solo (personal communication) for bringing this to our attention].}) filtering; see \apxref{sec:finv} for more detail. However, as demonstrated in \apxref{sec:nofinv}, invariance does \emph{not} extend to $m$-step GC for $m > 1$, unless $\Psi_{xy}(z) \equiv 0$ - equivalently, $\gc\bY\bX = 0$. In that case filter-invariance does hold for $m \ge 1$; that is, causal, invertible filtering will not induce \emph{spurious} Granger causalities at any prediction horizon.

VAR modelling is particularly suited to data-driven approaches to functional analysis (\apxref{sec:statinf}) and its applicability quite general. By the Wold decomposition theorem \citep{Hannan:1970}, any covariance-stationary stochastic process in discrete time has a moving-average (MA) representation. Further spectral conditions may be imposed so that the Wold MA representation may be inverted to yield a stable VAR representation \eqref{eq:var} \citep{Rozanov:1967}. We assume that these conditions apply for all discrete stochastic processes encountered in this study. We note that if there is nonlinear (delayed) feedback in the generative process, while this does not preclude VAR-based estimation of Granger causalities (provided the VAR representation criteria mentioned above hold), a linear model will not be parsimonious and transfer entropy (or a suitable nonlinear model-based version of Granger causality) may be preferable \citep{Barnett:teml:2012}. In general, though, VAR-based Granger causality has the advantages of simplicity, ease of estimation, a known sampling distribution and a natural spectral decomposition.

More recently, a theory of Granger causality has been developed for \emph{state-space} processes \citep{Barnett:ssgc:2015,Solo:2016}. The state-space approach offers some significant advantages from modelling, estimation and computational perspectives. This, as well as estimation, statistical inference and detection of (discrete-time) Granger causality from empirical time series data, is discussed in
\apxref{sec:statinf}.

\section{Distributed-lag vector autoregressive processes in continuous time} \label{sec:voudl}

Following the discussion in \secref{sec:intro} regarding the essentially continuous-time nature of biophysiological processes, in order to address the impact of subsampling we require appropriate continuous-time generative processes for which Granger causality may be defined. Accordingly, we start with an underlying analogue neurophysiological process $\bU(t)$ in continuous time\footnote{The unit of time is taken to be the same as for discrete-time processes.} $t$ and an \emph{observation function} $\xi(\cdot)$. The observed (multivariate) signal $\bX(t) = \xi(\bU(t))$ is then sampled at regular discrete time intervals. $\bU(t)$ may be stochastic (endogenous noise), as may be the observation function (exogneous, measurement noise), so that $\bX(t)$ is considered a continuous-time stochastic process. Our approach is to assume that $\bX(t)$ admits a continuous-time linear autoregressive representation.

The standard multivariate linear autoregressive model in continuous time is the vector Ornstein-Uhlenbeck (VOU) process \citep{OrnsteinUhlenbeck:1930,Doob:1953} defined by a linear stochastic differential equation (SDE)
\begin{equation}
	d\bX(t) = A \bX(t)\,dt + d\bW(t) \label{eq:vou}
\end{equation}
where $\bW(t)$ is a vector Wiener process. The process \eqref{eq:vou} must, however, be considered implausible as a model for an observed neurophysiological processes, since it fails to model delayed feedback at \emph{finite} time scales. To address this we generalise the VOU process to the CTVAR (continuous-time vector autoregressive) process described below.

Our construction closely mirrors that of the discrete-time VAR case (\secref{sec:varGC}). Thus we assume that the continuous-time, wide-sense stationary, stable, minimum-phase (and zero-mean) vector process $\bX \equiv \{\bX(t) \,|\, t \in \reals\}$ admits a moving-average representation \citep{CainesChan:1975,ComteRenault:1996}
\begin{equation}
	\bX(t) = \int_{u = 0}^\infty B(u) \,d\bW(t-u) \label{eq:ctvarma}
\end{equation}
where $\bW(t)$ is again a vector Wiener process\footnote{This might be generalised to continuous-time white noise processes as defined for the continuous-time Wold decomposition theorem \citep{Rozanov:1967}.}, the MA kernel $B(u)$ [with $B(0) = I$] is square-integrable and the integral is to be interpreted as an \emph{It\=o integral} \citep{Oksendal:2003}. In continuous time we define the lag operator $\zeta$ as follows: suppose that a complex-valued function $\cL(\zeta)$ may be written (uniquely) as a Laplace transform $\cL(\zeta) = \int_0^\infty L(u) \,e^{-\zeta u} \,du$. Then for a continuous-time process $\bU(t)$, we define $\cL(\zeta) \cdot \bU(t)$ as the It\=o integral $\int_{u = 0}^\infty L(u) \,d\bU(t-u)$,
and \eqref{eq:ctvarma} may be written as
\begin{equation}
	\bX(t) = \Psi(\zeta) \cdot \bW(t) \label{eq:cvmaconv}
\end{equation}
where
\begin{equation}
	\Psi(\zeta) \equiv \int_0^\infty B(u) e^{-\zeta u} \,du \label{eq:cvmagen}
\end{equation}
The minimum-phase property requires that $\dett{\Psi(\zeta)} \ne 0$ on the right half-plane $\re(\zeta) \ge 0$.

As in the discrete-time case, we assume that the MA representation \eqref{eq:ctvarma} may be inverted to yield a continuous-time vector autoregressive (CTVAR) representation as a stochastic linear integro-differential equation \citep{ComteRenault:1996}
\begin{equation}
	d\bX(t) = \int_0^\infty A(u) \bX(t-u)\,du\,dt + d\bW(t) \label{eq:ctvar}
\end{equation}
or
\begin{equation}
	\Phi(\zeta) \cdot \bX(t) = \bW(t) \label{eq:cvarconv}
\end{equation}
with
\begin{equation}
	\Phi(\zeta) \equiv \zeta I - \int_0^\infty A(u) e^{-\zeta u} \,du \label{eq:cvargen}
\end{equation}
To verify \eqreff{eq:cvarconv}{eq:cvargen}, note that $\Phi(\zeta)$ may be written as the Laplace transform of $\dot\delta(u)I - A(u)$,
where $\dot\delta(u)$ denotes the derivative, in the generalised function sense, of the delta function $\delta(u)$; \eqref{eq:cvarconv} then follows from the relation $\int_{-\infty}^\infty \dot\delta(u) \,\varphi(t-u) \,du = \dot\varphi(t)$ for any function $\varphi(u)$ \citep{FriedlanderJoshi:1998}. Stability requires that $\dett{\Phi(\zeta)} \ne 0$ on the right half-plane $\re(\zeta) \ge 0$, and in \apxref{sec:ctgenfun} we prove that $\Psi(\zeta) = \Phi(\zeta)^{-1}$ on the right half-plane $\re(\zeta) \ge 0$

The AR kernel $A(u)$ specifies causal, time-lagged coupling between nodes over a range of feedback delays $u$, while $d\bW(t)$ represents continuous-time white noise with (positive-definite) covariance matrix $\Sigma \,dt$, so that $\Sigma$ again represents residual noise \emph{intensity}. We assume $A(u)$ be be \emph{square-integrable}\footnote{This condition may be unnecessarily restrictive; we require at least that the CPSD of $\bX(t)$ (see below) exists \citep{Lighthill1958}.} and allow it to be a \emph{generalised function} \citep{FriedlanderJoshi:1998}, so it might, for example, include delta functions. The integral over $u$ in \eqref{eq:ctvar} is then taken to be a Lebesgue integral. Note that the VOU process \eqref{eq:vou} is a special case of \eqref{eq:ctvar} with $A(u) = A \delta(u)$ a delta function at ``infinitesimal lag'' $u = 0$. Analagous to the discrete-time case, we also assume that any vector \emph{sub}-process of $\bX(t)$ may be represented as a CTVAR\footnote{It seems plausible, although we have not established this rigorously, that this may follow from a similar boundedness condition on the CPSD to that described in \cite{Geweke:1982}.}.

Stochastic integro-differential equations similar to \eqref{eq:ctvar} have been studied \textit{in abstracto}, as models for various physical, engineering and biological phenomena, and (more along the present lines) in the econometrics literature \citep{Sims:1971,Geweke:1978,McCrorieChambers:2006}. They have not however, as far as we are aware, been deployed previously in the neurosciences. Our approach most closely resembles the ``CIMA'' processes presented in \cite{ComteRenault:1996}; our emphasis, however, is more on the autoregressive and (as we shall see later) \emph{predictive} aspects of the model.

In \apxref{sec:ctma} we show that the MA kernel $B(u)$ satisfies
\begin{subequations}
\begin{align}
	\dot B(u) &= \int_0^u A(s) B(u-s) \,ds \qquad u \ge 0 \label{eq:makern1} \\
	B(0)  &= I \label{eq:makern0}
\end{align} \label{eq:makern}%
\end{subequations}
where $\dot B(u)$ denotes differentiation from the right\footnote{We shall generally assume that appropriate derivatives exist wherever they appear in a formula.} with respect to $t$.

The autocovariance function of a stationary continuous-time vector stochastic process is defined as
\begin{equation}
	\Gamma(t) \equiv \cov{\bX(t+u)}{\bX(u)} \label{eq:cacov}
\end{equation}
which again, by stationarity, does not depend on $u$, and $\Gamma(-t) = \trans{\Gamma(t)}$. From the MA representation \eqref{eq:cvmaconv} an application of the \emph{It\=o isometry} \citep{Oksendal:2003} yields
\begin{equation}
	\Gamma(t) = \int_0^\infty B(t+u) \Sigma \trans{B(u)} \,du \qquad t \ge 0 \label{eq:cywma}
\end{equation}
From \eqref{eq:cywma} we may derive the continuous-time Yule-Walker equations for the process \eqref{eq:ctvar}
\begin{subequations}
\begin{align}
	\dot\Gamma(t) &= \int_0^\infty A(u) \Gamma(t-u) \,du \qquad t > 0 \label{eq:cyw1} \\
	\dot\Gamma(0) + \trans{\dot\Gamma(0)} &= -\Sigma \label{eq:cyw0}
\end{align} \label{eq:cyw}%
\end{subequations}
where $\dot\Gamma(t)$ denotes differentiation from the right.

Analogous to the discrete-time case, the CPSD for the process is defined by
\begin{equation}
	S(\lambda) \equiv \lim_{T \to \infty} \frac1{2T} \Covs{\hbX_T(\lambda)} \label{eq:cWKT}
\end{equation}
on $-\infty < \lambda < \infty$, now with $\hbX_T(\lambda) \equiv \int_{-T}^T \bX(t) e^{-2\pi i \lambda t} \,dt$ [\cf~\eqref{eq:WKT}], and the Wiener-Kintchine Theorem in continuous time again reads:
\begin{equation}
	S(\lambda) = \widehat\Gamma(\lambda) \label{eq:ccpsd}
\end{equation}
The continuous-time transfer function is again defined as
\begin{equation}
	 H(\lambda) \equiv \widehat B(\lambda) = \Psi(2\pi i\lambda) = \int_0^\infty B(u) e^{-2\pi i\lambda u} \,du \label{eq:ctfun}
\end{equation}
which may also be written as
\begin{equation}
	 H(\lambda) = \Phi(2\pi i\lambda)^{-1} = \bracr{2\pi i\lambda I -  \int_0^\infty A(u) e^{-2\pi i\lambda u} \,du}^{-1} \label{eq:ctfun1}
\end{equation}
and from \eqref{eq:cywma} it is not hard to establish the continuous-time spectral factorisation\footnote{This follows from the continuous-time versions of the Wiener-Kintchine and Convolution theorems, noting that \eqref{eq:cywma} may be written as $\Gamma(t) = (\sB * \trans\sB)(t)$ where $\sB(u) \equiv B(u)L$, with $L$ a matrix square root of $\Sigma$ satisfying $L\trans L = \Sigma$ (by positive-definiteness, such an $L$ exists).}
\begin{equation}
	S(\lambda) = H(\lambda) \Sigma \ctran{H(\lambda)} \label{eq:cspecfac}
\end{equation}
We note that $H(\lambda)$ satisfies
\begin{equation}
	\lim_{|\lambda| \to \infty} 2\pi i\lambda \, H(\lambda) = I \label{eq:hlim}
\end{equation}
so that from \eqref{eq:cspecfac}
\begin{equation}
	\lim_{|\lambda| \to \infty} 4\pi^2\lambda^2 \,S(\lambda) = \Sigma \label{eq:slim}
\end{equation}
\ie, $S(\lambda)$ decays as $\lambda^{-2}$, as $|\lambda| \to \infty$. We conjecture that, analagous to the discrete-time case, given a continuous-time CPSD $S(\lambda)$ satisfying suitable regularity conditions, there exists a unique $\Psi(\zeta)$ holomorphic on $\re(\zeta) \ge 0$ with $\lim_{|\omega| \to \infty} \, i\omega\Psi(i\omega) = I$ and positive-definite $\Sigma$, such that \eqref{eq:cspecfac} is satisfied for $H(\lambda) = \Psi(2\pi i\lambda)$ on $-\infty < \lambda < \infty$.

\subsection{Subsampling a CTVAR process} \label{sec:voudlds}

We next examine some properties of the discrete-time processes $\bX(\Dt)$ obtained by subsampling a CTVAR process $\bX$ at fixed time intervals $\Dt$; \ie, $\bX_k(\Dt) \equiv \bX(k\Dt)$. It is these $\Dt$-subsampled processes which stand as models for discretely-sampled neurophysiological recordings of an underlying biophysiological process (\secref{sec:intro}). A subtlety which we must address is that, while a $\Dt$-subsampling of a (stable, minimum-phase) CTVAR is itself always stable, there is no guarantee that it will be minimum-phase for all $\Dt$ - see \eg, \cite{AstromEtal:1984}. We thus assume that the minimum-phase condition for $\Dt$-subsamplings holds as necessary [in worked examples (\cf~\secref{sec:ouminlag:subsamp}) it must be tested explicitly], and that in particular (\cf~\secref{sec:gcouc} below) it holds in the limit of fine subsampling; that is, there exists a sampling interval $\Dt_0$ such that for any $\Dt$-subsampling with $0 < \Dt \le \Dt_0$, $\bX(\Dt)$ is minimum phase.

With a view to calculation of (multistep) Granger causalities (\secref{sec:gc}), we require expressions for the transfer function, residual noise intensity, MA coefficients and CPSD of the $\Dt$-subsampled processes. The crucial observation is that the autocovariance sequence $\Gamma(\Dt)$ of the subsampled process $\bX(\Dt)$ is just $\Gamma_k(\Dt) = \Gamma(k\Dt)$, where $\Gamma$ is the autocovariance function of the original continuous-time process - this follows immediately from \eqref{eq:cacov} and \eqref{eq:acov}. Recall that for calculation of time-domain multistep Granger causalities for a discrete-time process, we require the residual noise intensity matrices and MA coefficients of the process itself and also of subprocesses. In the frequency domain we require, in addition, the transfer function. Analytically, while in principle the $\Dt$-subsampled VAR parameters might be derived from the autocovariance sequence via the discrete-time Yule-Walker equations \eqref{eq:yw} (\cf\ our remarks in \secref{sec:varGC} regarding the multiple representations for a VAR process), in practice this is generally intractable, and it is more convenient to calculate them from the discrete-time CPSD by spectral factorisation\footnote{Solving for $A_k,\Sigma$ from \eqref{eq:yw} involves a matrix deconvolution, which is generally difficult to perform analytically. In the frequency domain, the Convolution Theorem---which underlies the spectral factorisation formula \eqref{eq:specfac}---renders the deconvolution more tractable. In continuous time, however, the integro-differential Yule-Walker equations \eqref{eq:cyw} may well be more tractable (\cf~\secref{sec:ouminlag:ctime}).}.

Given a CTVAR specified by an autoregressive coefficients kernel $A(u)$ and residuals covariance matrix $\Sigma$, a procedure for analytic calculation of multistep Granger causalities for the discrete-time $\Dt$-subsampled process is described in \tabref{tab:gcanaproc}.
\begin{table} \label{tab:gcanaproc}
\fbox{%
\parbox{0.965\columnwidth}{%
\small
\begin{enumerate}
	\item Calculate the continuous-time MA kernel $B$ by direct solution of \eqref{eq:makern}. \label{it:dsproc1d}
	\item Calculate the continuous-time autocovariance function $\Gamma$ as follows: either
	\begin{enumerate}
		\item Calculate the continuous-time transfer function $H$ \eqref{eq:ctfun}. \label{it:dsproc1a}
		\item Calculate the continuous-time CPSD $S$ \eqref{eq:cspecfac}. \label{it:dsproc1b}
		\item Calculate $\Gamma$ by inverse Fourier transform \eqref{eq:ccpsd}. \label{it:dsproc1c}
	\end{enumerate}
	or
	\begin{enumerate}[resume]
		\item Calculate $\Gamma$ by integration \eqref{eq:cywma}. \label{it:dsproc1e}
	\end{enumerate}
	or
	\begin{enumerate}[resume]
		\item Calculate $\Gamma$ by direct solution of the continuous-time Yule-Walker equations \eqref{eq:cyw}. \label{it:dsproc1f}
	\end{enumerate} \label{it:dsproc1}
\item Calculate the discrete-time subsampled process autocovariance sequence $\Gamma(\Dt)$ by $\Gamma_k(\Dt) = \Gamma(k\Dt)$. \label{it:dsproc2}
	\item Calculate the subsampled process CPSD $S(\Dt)$ by discrete-time Fourier transform of $\Gamma(\Dt)$ \eqref{eq:cpsd}. \label{it:dsproc3}
	\item Calculate the subsampled process transfer function $H(\Dt)$ and residuals intensity $\Sigma(\Dt)$ by discrete-time spectral factorisation of $S(\Dt)$, for both full and (time-domain only) reduced models \eqref{eq:specfac}. \label{it:dsproc4}
	\item Time domain ($1$-step): calculate Granger causality from full and reduced subsampled residuals intensities \eqref{eq:gc}.
	\item Time domain ($m$-step):
	\begin{enumerate}[resume]
		\item Calculate subsampled MA coefficients $B_k(\Dt)$ up to $k = m-1$ by inverse Fourier transform of the subsampled transfer function $H(\Dt)$, for both full and reduced models \eqref{eq:trfun}.
		\item Calculate MSPEs $\MSE_m(\Dt)$ from $\Sigma(\Dt)$ and the $B_k(\Dt)$, for both full and reduced models \eqref{eq:varpredmse}.
		\item Calculate subsampled $m$-step Granger causality from the full and reduced MSPEs \eqref{eq:hgc}. \label{it:dsproc6}
	\end{enumerate}
	\item Frequency domain: calculate frequency domain Granger causality from (full model) $\Sigma(\Dt)$, $S(\Dt)$ and $H(\Dt)$ \eqref{eq:sgc}. \label{it:dsproc5}
\end{enumerate}
}%
}%
\caption{Procedure for analytical calculation of multistep time-domain and/or spectral Granger causalities for a $\Dt$-subsampled CTVAR process from known CTVAR parameters $A(u),\Sigma$.}
\end{table}
In \secref{sec:ouminlag:ctime} below we follow precisely this procedure for a non-trivial analytic example.

In \apxref{sec:doulim} we establish firstly an asymptotic expansion for the CPSD of the subsampled process in the limit $\Delta \to 0$
\begin{equation}
	S(\lambda;\Dt) = S(\lambda) + \tfrac1{12}\Dt^2 \Sigma + \tfrac1{720}\Dt^4 (\Omega+12\pi^2\lambda^2\Sigma) + \bigO{\Dt^5} \label{eq:dcpsdlim}
\end{equation}
where $\Omega \equiv \dddot\Gamma(0) + \trans{\dddot\Gamma(0)}$, and also the scaling relations \citep{Zhou:2014}
\begin{subequations}
\begin{align}
	H(\lambda;\Dt) &= H(\lambda)  + \bigO\Dt \label{eq:dtrfunlim} \\
	\Sigma(\Dt) &= \Sigma  + \bigO\Dt , \label{eq:drescovlim}
\end{align} \label{eq:dlim}%
\end{subequations}
while from \eqref{eq:ywma} and \eqref{eq:cywma} we have:
\begin{equation}
	B_k(\Dt) = B(k\Dt) +  \bigO\Dt \label{eq:macoeflim}
\end{equation}

\subsection{Subsampling a VOU process} \label{sec:vouss}

The special case of subsampling a vector Ornstein-Uhlenbeck process, \ie, where $A(u) = A\delta(u)$, may be solved exactly. The Yule-Walker equation \eqref{eq:cyw1} becomes the ordinary differential equation $\dot\Gamma(t) = A\Gamma(t)$, with solution
\begin{equation}
	\Gamma(t) = e^{At} \Gamma(0) \qquad t \ge 0 \label{eq:vougam}
\end{equation}
and from the initial condition \eqref{eq:cyw0}, $\Gamma(0)$ satisfies the continuous-time Lyapunov equation
\begin{equation}
	A \Gamma(0) + \Gamma(0) \trans A = -\Sigma \label{eq:vougam0}
\end{equation}
The autocovariance sequence for the $\Dt$-subsampled process is thus
\begin{equation}
	\Gamma_k(\Dt) = e^{\Dt A k} \,\Gamma(0) \label{eq:voussgam}
\end{equation}
Now it is easily calculated from the discrete-time Yule-Walker equations \eqref{eq:yw} that the discrete-time VAR(1) process $\bX_k = A \bX_{k-1} + \beps_k$ has autocovariance sequence $\Gamma_k = A^k \Gamma_0$, where $\Gamma_0$ satisfies the discrete-time Lyapunov equation $\Gamma_0 - A \Gamma_0 \trans A = \Dt\Sigma$. Since a VAR process is uniquely identified by its autocovariance sequence, we thus find that the subsampled process $\bX(\Dt)$ is VAR(1) (which underlines the unsuitability of VOU processes as models for neurophysiological data). The ($1$-lag) coefficient matrix and residual noise intensity are given respectively by
\begin{subequations}
\begin{align}
	A(\Dt) &= e^{\Dt A} \label{eq:vouss:sscoeff} \\
	\Sigma(\Dt) &= \Dt^{-1} \bracs{\Gamma(0) - e^{\Dt A} \Gamma(0) e^{\Dt \trans A}} \label{eq:vouss:sig}
\end{align}%
\end{subequations}
Note that in general a subprocess of a VOU process will \emph{not} be VOU, nor will a subsampled subprocess be VAR(1).

\subsection{Granger causality for CTVAR processes} \label{sec:gcouc}

It is not immediately clear how we should \emph{define} Granger causality for continuous-time processes in general, and for CTVAR processes (considered as natural continuous-time analogues of VAR processes) in particular. As we shall see, if we attempt to calculate Granger causality at an ``infinitesimal'' prediction horizon, then prediction errors becomes negligible and, in particular, full and reduced prediction errors decay to zero \emph{at the same rate} \citep{RenaultSzafarz:1991,ComteRenault:1996,RenaultEtal:1998}; thus Granger causality vanishes in the infinitesimal horizon limit. This suggests that we consider prediction at \emph{finite} time horizons; that is, a Granger causality measure $\gc\bY\bX(h)$ based on a prediction horizon a finite time $h$ into the future \citep{ComteRenault:1996,FlorensFougere:1996}. We would also like continuous-time GC to be, in a precise sense, the limiting case of discrete-time GC under increasingly fine subsampling.

We are thus lead to consider optimal prediction of $\bX(t+h)$ given the history $\bX^-(t) \equiv \{\bX(s) \,|\, s \le t\}$ of the process $\bX$ up to and including time $t$. The orthogonal projection $\cexpect{\bX(t+h)}{\bX^-(t)}$ may be expressed as the limiting case, as $\Dt \to 0$, of the expectation of $\bX(t+h)$ conditioned on a $\Dt$-subsampling of the history $\bX^-(t)$:
\begin{multline}
	\cexpect{\bX(t+h)}{\bX^-(t)} = \\ \lim_{\Dt \to 0} \cexpect{\bX(t+h)}{\bX(t),\bX(t-\Dt),\bX(t-2\Dt),\ldots}
\end{multline}
Then, setting $\Dt = h/m$, this is just the limit as $m \to \infty$ of the optimal $m$-step prediction of the $h/m$-subsampled process $\bX(h/m)$---recall that by assumption (\secref{sec:voudlds}) it has a stable, minimum-phase VAR representation, at least for large enough $m$---and from \eqref{eq:varpred} and \eqref{eq:macoeflim} we obtain in the limit
\begin{equation}
	\cexpect{\bX(t+h)}{\bX^-(t)} = \int_{u = h}^\infty B(u) \,d\bW(t+h-u)
\end{equation}
An application of the It\=o isometry then yields the continuous-time MSPE
\begin{align}
	\MSE(h) &\equiv \Covs{\cexpect{\bX(t+h)}{\bX^-(t)} - \bX(t+h)} \notag \\
	&= \int_0^h B(u) \Sigma \trans{B(u)} \,du, \qquad h \ge 0 \label{eq:ctpredmse}
\end{align}
Alternatively, we might have defined the continuous-time MSPE at horizon $h$ as the limit of its subsampled counterpart:
\begin{equation}
	\MSE(h) \equiv \lim_{m \to \infty} \MSE_m(h/m), \qquad h \ge 0 \label{eq:ctpredmsedef}
\end{equation}
where $\MSE_m(h/m)$ denotes the $m$-step MSPE of the $h/m$-subsampled process. By \eqref{eq:varpredmse}, \eqref{eq:drescovlim} and \eqref{eq:macoeflim} the definitions coincide. Note that convergence in \eqref{eq:ctpredmsedef} is from above: for fixed $m \ge 1$ and any integer $r > 1$, $\MSE_{rm}(h/rm) \le \MSE_m(h/m)$, since the corresponding orthogonal projections predict the process at the same horizon (\ie, time $h$ into the future), but the $rm$-step prediction is based on a superset of the historic predictor set of the $m$-step prediction. Since $\MSE_m(h/m) \ge 0$ for any $m \ge 1$, the limit \eqref{eq:ctpredmsedef} thus exists. From \eqref{eq:ctpredmse} $\MSE(h)$ satisfies the ordinary differential equations
\begin{subequations}
\begin{align}
	\dot\MSE(h) &= B(h) \Sigma \trans{B(h)} \qquad h \ge 0 \\
	\MSE(0) &= 0
\end{align} \label{eq:msedot}%
\end{subequations}

\begin{figure*}
\begin{center}
\includegraphics{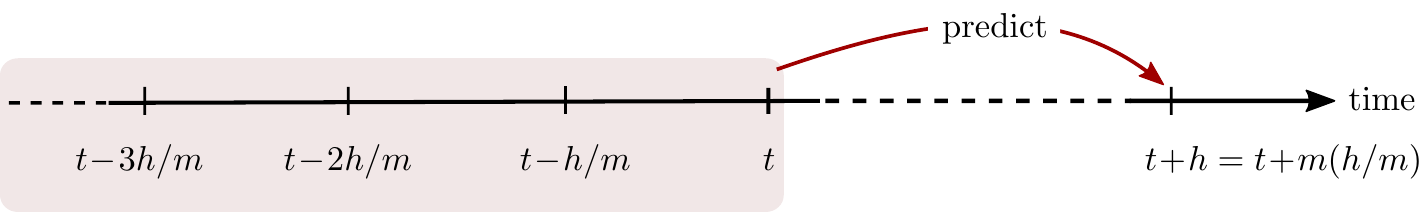}
\end{center}
\caption{Illustration of prediction underlying eq.~\eqref{eq:cgc1}: $\gc\bY\bX(h)$ is the limit of the subsampled discrete-time $m$-step GC $\gc{\bY(h/m)}{\bX(h/m),m}$ at fixed prediction horizon $h = m(h/m)$ under progressively finer subsampling ($m \to \infty$). Note that the historic predictor set $\{\bX(s),\bY(s) \,|\,s = t,t-h/m,t-2h/m,t-3h/m,\ldots\}$ becomes progressively more detailed as $m$ increases, approaching the continuous-time predictor set $\{\bX(s),\bY(s) \,|\, s \le t\}$ as $m \to \infty$.} \label{fig:ctgc}
\end{figure*}

For a joint CTVAR process $\trans{[\trans\bX \trans\bY]}$, we now define Granger causality at horizon $h$ in continuous time analagously to the discrete-time case \eqref{eq:hgc} as
\begin{equation}
	\gc\bY\bX(h) \equiv \log\frac{|\MSE'_{xx}(h)|}{|\MSE_{xx}(h)|}, \qquad h \ge 0 \label{eq:cgc}
\end{equation}
where $\MSE'_{xx}(h)$ denotes the continuous-time MSPE at horizon $h$ for the subprocess $\bX$ (recall that by assumption $\bX$ has a CTVAR representation). From \eqref{eq:hgc} and \eqref{eq:ctpredmsedef} we have
\begin{equation}
	\gc\bY\bX(h) = \lim_{m \to \infty} \gc{\bY(h/m)}{\bX(h/m),m}, \qquad h \ge 0  \label{eq:cgc1}
\end{equation}
so that continuous-time GC may be \emph{defined} as the limit of discrete-time GC under progressively finer subsampling, whilst holding the prediction horizon $h$ fixed (\figref{fig:ctgc}).

Clearly $\gc\bY\bX(h) \ge 0$ always. We now show that $\gc\bY\bX(h) \to 0$ linearly as $h \to 0$  [\cf~\cite{Zhou:2014}]. From \eqref{eq:msedot} we have $\MSE(0) = 0$ and $\dot\MSE(0) = \Sigma$, so that from \eqref{eq:cgc}
\begin{equation}
	\gc\bY\bX(h) = \log\frac{\big|h\Sigma'_{xx} + \bigO{h^2}\big|}{\big|h\Sigma_{xx} + \bigO{h^2}\big|} \label{eq:cgclijm}
\end{equation}
as $h \to 0$. Now the CPSD of $\bX$ may be written in two ways as
\begin{equation}
	S_{xx}(\lambda) = \bracs{H(\lambda) \Sigma H(\lambda)^*}_{xx} = H'_{xx}(\lambda) \Sigma'_{xx} H'_{xx}(\lambda)^* \label{eq:credspecfac}
\end{equation}
where $H'_{xx}(\lambda), \Sigma'_{xx}$ denote respectively the transfer function and residual noise intensity associated with the reduced CTVAR. Multiplying through by $\lambda^2$ and letting $\lambda \to \infty$, from \eqref{eq:hlim} we obtain $\Sigma'_{xx} = \Sigma_{xx}$ and by \eqref{eq:cgclijm} we see that [as noted by \cite{RenaultSzafarz:1991,FlorensFougere:1996} and \cite{ComteRenault:1996}], $\gc\bY\bX(h) \to 0$ as $h \to 0$.

Intuitively, this result may be thought of as follows: insofar as the transfer function $H(\lambda)$ represents the input $\to$ output response of the system, \eqref{eq:hlim} indicates that on short timescales ($\lambda \to 0$), the off-diagonal elements of $H(\lambda)$ (cross-response) decay to zero faster than the on-diagonal elements (self-response). Thus at short predictive time scales, conditional on their past joint history, the variables $\bX,\bY$ effectively decouple.

From \eqref{eq:ctpredmse} and \eqref{eq:cywma}, se see that both $\MSE_{xx}(h)$ and $\MSE'_{xx}(h) \to \Gamma_{xx}(0)$ as $h \to \infty$, so that $\gc\bY\bX(h) \to 0$  as $h \to \infty$. Thus, unless identically zero, $\gc\bY\bX(h)$ will attain a maximum at some finite horizon $0 < h < \infty$.

In contrast to  $\gc\bY\bX(0)$, the \emph{zero-horizon Granger causality rate}\footnote{The Granger-causal concept underlying this quantity has been described in the econometrics literature as ``local causality'' or ``instantaneous causality''. Here we do not use the former term, since ``local'' is more commonly associated with \emph{spatial} rather than \emph{temporal} proximity, nor the latter, to avoid confusion with what \cite{Geweke:1982} terms ``instantaneous feedback'', an entirely distinct concept.}
\begin{equation}
	\rgc\bY\bX \equiv \lim_{h \to 0} \frac1h \gc\bY\bX(h) = \lim_{\Dt \to 0} \frac1\Dt \gc{\bY(\Dt)}{\bX(\Dt)} \label{eq:rgc}
\end{equation}
[the last equality follows from \eqref{eq:cgclijm} and \eqref{eq:drescovlim}] will generally be non-zero. Setting
\begin{equation}
	\DDC \equiv \tfrac12 \ddot\MSE(0) = \tfrac12 \big[\dot B(0) \Sigma + \Sigma \trans{\dot B(0)}\!\big] \label{eq:ddc}
\end{equation}
we may calculate
\begin{equation}
	\rgc\bY\bX = \trace{\Sigma_{xx}^{-1} \bracs{\DDC'_{xx} - \DDC_{xx}}} \label{eq:rgc1}
\end{equation}
where $\DDC'_{xx}$ denotes the corresponding quantity for the reduced CTVAR. $\rgc\bY\bX$ may be considered an information transfer rate, measured in bits or nats per unit time.

While (as for the discrete-time, multistep case) we do not have a workable definition for spectral GC $\sgc\bY\bX(\lambda;h)$ at finite prediction horizon $h$, we define at least the zero-horizon spectral GC in continuous time (again measured in bits or nats) as
\begin{equation}
	\sgc\bY\bX(\lambda;0) \equiv \log\frac{\dett{S_{xx}(\lambda)}}{\dett{S_{xx}(\lambda) - H_{xy}(\lambda) \Sigma_{yy|x} H_{xy}(\lambda)^*}} \label{eq:csgc}
\end{equation}
In contrast to the time-domain GC, spectral GC does not generally vanish at zero prediction horizon; for any $\lambda$, the pointwise limit as $\Dt \to 0$ of the $\Dt$-subsampled spectral GC is equal to $\sgc\bY\bX(\lambda;0)$:
\begin{equation}
	\lim_{\Dt \to 0} \sgc{\bY(\Dt)}{\bX(\Dt)}(\lambda) = \sgc\bY\bX(\lambda;0) \label{eq:csgc1}
\end{equation}
This follows from the discrete-time spectral GC definition \eqref{eq:sgc} via \eqref{eq:dcpsdlim}, \eqref{eq:dtrfunlim} and \eqref{eq:drescovlim}. From \eqref{eq:gcint} we then obtain a spectral decomposition for the continuous-time zero-horizon GC rate:
\begin{equation}
    \rgc\bY\bX = \int_{-\infty}^\infty \sgc\bY\bX(\lambda; 0) \,d\lambda \label{eq:cgcdecomp}
\end{equation}

It is not quite obvious that $\Psi_{xy}(\zeta) \equiv 0 \implies \gc\bY\bX(h) = 0$ for all $h > 0$. This may be seen as follows: $\Psi_{xy}(\zeta) \equiv 0$ implies that the MA kernel $B(u)$ and transfer function $H(\lambda)$ are lower block-triangular. From \eqref{eq:cspecfac} it follows that the CPSD of $\bX$ is given by $S_{xx}(\lambda) = \bracs{H(\lambda) \Sigma H(\lambda)^*}_{xx} = H_{xx}(\lambda) \Sigma_{xx} H_{xx}(\lambda)^*$, so that  [\cf~\eqref{eq:credspecfac}] $\Sigma'_{xx} = \Sigma_{xx}$ and, since the MA kernel is the inverse Fourier transform of the transfer function, $B'_{xx}(u) = B_{xx}(u)$. From \eqref{eq:ctpredmse} and \eqref{eq:cgc} it then follows that $\gc\bY\bX(h) = 0$ for any $h$, and thence that $\rgc\bY\bX = 0$. By a result of \citet[][Prop.~17]{ComteRenault:1996} the converse also holds; that is, $\rgc\bY\bX = 0 \implies \Psi_{xy}(\zeta) \equiv 0$, so that we may state\footnote{$\rgc\bY\bX = 0$ is equivalent to what \citet{ComteRenault:1996} describe as ``local noncausality'', while $\gc\bY\bX(h) = 0 \;\forall h > 0$ corresponds to ``global noncausality''. The former is shown to be equivalent to $\Phi_{xy}(\zeta) \equiv 0$; in the \emph{unconditional} GC case considered here, this is equivalent to $\Psi_{xy}(\zeta) \equiv 0$.  We note also that \cite{CainesChan:1975}, regarding some results which would seem to support this result (at least for rational transfer functions), remark that: ``[\ldots] the definitions and results in this paper are also applicable to continuous time processes'', where by ``continuous time processes'' they refer explicitly to processes of the form \eqref{eq:ctvarma}.}
\begin{equation}
    \rgc\bY\bX = 0 \iff \Psi_{xy}(\zeta) \equiv 0 \iff \gc\bY\bX(h) = 0 \;\forall h > 0 \label{eq:ctgcequiv}
\end{equation}
This result may be considered a continuous-time analogue of \eqref{eq:gcvanish}. In contrast to the discrete-time case, where it is possible that $\gc\bY\bX > 0$ but $\gc\bY{\bX,m} = 0$ for some $m > 1$ (\secref{sec:gc} and \apxref{sec:nofinv}), it is not clear whether we may have $\rgc\bY\bX > 0$ but $\gc\bY\bX(h) = 0$ for some $h > 0$ (we have not found any examples of such behaviour, either analytically or numerically).

In general, there is no reason to suppose that $\gc\bY\bX(h) \equiv 0$ will imply the vanishing of $\gc{\bY(\Dt)}{\bX(\Dt)}$ for a $\Dt$-subsampling; that is \citep{ComteRenault:1996}, \emph{subsampling a CTVAR may induce spurious Granger causality}. We remark that it is non-trivial to verify this phenomenon analytically by example (\cf~\secref{sec:ouminlag} below). Indeed, it is not hard to see that spurious causality cannot occur for a subsampled VOU process \citep[][Prop.~21]{ComteRenault:1996}. In this case (\secref{sec:vouss}), we have $\Psi_{xy}(\zeta) \equiv 0 \iff A_{xy} = 0$, where the VOU AR kernel is $A(u) = A\delta(u)$, and \eqref{eq:vouss:sscoeff} implies immediately that $A_{xy}(\Dt) = 0$ where $A(\Dt)$ is the VAR(1) AR coefficient matrix for the $\Dt$-subsampled process, so that $\gc{\bY(\Dt)}{\bX(\Dt)} = 0$ for any $\Dt$. Furthermore, the analysis of higher-order SDEs in \citet[][Sec.~3]{ComteRenault:1996} would appear to imply that for a $2 \times 1$-dim (bivariate) CTVAR, spurious causality cannot arise (\cf~\secref{sec:ouminlag:subsamp}). In general, it is possible that $\gc{\bY(\Dt)}{\bX(\Dt)} > \gc YX(\Dt)$ for some $\Dt$ values (\cf~\secref{sec:ouminlag:subsamp}, \figref{fig:gc_ds_x}).

Our discussion (\secref{sec:gc}) regarding filter-invariance of GC in discrete time suggests that, for $h > 0$, $\gc\bY\bX(h)$ will not in general be invariant under a continuous-time causal invertible filter $\cG(\zeta) = \int_0^\infty G(u) \,e^{-\zeta u} \,du$ with $\lim_{|\omega| \to \infty} \cG(i\omega) = I$ and $\cG_{xy}(\zeta) \equiv 0$; this is indeed the case - see \secref{sec:ouminlag:ctime} below for an example. As in the discrete-time case, filter-invariance does hold if  $\Psi_{xy}(z) \equiv 0$, so that again causal, invertible filtering will not induce spurious causality at any prediction horizon. It may also be confirmed that filter invariance always holds at zero prediction horizon\footnote{The argument of \apxref{sec:finv} goes through verbatim for $\sgc\bY\bX(\lambda;0)$; then \eqref{eq:cgcdecomp} establishes invariance for $\rgc\bY\bX$. Alternatively, we may take the limiting case $\Dt \to 0$ in \eqref{eq:rgc} under a suitable  discretisation of $\cG(\zeta)$.}; that is, $\rgc\bY\bX$ and $\sgc\bY\bX(\lambda;0)$ are invariant under causal, invertible filtering.

\subsection{Estimation and inference of Granger causalities for subsampled continuous-time data} \label{sec:gcdsest}

In an empirical setting, given neural data in the form of a discrete subsampling of an underlying continuous-time neurophysiological process, our standpoint is that the objective of GC-based functional analysis is to estimate as best we can (and perform statistical inference about) Granger causalities \emph{for the underlying neurophysiological process}. That is, having access only to a $\Dt$-subsampling of a joint continuous-time process $\trans{[\trans\bX \trans\bY]}$, our aim is to estimate as well as possible the continuous-time Granger causalities $\gc\bY\bX(h)$ at prediction horizons of interest, which we regard as reflecting ``true'' directed functional---as distinct from mechanistic \citep{BarrettBarnett:2013}---relationships between the neurophysiological variables at various time scales. This suggests, on the basis of the preceeding analysis, two possible avenues
\begin{enumerate}
\item Estimate a CTVAR for the underlying process from the subsampled data and calculate continuous-time GC directly.
\item Calculate discrete-time GC based on a VAR estimate of the subsampled data.
\end{enumerate}
The second is, of course, the standard route for GC-based functional connectivity analysis. As we have seen, in the limit that the sample increment $\Dt \to 0$, then \emph{on a theoretical level}---\ie, given \emph{exact} models---cases 1 and 2 converge to the same result. With limited data, however, several issues arise: regarding case 1, existing theory for the identification of continuous-time models from subsampled data \citep{Astrom:1969,LarssonEtal:2006,GarnierWang:2008} in general does not extend to distributed-lag multivariate stochastic integro-differential equation models\footnote{Note that the parameter space for a general CTVAR \eqref{eq:ctvar} is infinite-dimensional. Existing theory appears to consider at best continuous-time VARMA models/higher-order SDEs, and does not accommodate distributed lags. A viable CTVAR approach might be to limit the model to a finite number of point lags \citep[see \eg][and references therein]{McKetterickGiuggioli:2014}; but even in that restricted case we are not aware of any useful results on system identification from discretely-sampled data.}. In any case, numerical computation of GCs from (known) CTVAR parameters is likely to be nuch harder than for VARs, where standard algorithms are available (\apxref{sec:statinf}).

In neither case is it clear how the relationship between sample rate and time scales underpinning the underlying neurophysiological process interact with GC inference - the principal theme of this study. In this paper we do not view CTVAR models as a \emph{constructive} model with regard to Granger-causal inference; that is, we reject approach 1 above as a practical alternative (at least under the current theoretical background). Rather, we adopt approach 2 and regard the CTVAR model as an appropriate \emph{analytical} tool for examining the effects of subsampling on Granger-causal inference. This approach is exemplified in the next section, where we solve, analytically, arguably the simplest non-trivial scenario, which demonstrates a basic mode of interaction between neural and sampling time scales.

A further question must also be addressed: given a continuous-time process accessible only via a $\Dt$-subsampling, then in ascertaining \eg, whether we have obtained a spurious causality or failed to detect a non-zero causality what should we consider to be the ``ground truth'' Granger causality? Here we take the pragmatic view that, since our prediction horizon is constrained by the sampling interval $\Dt$, the ``true causality'' is the underlying continuous-time GC at prediction horizon $\Dt$; that is, an empirical estimate $\egc{\bY(\Dt)}{\bX(\Dt)}$ of discrete-time $\Dt$-subsampled GC should be compared against the continuous-time GC $\gc\bY\bX(\Dt)$. This makes sense since $\gc{\bY(\Dt)}{\bX(\Dt)}$ and $\gc\bY\bX(\Dt)$ are both based on prediction of $\bX(t+\Dt)$ by histories of $\bX,\bY$ up to and including time $t$, the difference being that the former is based on sparse, discrete (\ie, $\Dt$-subsampled) histories (\cf~\figref{fig:ctgc}).

It might be argued that, for completeness, we should estimate and perform statistical inference for $m$-step $\Dt$-subsampled GC; \ie, we should estimate $\egc{\bY(\Dt)}{\bX(\Dt),m}$, to be compared against $\gc\bY\bX(m\Dt)$, for $m = 1,2,\ldots$. For simplicitly we omit this analysis for our worked example (\secref{sec:ouminlag}), although we do consider the situation where, for a fixed quantity of data, we have a \emph{choice} of sampling interval $\Dt$ (\ie, we may downsample). Other possibilities worth consideration include comparison of the ``empirical GC rate'' $\frac1\Dt \egc{\bY(\Dt)}{\bX(\Dt)}$, at least for small $\Dt$, with the zero-horizon GC rate $\rgc\bY\bX$, or comparison of an estimated ``total GC'' $\Dt\sum_{m = 1}^\infty \egc{\bY(\Dt)}{\bX(\Dt),m}$ against the corresponding continuous-time quantity $\int_0^\infty \gc\bY\bX(h) \,dh$.

\section{Subsampling analysis for a minimal CTVAR process with finite causal delay} \label{sec:ouminlag}

Having established a consistent theoretical framework for the analysis of Granger causality for distributed-lag continuous-time processes and time series extracted by subsampling, we now work through a detailed Granger-causal subsampling analysis of a minimal (but certainly non-trivial) CTVAR process with finite-time feedback delay. This example is solvable analytically---in both discrete and continuous time---and serves to highlight some key modes of interaction between sampling frequency, causal delay and statistical power of Granger-causal inference. (Of course we should be cautious in assuming generalisation of results obtained here to more realistic neurophysiological models.) Analytical results are compared with detailed simulation. Although we concentrate on time-domain GCs (and, in the subsampled case, $1$-step GC), for illustrational purposes we include some results on spectral GC.

We consider a CTVAR process \eqref{eq:ctvar} with AR kernel
\begin{equation}
	A(u) \equiv -A \,\delta(u) + C \,\delta(t-\tau) \label{eq:mctv:ker}
\end{equation}
and residual noise intensity $\Sigma$, where
\begin{equation}
	A \equiv \begin{bmatrix} a & 0 \\ 0 & b \end{bmatrix}, \qquad
	C \equiv \begin{bmatrix} 0 & c \\ 0 & 0 \end{bmatrix}, \qquad
	\Sigma \equiv \begin{bmatrix} 1 & \rho \\ \rho & 1 \end{bmatrix}
\end{equation}
\ie,
\begin{subequations}
\begin{align}
	dX(t) &= -a \,X(t)\,dt + c Y(t-\tau)\,dt + dW_x(t) \label{eq:mctv:x} \\
	dY(t) &= -b \,Y(t)\,dt \hspace{57.5pt} + dW_y(t) \label{eq:mctv:y}
\end{align} \label{eq:mctv}%
\end{subequations}
The process $Y(t)$ is a standard Ornstein-Uhlenbeck process, which drives the process $X(t)$ at a fixed delay of $\tau$. $1/a, 1/b$ represent the (exponential decay) relaxation time of the $x$ and $y$ nodes in the absence of input, while $c$ controls the strength of feedback from node $y$ to node $x$ at delay $\tau$. For simplicity we assume $b \ne a$ (the special case $b = a$, which we may confirm behaves qualitatively similarly, may be solved along similar lines). The residual (instantaneous) correlation is $-1 < \rho < 1$.  We henceforth refer to \eqref{eq:mctv} as the ``minimal CTVAR''.

Processes like \eqref{eq:mctv} (and non-linear generalisations)  have been widely studied in the literature under the name \emph{Stochastic Delay-Differential Equations} (SDDEs) \citep{Longtin:2010}. Usually, though, only the univariate case is considered \citep[but see \eg][]{McKetterickGiuggioli:2014} and the emphasis is generally on questions of stability, convergence, approximation, numerical simulation and perturbation theory, rather than Granger-causal analysis of stationary, stable systems.

\textit{Notation}: In this section, for compactness we shall use the scaled frequency $\omega \equiv 2\pi\lambda$ in continuous time and the angular frequency $\omega \equiv 2\pi\Dt\lambda$ in discrete time for mathematical analyses, where $\Dt$ is as usual the sample interval. However, for displaying results we always convert back to ordinary frequency $\lambda$, so that spectral quantities scale appropriately with sample interval.

For the various plots and simulations in this section, we shall (unless otherwise stated) use the reference parameters:
\begin{equation}
	1/a = 5, \quad 1/b = 6, \quad 1/c = 8, \quad \tau = 30 \label{eq:refparms}
\end{equation}
(all times in milliseconds), while the residuals correlation coefficint $\rho$ will take stated values.

\subsection{Continuous-time analysis} \label{sec:ouminlag:ctime}

From \eqref{eq:mctv:ker} using definition \eqref{eq:cvargen} we have
\begin{equation}
	\Phi(\zeta) = \begin{bmatrix}
		a+\zeta & \displaystyle -c e^{-\tau\zeta} \\
		0 & b+\zeta
	\end{bmatrix} \label{eq:mvtvar:cvargen}
\end{equation}
which we invert to obtain
\begin{equation}
	\Psi(\zeta) = \begin{bmatrix}
		\displaystyle \frac1{a+\zeta} & \displaystyle \frac{c e^{-\tau\zeta}}{(a+\zeta)(b+\zeta)} \\[1em]
		0 & \displaystyle \frac1{b+\zeta}
	\end{bmatrix} \label{eq:mvtvar:cvmagen}
\end{equation}
from which we see that the minimum-phase condition is satisfied, and stability requires just that $a,b > 0$. From \eqref{eq:makern} it is straightforward to calculate the MA kernel as
\begin{equation}
	B(u) =  e^{-Au} + \heavi(u-\tau) \, q(u-\tau) \,C \label{eq:mctvar:makern}
\end{equation}
where $\heavi(u)$ is the Heaviside step function equal to $0$ for $u < 0$ and $1$ for $u \ge 0$, and
\begin{equation}
	q(u) \equiv  \frac1{b-a}\bracr{e^{-au}-e^{-bu}}
\end{equation}
From \eqref{eq:mvtvar:cvmagen} we have
\begin{equation}
	H(\omega) = \begin{bmatrix}
		\displaystyle \frac1{a+i\omega} & \ \displaystyle \frac{c e^{-i\tau\omega}}{(a+i\omega)(b+i\omega)} \\[1em]
		\displaystyle 0 & \ \displaystyle \frac1{b+i\omega}
	\end{bmatrix} \label{eq:mctv:tfun}
\end{equation}
and from \eqref{eq:cspecfac} we obtain the continuous-time CPSD
\begin{subequations}
\begin{align}
	S_{xx}(\omega) &= \frac{b^2 + 2\rho c \,\upsilon(\omega;b,\tau) + c^2 + \omega^2}{(a^2+\omega^2)(b^2+\omega^2)}\label{eq:mctv:cpsdxx} \\
	S_{xy}(\omega) &= \frac{(a-i\omega) \bracs{c e^{-i\tau\omega} + \rho (b+i\omega)}}{(a^2+\omega^2)(b^2+\omega^2)} \label{eq:mctv:cpsdxy} \\
	S_{yy}(\omega) &= \frac1{b^2+\omega^2} \label{eq:mctv:cpsdyy}
\end{align} \label{eq:mctv:cpsd}%
\end{subequations}
where
\begin{equation}
	\upsilon(\omega;b,\tau) \equiv b \cos \tau\omega - \omega \sin\tau\omega \label{eq:upsdef}
\end{equation}
\begin{figure*}
\begin{center}
\includegraphics{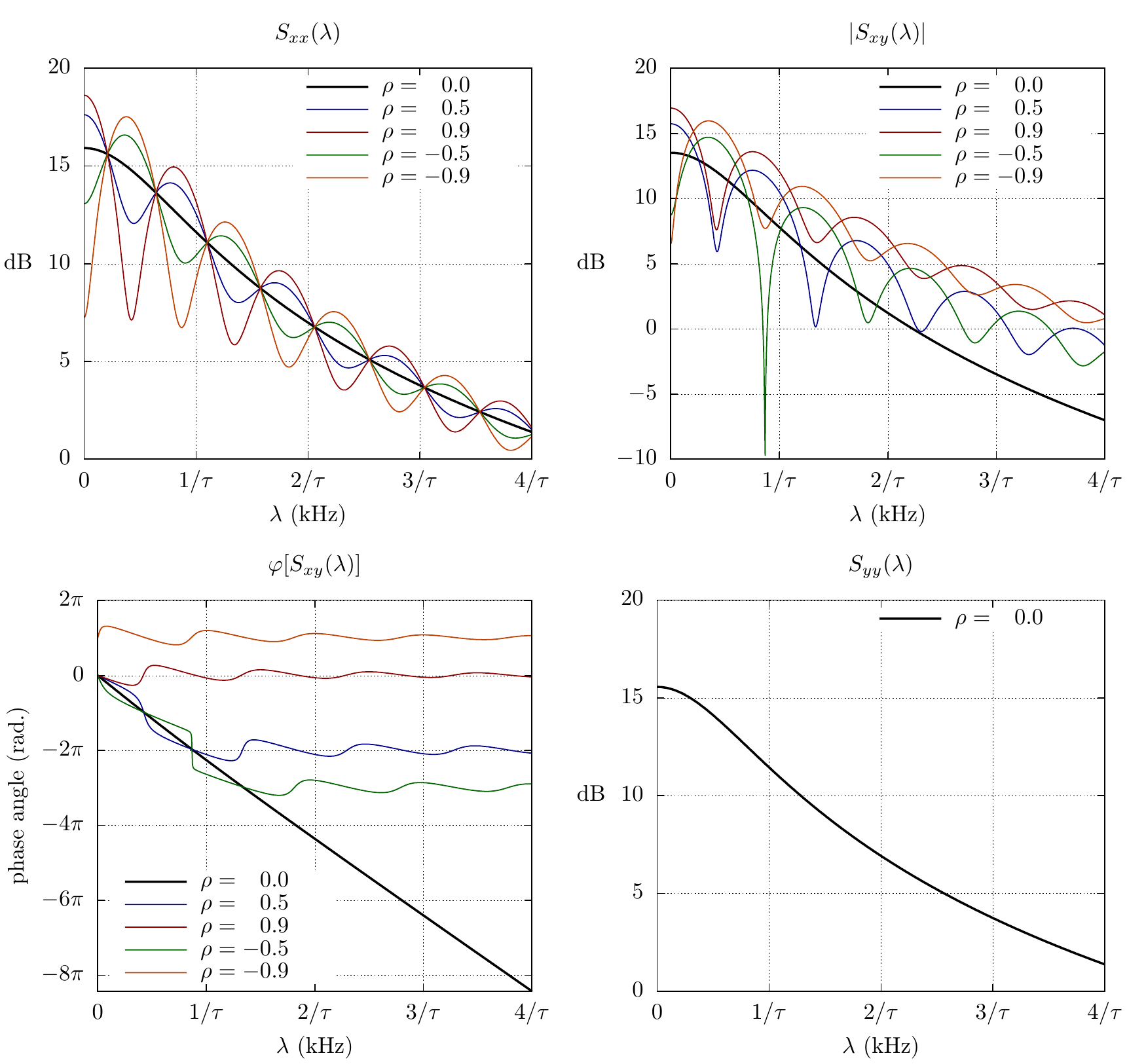}
\end{center}
\caption{The CPSD \eqref{eq:mctv:cpsd} for the minimal CTVAR \eqref{eq:mctv} with reference parameters \eqref{eq:refparms} plotted (in dB) against ordinary frequency $\lambda$ (kHz), for a few values of the residual noise correlation parameter $\rho$. The upper right figure plots the modulus, and the lower left figure the (unwrapped) phase angle, of the $xy$-cross-power term \eqref{eq:mctv:cpsdxy}. Note that the $y$-autospectrum  \eqref{eq:mctv:cpsdyy} (bottom right) does not depend on $\rho$.} \label{fig:mctv_cpsd}
\end{figure*}%
\eqref{eq:mctv:cpsd} is plotted in \figref{fig:mctv_cpsd} for the reference parameters \eqref{eq:refparms} and a few values of $\rho$. We see that power attenuates as $\omega^{-2}$ [\cf~\eqref{eq:slim}], for $\rho \ne 0$ the $xx$-compenent oscillates with period $\tau$, while for $\rho = 0$ the feedback delay $\tau$ appears only in the phase angle of the $xy$-component.

The autocovariance function is most easily calculated from $B(t)$ using \eqref{eq:cywma}; alternatively, we might solve the Yule-Walker equations \eqref{eq:cyw}, or invert the continuous-time Fourier transform \eqref{eq:ccpsd}. For convenience we define the dimensionless quantities
\begin{equation}
	\theta \equiv \frac c{b-a}, \qquad \eta \equiv \frac c{a+b}
\end{equation}
and we may calculate that for $t \ge 0$,
\begin{equation}
	\Gamma(t) = \Gamma^{(0)}(t) + \heavi(\tau-t) \,\Gamma^{(1)}(\tau-t) + \heavi(t-\tau) \,\Gamma^{(2)}(t-\tau) \label{eq:mctv:gamma}
\end{equation}
where
\begin{subequations}
\begin{align}
	\Gamma^{(0)}_{xx}(t) &= \frac{1+\theta\eta}{2a} e^{-at} - \frac{\theta\eta}{2b} e^{-bt} + \rho\frac\eta{2a} e^{-a(t+\tau)} \\
	\Gamma^{(0)}_{xy}(t) &= \rho\frac1{a+b} e^{-at} \\
	\Gamma^{(0)}_{yx}(t) &= \rho\frac1{a+b} e^{-bt} + \frac\eta{2b} e^{-b(t+\tau)} \\
	\Gamma^{(0)}_{yy}(t) &= \frac1{2b} e^{-bt} \label{eq:mctv:xgammayy} \\
	\notag \\
	\Gamma^{(1)}_{xx}(t) &= \rho\frac\eta{2a} e^{-at} \\
	\Gamma^{(1)}_{xy}(t) &= \;\,\frac\eta{2b} e^{-bt} \\
	\notag \\
	\Gamma^{(2)}_{xx}(t) &= \rho\theta\bracr{\frac1{2a} e^{-at} - \frac1{a+b} e^{-bt}} \\
	\Gamma^{(2)}_{xy}(t) &= \;\,\theta\bracr{\frac1{a+b} e^{-at} - \frac1{2b} e^{-bt}}
\end{align} \label{eq:mctv:xgamma}%
\end{subequations}
and all other entries vanish.
\begin{figure*}
\begin{center}
\includegraphics{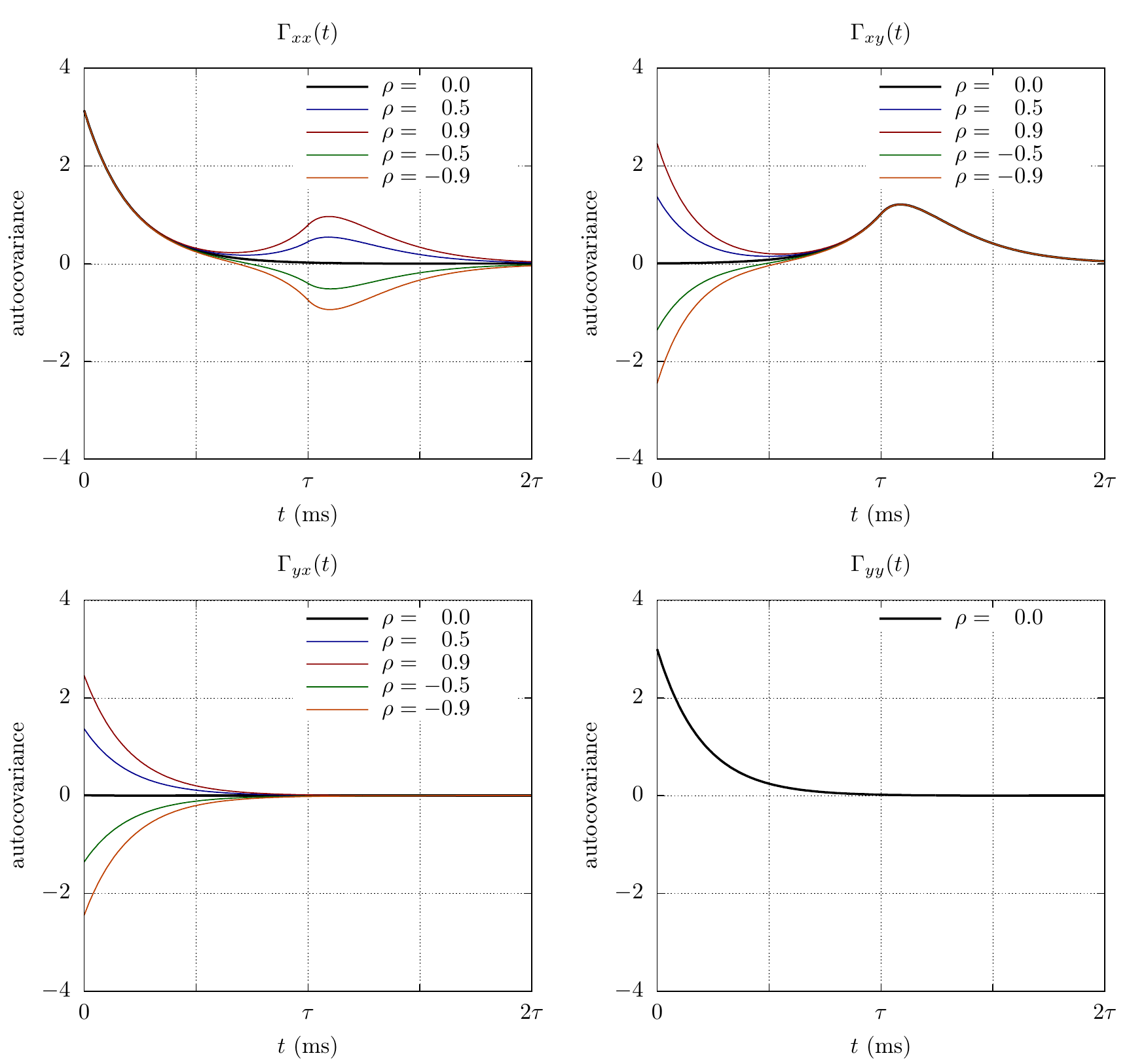}
\end{center}
\caption{The autocovariance function \eqref{eq:mctv:gamma} for the minimal CTVAR \eqref{eq:mctv} with reference parameters \eqref{eq:refparms} plotted against lag time $t$ (ms), for a few values of the residual noise correlation parameter $\rho$. Note that the $y$-autocovariance  \eqref{eq:mctv:xgammayy} (bottom right) does not depend on $\rho$.} \label{fig:mctv_autocov}
\end{figure*}%
$\Gamma(t)$ is plotted for a few values of $\rho$ in \figref{fig:mctv_autocov}. We see that the $xx$- and $yy$-components have local peaks just after $t = \tau$ and autocovariance decays exponentially for large $t$.

To calculate the continuous-time Granger causality $\gc YX(h)$ [clearly $\gc XY(h)$ vanishes identically, since $H(\omega)$ is upper-triangular] we firstly require $\MSE_{xx}(h)$, which may be calculated from \eqref{eq:mctvar:makern} by straightforward integration as per \eqref{eq:ctpredmse}. We find:
\begin{equation}
	\MSE_{xx}(h) = \frac1{2a}\bracr{1-e^{-2ah}} + \heavi(h-\tau) r(h-\tau) \label{eq:minmod:msex}
\end{equation}
where
\begin{multline}
	r(h) \equiv 2\rho\theta \, e^{-a\tau} \bracc{\frac1{2a}\bracr{1-e^{-2ah}} - \frac1{a+b}\bracs{1-e^{-(a+b)h}}} \\
	+ \theta^2 \left\{ \frac1{2a}\bracr{1-e^{-2ah}} - \frac2{a+b}\bracs{1-e^{-(a+b)h}} \right. \\
	\left. + \frac1{2b}\bracr{1-e^{-2bh}} \right\}
\end{multline}
For $\MSE'_{xx}(h)$, we need to solve the reduced spectral factorisation problem. It may be verified with \eqref{eq:mctv:cpsdxx} that
\begin{multline}
	H'_{xx}(\omega) = \\ \frac{\sqrt{\big(1-\rho^2\big) c^2 + \big(b + \rho c \cos\tau\omega\big)^2} + i(\omega - \rho c \sin\tau\omega)}{(a+i\omega) (b+i\omega)} \label{eq:minmod:redspecfac}
\end{multline}
satisfies $S_{xx}(\omega) = H'_{xx}(\omega) H'_{xx}(\omega)^*$ (we know that $\Sigma'_{xx} = \Sigma_{xx} = 1$). Then, at least in principal, we may calculate the reduced MA kernel as the inverse Fourier transform\footnote{The factor of $1/\pi$ arises from the use of scaled frequency $\omega = 2\pi\lambda$; \cf~\eqref{eq:cgcdecomp}.} $B'_{xx}(t) = \frac1{2\pi} \int_{-\infty}^\infty H'_{xx}(\omega) \,e^{it\omega}\,d\omega$. In the general case $\rho \ne 0, \tau > 0$ the inverse transform appears to be analytically intractable, although $B'_{xx}(u)$ may be calculated numerically. To this end we note that for large $\omega$, $H'_{xx}(\omega) e^{i\omega u}$ is dominated by $(\sin \omega u)/\omega$, and we find that $B'_{xx}(u) \approx \frac1\pi \int_0^W H'_{xx}(\omega) \,e^{i\omega u} \,d\omega + \frac12 - \frac1\pi\Si(Wu)$  with $W \gg \max(a,b,|\rho c|)$, where $\Si(x) \equiv \int_0^x (\sin\xi)/\xi \,d\xi$ is the sine integral function and the integral may be approximated by numerical quadrature\footnote{To avoid aliasing artefacts, for numerical quadrature we should ensure that for each $u$ we have $d\omega \le 2\pi/u$.}. We may then approximate $\MSE'_{xx}(h)$ from \eqref{eq:ctpredmse}, again by numerical quadrature, and $\gc YX(h)$ is calculated as per \eqref{eq:cgc}.

The zero-horizon spectral GC \eqref{eq:csgc} may be calculated from \eqref{eq:mctv:tfun} and \eqref{eq:mctv:cpsd}, noting that $\Sigma_{yy|x} = 1-\rho^2$. We have
\begin{equation}
	\sgc YX(\omega;0) = \log\bracr{1+\frac{\bracr{1-\rho^2}c^2}{b^2 + 2\rho c \,\upsilon(\omega;b,\tau) + \rho^2 c^2 + \omega^2}} \label{eq:minmod:csgcyx}
\end{equation}
The zero-horizon GC rate $\rgc YX$ may then in principal be obtained from the spectral decomposition \eqref{eq:cgcdecomp} by integrating \eqref{eq:minmod:csgcyx}. Again the integral appears analytically intractable, but may be calculated numerically. For large $\omega$, the denominator of the fraction in \eqref{eq:minmod:csgcyx} is dominated by the $\omega^2$ term, from which we may calculate that $\rgc YX \approx \frac1\pi \int_0^W \sgc YX(\omega;0) \,d\omega + \frac1\pi \big(1-\rho^2\big)c^2/W$ with $W \gg \max(b,|\rho c|)$ and the integral may again be approximated by numerical quadrature.

Full analytical calculation of $\gc YX(h)$ and $\rgc YX$ \emph{is} tractable in the special cases $\tau = 0$ and $\rho = 0$. Recall (\secref{sec:voudl}) that we do not consider zero feedback delay to be plausible for neurophysiological processes. While it seems unlikely that residuals correlation will be completely absent, we present the $\rho = 0$ case analytically both in continuous time and under subsampling. In particular,
\begin{equation}
	H'_{xx}(\omega) = \frac{k+i\omega}{(a+i\omega) (b+i\omega)} \label{eq:minmod:redspecfac_r0}
\end{equation}
with $k \equiv \sqrt{b^2+c^2}$, and performing the inverse Fourier transform, we may calculate that
\begin{equation}
	B'_{xx}(u) = \frac{(k-a) e^{-au} -(k-b) e^{-bu}}{b-a} \label{eq:minmod:redvma_r0}
\end{equation}
We may then calculate $\MSE'_{xx}(h)$ by integration from \eqref{eq:ctpredmse}:
\begin{multline}
	\MSE'_{xx}(h) =  \frac1{(b-a)^2} \left\{ \frac{(k-a)^2}{2a}\bracr{1-e^{-2ah}} \right. \\
	\left. - \frac{2(k-a)(k-b)}{a+b}\bracs{1-e^{-(a+b)h}} + \frac{(k-b)^2}{2b}\bracr{1-e^{-2bh}} \right\}
	\label{eq:minmod:msexr}
\end{multline}
$\gc YX(h)$ may then be calculated according to \eqref{eq:cgc} with \eqreff{eq:minmod:msex}{eq:minmod:msexr}. From the resulting expression, we may confirm that for $h \gg \tau$, $\gc YX(h)$ decays exponentially with $h$, with exponent $\min(a,b)$.

From \eqref{eq:rgc} or \eqref{eq:rgc1} we may calculate
\begin{equation}
	\rgc YX = \sqrt{b^2+c^2}-b
\end{equation}
and from \eqref{eq:csgc} we have
\begin{equation}
	\sgc YX(\omega;0) = \log\bracr{1+\frac{c^2}{b^2 + \omega^2}} \label{eq:minmod:csgcyx_r0}
\end{equation}
Note that, unlike in the general $\rho \ne 0$ case, the feedback delay time $\tau$ does not appear in these expressions. (We would, however, expect to see dependence on causal delay if more than a single delayed feedback were present.)

\begin{figure}
\begin{center}
\includegraphics{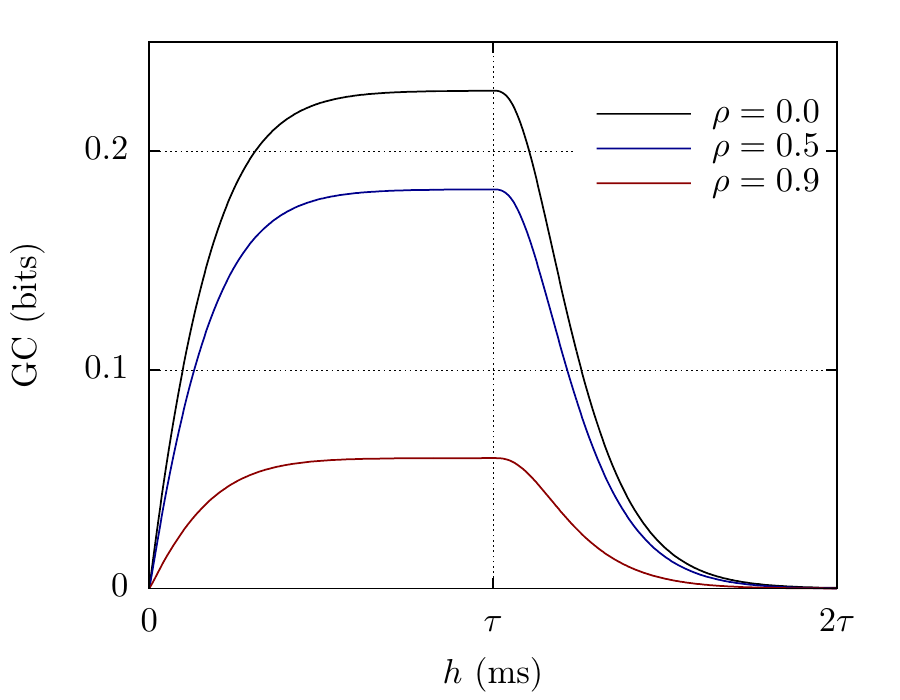}
\end{center}
\caption{Granger causality $\gc YX(h)$  for the minimal CTVAR \eqref{eq:mctv} with reference parameters \eqref{eq:refparms} plotted against prediction horizon $h$ (ms), for a few values of the residual noise correlation parameter $\rho$. Values for $\rho < 0$ (not displayed) are close to their positive counterparts.} \label{fig:mctv_gc}
\end{figure}

\figref{fig:mctv_gc} plots $\gc YX(h)$ against prediction horizon $h$ for a few values of $\rho$.
We see that for small $h$, $\gc YX(h)$ rises approximately linearly with slope $\rgc YX$ [\cf~\eqreff{eq:cgclijm}{eq:rgc}]. It then flattens out and attains a maximum just beyond $h = \tau$, before decaying exponentially as described above. Overall, $\gc YX(h)$ is reduced for larger values of $|\rho|$ ($\rho < 0$ plots are not displayed; for our reference parameters they lie very close to the corresponding positive $\rho$ plots).

We also make the following observation: suppose that we define a filter $\cG(\zeta) = \begin{bmatrix} \displaystyle \frac{a+\zeta}{g+\zeta} & 0 \\ 0 & 1 \end{bmatrix}$, which is stable, minimum-phase and invertible provided $g > 0$. Now applying the filter to \eqref{eq:mvtvar:cvmagen}, we see that the filtered CTVAR is identical to the original CTVAR \eqref{eq:mctv}, but with $a$ replaced by $g$ (\cf~\apxref{sec:nofinv}). Since $\gc YX(h)$ depends on $a$ [\cf~ \eqreff{eq:minmod:msex}{eq:minmod:msexr}], we have demonstrated the non-invariance of finite-horizon GC in the continuous-time case, as claimed at the end of \secref{sec:gcouc}. Since the zero-horizon spectral GC \eqref{eq:minmod:csgcyx} does not depend on $a$, we see that, as expected, zero-horizon GC, in both time and frequency domains, is invariant under $\cG(\zeta)$.

\subsection{Subsampling analysis} \label{sec:ouminlag:subsamp}

A procedure for calculating $\Dt$-subsampled Granger causalities analytically from the CTVAR parameters is described in \secref{sec:voudlds}, \tabref{tab:gcanaproc}. We have already \eqref{eq:mctv:gamma} performed the first stage, calculation of the continuous-time autocovariance function $\Gamma$ (\tabref{tab:gcanaproc}, steps~\ref{it:dsproc1d},\ref{it:dsproc1e}), while step~\ref{it:dsproc2}, subsampling $\Gamma$, is trivial. Step~\ref{it:dsproc3}, calculation of the discrete-time CPSD, is straightforward, if laborious.

Step~\ref{it:dsproc4}, spectral factorisation of the discrete-time CPSD, is particularly demanding in the general case. In lieu of an analytical factorisation, we employed the following numerical method: since the autocovariance sequence $\Gamma(\Dt)$ of the subsampled process is just the $\Dt$-subsampling of the continuous-time autocovariance function \eqref{eq:mctv:gamma}, it is easily calculated. We then use Whittle's time-domain multivariate spectral factorisation algorithm \citep{Whittle:1963}, which takes as input a discrete-time autocovariance sequence and yields the corresponding VAR parameters as output\footnote{An alternative approach would be to calculate the CPSD analytically (\tabref{tab:gcanaproc}, step~\ref{it:dsproc3}) and apply Wilson's frequency-domain multivariate spectral factorisation algorithm \citep{Wilson:1972}.} (the autocovariance sequence was truncated at sufficient lags for all autocovariances to have decayed below machine precision). Here, these calculations were performed using the Multivariate Granger Causality (MVGC) Matlab\textsuperscript{\textregistered} toolbox \citep{Barnett:mvgc:2014}. An interesting result of this experiment, was that, even with $\rho \ne 0$, subsampled GC in the non-causal direction, $\gc{X(\Dt)}{Y(\Dt)}$, was seen to be zero for any sampling interval $\Dt$. That is, \emph{for our minimal CTVAR, there is no spurious GC}. This is somewhat surprising as it is known that, as previously noted, subsampling is in general likely to induce spurious causalities in both discrete and continuous time\footnote{But see also our remarks  in \secref{sec:gcouc} regarding non-occurrence of subsampling-induced spurious causality in $2$-dim bivariate CTVAR processes.}.

\begin{figure*}
\begin{center}
\includegraphics{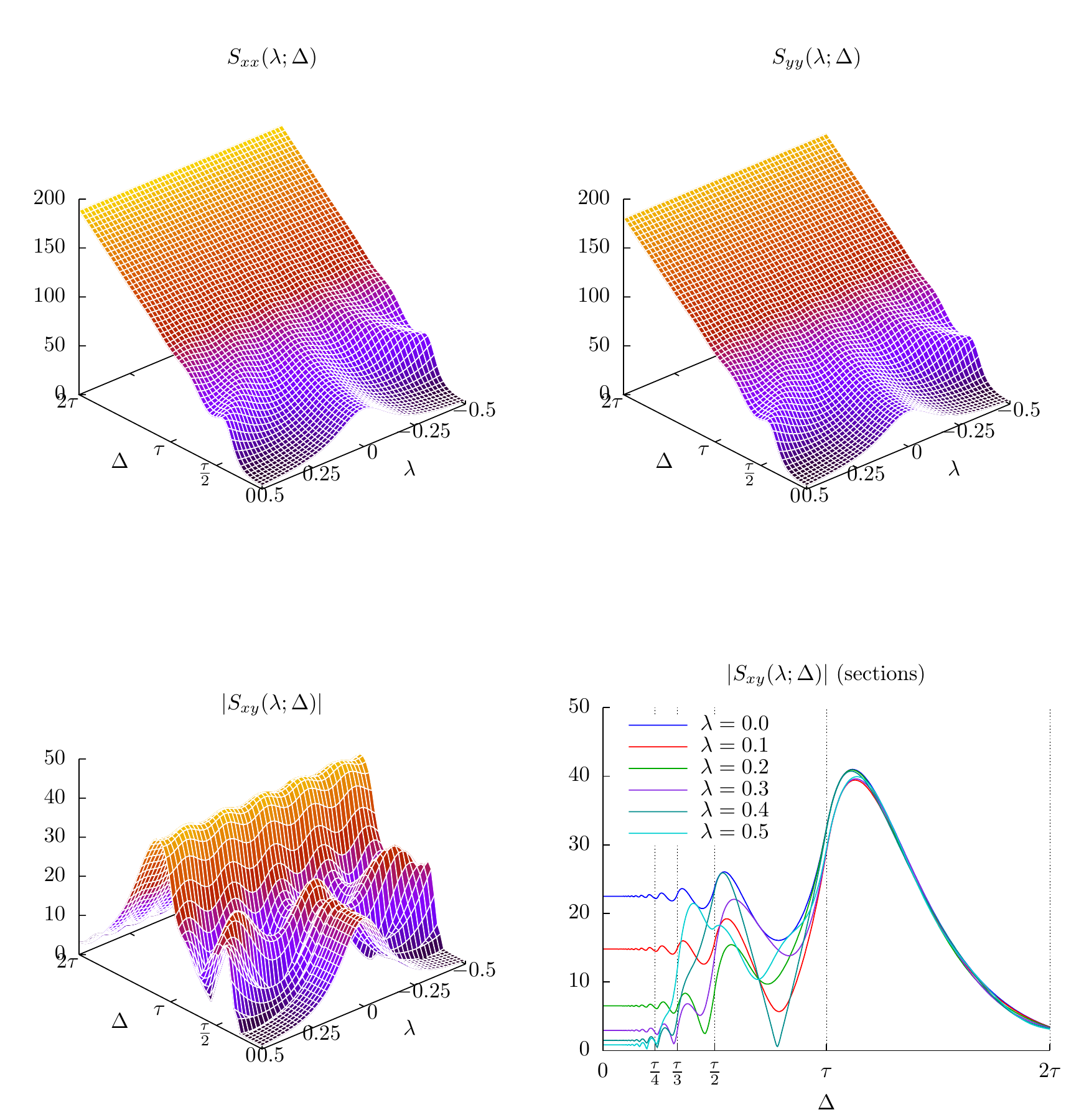}
\end{center}
\caption{The CPSD $S(\lambda;\Dt)$ \eqref{eq:oudscpsd} for the subsampled minimal CTVAR \eqref{eq:mctv} with reference parameters \eqref{eq:refparms} and $\rho = 0$, plotted against ordinary frequency $\lambda$ and sample interval $\Dt$. In all plots the $\Dt = 0$ spectral power corresponds to the continuous-time CPSD \eqref{eq:mctv:cpsd} [\cf~\figref{fig:mctv_cpsd}].} \label{fig:cpsd_ds}
\end{figure*}

We thus concentrate on the detectability of GC under subsampling in the causal $Y \to X$ direction. Since non-zero $\rho$ was not found to have a profound effect on $\gc{Y(\Dt)}{X(\Dt)}$, from here on we set $\rho = 0$; the subsampling problem may then be solved entirely analytically for $\tau \ge 0$. We start, again, by $\Dt$-subsampling the continuous-time autocovariance function \eqref{eq:mctv:gamma} to obtain $\Gamma(\Dt)$. We may then calculate\footnote{The only awkward case is the cross-power term $S_{xy}(\Dt)$, which we calculate in \apxref{sec:Sxy}.} the CPSD $S(\Dt)$ of the subsampled process by discrete-time Fourier transform \eqref{eq:cpsd} of $\Gamma(\Dt)$. For convenience, we define the following quantities:
\begin{subequations}
\begin{align}
	\alpha &\equiv e^{-a\Dt} &&& q     &\equiv \lceil\tau/\Dt\rceil\\
	\beta  &\equiv e^{-b\Dt} &&&\kappa &\equiv \lceil\tau/\Dt\rceil-\tau/\Dt \label{eq:kappa} \\
	\gamma &\equiv e^{b\Dt} = \beta^{-1}
\end{align}%
\end{subequations}
where $ \lceil x \rceil \equiv \min\{n \in \posints \,|\, n \ge x\}$ denotes the ceiling function. Note that (i) $q \ge 1$, with $q = 1 \iff \Dt \ge \tau$ and (ii) $0 \le \kappa < 1$, with $\kappa = 0 \iff \Dt$ divides $\tau$ exactly. We also define
\begin{subequations}
\begin{align}
		u_1 &\equiv \displaystyle \frac{1+\theta\eta}{2a} \,(1-\alpha^2) &&& v_1 &\equiv \displaystyle \frac{\theta+\eta}{2b} \,\alpha^\kappa = \dfrac{\theta\eta}c \,\alpha^\kappa\\
		u_2 &\equiv \ \ \ \, \displaystyle -\frac{\theta\eta}{2b} \,(1-\beta^2) &&& v_2 &\equiv \ \displaystyle -\frac\theta{2b} \,\beta^\kappa\\
		w &\equiv \ \ \ \ \ \,\, \displaystyle \frac1{2b} \,(1-\beta^2) &&& v_3 &\equiv \; \displaystyle -\frac\eta{2b} \,\gamma^\kappa
\end{align}%
\end{subequations}
We then find (for compactness we express spectral quantities in terms of  $z = e^{-i\omega}$ on the circle $|z| = 1$ in the complex plane):
\begin{subequations}
\begin{align}
	S_{xx}(z;\Dt) &= \frac{u_1}\mfasq + \frac{u_2}\mfbsq \label{eq:oudscpsd:xx} \\
	S_{xy}(z;\Dt) &= \bracr{\frac{v_1}\mfa +\frac{v_2}\mfb  + \frac{v_3}{1-\gamma z}} z^q \\
	S_{yy}(z;\Dt) &= \frac{w}\mfbsq
\end{align} \label{eq:oudscpsd}%
\end{subequations}
Note that $S(z;\Dt)$ depends on the causal delay $\tau$ only via the cross-power term, through dependency on $q$ and $\kappa$. \figref{fig:cpsd_ds} plots the CPSD \eqref{eq:oudscpsd} against ordinary frequency $\lambda$ and subsampling interval $\Dt$, for the reference parameters \eqref{eq:refparms}.
In all plots the $\Dt = 0$ spectral power corresponds to the continuous-time CPSD \eqref{eq:mctv:cpsd} (\figref{fig:mctv_cpsd}). Note in particular the cross-spectral power $|S_{xy}|$, which undergoes a series of increasingly rapid oscillations with decreasing $\Dt$, peaking just past the values $\tau, \tau/2, \tau/3,\ldots$, reflecting the oscillations in $\kappa$ \eqref{eq:kappa} as the sample frequency $1/\Dt$ resonates with the causal delay frequency $1/\tau$ (\figref{fig:cpsd_ds}, bottom right).

Factorisation of $S(\lambda;\Dt)$ for both full and reduced regressions, is non-trivial and detailed in \apxref{sec:minoulagdsgc}. Results may be summarized as follows:
\begin{enumerate}
\item The joint subsampled process $X(\Dt),Y(\Dt)$ is VARMA($2,q+1$), the subprocess $X(\Dt)$ is VARMA($2,1$) and the subprocess $Y(\Dt)$ is VAR($1$).
\item The residuals covariance matrix $\Sigma(\Dt) \equiv \begin{bmatrix} \sigma_{xx} & \sigma_{xy} \\ \sigma_{xy} & \sigma_{yy} \end{bmatrix}$ of the full regression of the joint process is specified by
\begin{subequations}
\begin{align}
	\sigma_{xy} &= \left\{
		\begin{array}{rcl}
			 -\beta P && q = 1 \\
			0 && q > 1
		\end{array}
	\right. \\
	\sigma_{yy} &= w \\
	\sigma_{xx} &= \frac{D + \sigma_{xy}^2}{\sigma_{yy}} \label{eq:minmod:sigmaxx}
\end{align} \label{eq:minmod:sigmas}%
\end{subequations}
where $P$ is given by  \eqref{eq:Pdef} and $D$ by \eqref{eq:4degDsol} of \apxref{sec:minoulagdsgc}.
\item The residuals variances $\Sigma'_{xx}(\Dt) \equiv \sigma'_{xx}$ and $\Sigma'_{yy}(\Dt) \equiv \sigma'_{yy}$ of the reduced regressions of $X(\Dt), Y(\Dt)$ respectively are given by
\begin{subequations}
\begin{align}
		\sigma'_{xx} &= \varphi + \sqrt{\varphi^2-\psi^2} \label{eq:minmod:sigmarxx} \\
		\sigma'_{yy} &= \sigma_{yy} = w
\end{align}%
\end{subequations}
where $\varphi,\psi$ are given by \eqref{eq:phipsi} of \apxref{sec:minoulagdsgc}.
\end{enumerate}
It was also verified, for all parameter values and subsampling intervals examined, that the joint $\Dt$-subsampled process was minimum-phase\footnote{We conjecture that the joint process is in fact minimum-phase for all parameters and subsampling intervals---indeed, we suspect that some results of \cite{AstromEtal:1984} for univariate rational transfer functions may generalise to cover our case---but have not succeeded in proving this conclusively due to the extreme algebraic complexity of the relevant condition.} [see \apxref{sec:minoulagdsgc}, final paragraph; the sub-processes $X(\Dt)$ and $Y(\Dt)$ are always minimum-phase].

Our key result, analytical expressions for the directional and instantaneous Granger causalities of the subsampled process in the time-domain, follow immediately:
\begin{subequations}
\begin{align}
	\gc{X(\Dt)}{Y(\Dt)}  &= \log\bracr{\frac{\sigma'_{yy}}{\sigma_{yy}}} = 0 \label{eq:minmod:gcxy} \\
	\gc{Y(\Dt)}{X(\Dt)}  &= \log\bracr{\frac{\sigma'_{xx}}{\sigma_{xx}}} \label{eq:minmod:gcyx} \\
	\igc{X(\Dt)}{Y(\Dt)} &= -\log\bracr{1-\trho^2} \label{eq:minmod:igcxy}
\end{align} \label{eq:minmod:gc}%
\end{subequations}
where $\displaystyle \trho = \sigma_{xy}\,\big/\!\sqrt{\sigma_{xx} \sigma_{yy}}$ is the residuals correlation coefficient of the subsampled processes [\cf\ \secref{sec:gc}, eq.~\eqref{eq:igc}]. Note that, as might be expected with $\rho = 0$, $\gc{X(\Dt)}{Y(\Dt)}$ vanishes for any  $\Dt$, so that subsampling does not induce spurious causality in the non-causal $X \to Y$ direction. For completeness, we supply the formula for the frequency-domain GC
\begin{multline}
	\sgc{Y(\Dt)}{X(\Dt)}(z) = \\ -\log\bracs{1-(1-\rho^2) \frac{\big|\sigma_{xy} B_{xx}(z) + L(z) z^{q-1}\big|^2}{D|B_{xx}(z)|^2 + |L(z)|^2}}  \label{eq:minmod:sgc}
\end{multline}
with $B_{xx}(z)$ given by \eqref{eq:Bxx1} and $L(z)$ by \eqref{eq:Lofz} of \apxref{sec:minoulagdsgc}, while $\sgc{X(\Dt)}{Y(\Dt)}(z)$ is again identically zero.

\begin{figure}
\begin{center}
\includegraphics{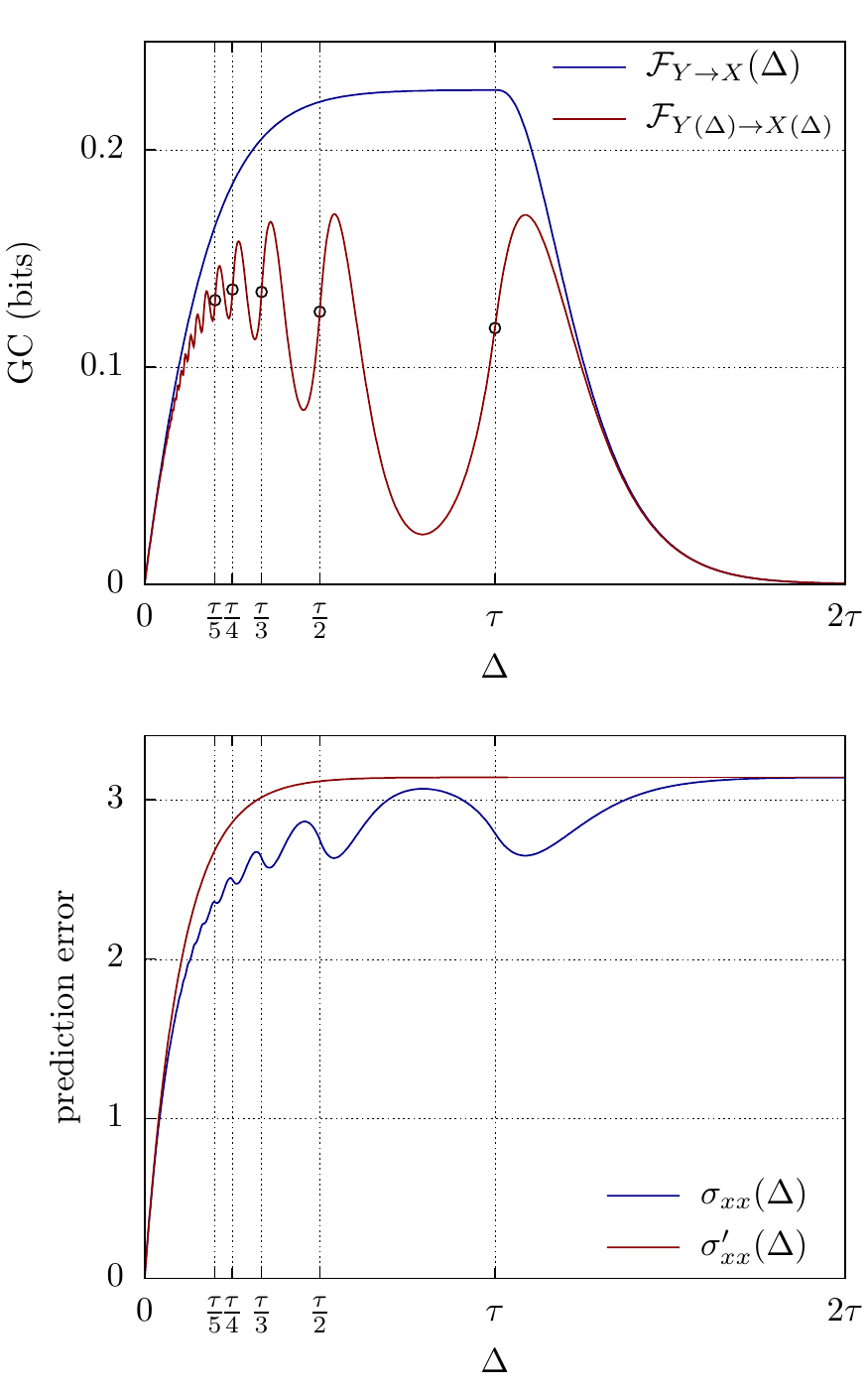}
\end{center}
\caption{\textit{Top figure:} $\gc YX(\Dt)$ (blue, eqs.~\ref{eq:minmod:msex},\ref{eq:minmod:msexr}) and $\gc{Y(\Dt)}{X(\Dt)}$ (red, eq.~\ref{eq:minmod:gcxy}) plotted against sample interval $\Dt$ for the subsampled CTVAR process \eqref{eq:mctv} with $\rho = 0$ and reference parameters \eqref{eq:refparms}. Black circles mark inflexion points of $\gc{Y(\Dt)}{X(\Dt)}$. \textit{Bottom figure:} subsampled process full prediction error $\sigma_{xx}(\Dt)$ (blue, eq.~\ref{eq:minmod:sigmaxx}) and reduced prediction errors $\sigma'_{xx}(\Dt)$ (red, eq.~\ref{eq:minmod:sigmarxx}) plotted against $\Dt$.} \label{fig:gc_ref_ds}
\end{figure}

In \figref{fig:gc_ref_ds} (top figure), the $\Dt$-subsampled discrete-time GC $\gc{Y(\Dt)}{X(\Dt)}$ calculated according to \eqref{eq:minmod:gcyx} is plotted, along with the continuous-time GC $\gc YX(\Dt)$ as calculated from \eqreff{eq:minmod:msex}{eq:minmod:msexr}---that is, for the same prediction horizon (\cf\ our remarks in \secref{sec:gcdsest})---against $\Dt$ for $\rho = 0$ and reference parameters \eqref{eq:refparms}.
The pattern of distortion induced by subsampling is clear from the top figure: $\gc{Y(\Dt)}{X(\Dt)}$ sits under the envelope of the ``true'' GC $\gc YX(\Dt)$, and closely follows both its linear rise from $\Dt = 0$ and also its exponential decay after peaking just beyond $\Dt = \tau$ [it is straightforward (if tedious) to show that in the limit of large sample interval, $\gc{Y(\Dt)}{X(\Dt)}$ decays with the same exponent as $\gc YX(\Dt)$ as $\Dt \to \infty$].

As for the subsampled cross-spectral power $S_{\!xy}(\lambda;\Dt)$ (\figref{fig:cpsd_ds}), we see increasingly rapid oscillations near the points $\Dt = \tau, \tau/2, \tau/3, \ldots$, which again arise as $\kappa$ \eqref{eq:kappa} oscillates between $0$ and $1$, reflecting resonance between sampling and delayed feedback frequencies. Note that $\kappa$ affects only the full prediction error $\sigma_{xx}(\Dt)$ \eqref{eq:minmod:sigmaxx}, while the reduced prediction errors $\sigma'_{xx}(\Dt)$ \eqref{eq:minmod:sigmarxx} increases monotonically (\figref{fig:gc_ref_ds}, bottom figure): this reflects the fact that (for $\rho = 0$) the causal delay $\tau$ affects only the cross-power term of the CPSD, in both continuous time \eqref{eq:mctv:cpsd} and subsampled \eqref{eq:oudscpsd}. In this case the positive-slope inflexion points of $\gc{Y(\Dt)}{X(\Dt)}$ lie almost exactly at $\Dt = \tau, \tau/2, \tau/3, \ldots$ (although this is not always the case; \cf~\figref{fig:gc_ds_x} below.) Note that these oscillations are quite distinct from those reported in \cite{Zhou:2014}, which are of constant period, and are ascribed to periodicity in the time series. Since, as may be seen from the power spectra (\figref{fig:cpsd_ds}), there is no strong periodic behaviour in the (subsampled) minimal CTVAR, we do not see the Zhou oscillations here (see also \secref{sec:disc}). A striking feature is that for $\tau/2 < \Dt < \tau$ there is strong distortion--- a pronounced dip---in $\gc{Y(\Dt)}{X(\Dt)}$.

\begin{figure*}
\begin{center}
\includegraphics{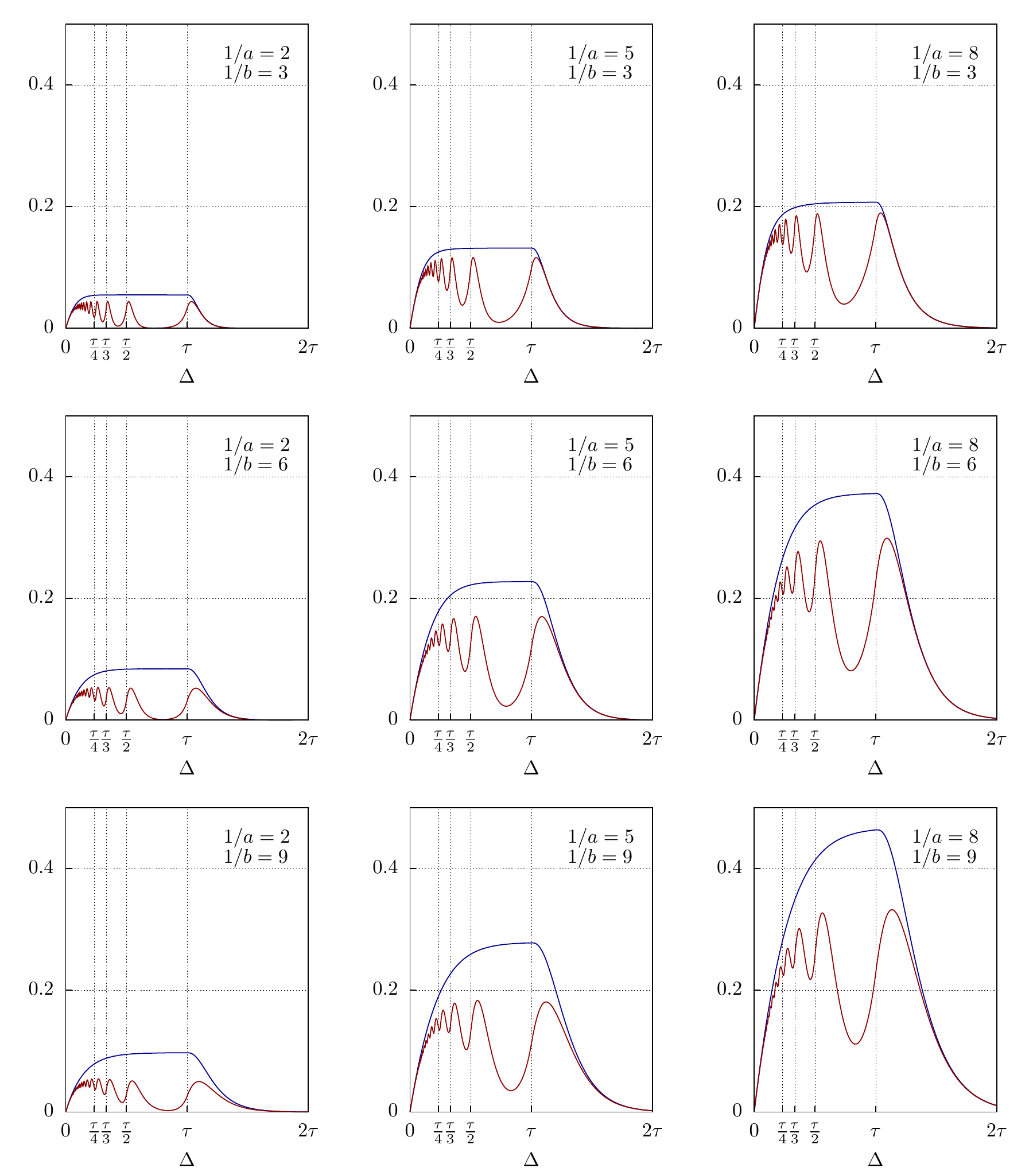}
\end{center}
\caption{$\gc YX(\Dt)$ (blue) and $\gc{Y(\Dt)}{X(\Dt)}$ (red) plotted against sample interval $\Dt$ for the subsampled CTVAR process \eqref{eq:mctv} with $\rho = 0$. Parameters are as in \eqref{eq:refparms} except for $a$ and $b$: $1/a$ takes the values $2,5,8$ from left to right, while $1/b$ takes the values $3,6,9$ from top to bottom (the centre figure thus corresponds to the exact reference parameters).} \label{fig:gc_ds}
\end{figure*}

In \figref{fig:gc_ds}, $\gc{Y(\Dt)}{X(\Dt)}$ calculated according to \eqref{eq:minmod:gcyx} is plotted, along with the continuous-time GC $\gc YX(\Dt)$ for $\rho = 0$, against $\Dt$ for a range of node relaxation time parameters $1/a,1/b$ around the reference values \eqref{eq:refparms}.
We see that the dip at $\tau/2 < \Dt < \tau$ closely approaches zero for small values of $1/a$ (\ie, fast relaxation of the $X$ variable). As $1/a$ increases (left to right), the $\gc YX(\Dt)$ plateaux becomes more peaked towards $\Dt = \tau$. As the $Y$ relaxation time $1/b$ increases (top to bottom) we see that the subsampled GC $\gc{Y(\Dt)}{X(\Dt)}$ ``pulls away'' from the continuous-time GC, indicating a higher degree of distortion.

To further explore the parameter space of the model, in \figref{fig:gc_ds_x}, $\gc{Y(\Dt)}{X(\Dt)}$ and $\gc YX(\Dt)$ are again plotted against $\Dt$ for $\rho = 0$, for a selection of more extreme (if arguably biophysically implausible) parameters.
\begin{figure}
\begin{center}
\includegraphics{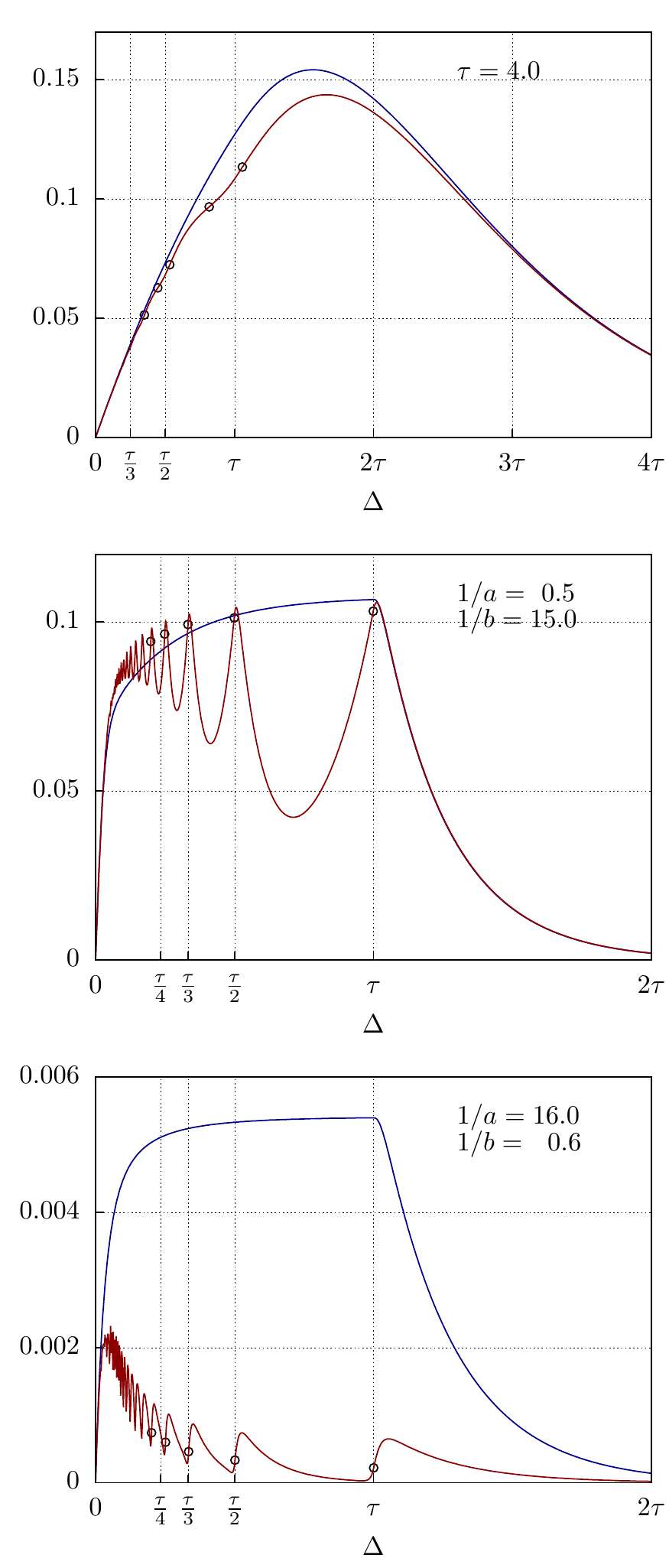}
\end{center}
\caption{$\gc YX(\Dt)$ (blue) and $\gc{Y(\Dt)}{X(\Dt)}$ (red) plotted against sample interval $\Dt$ for the subsampled CTVAR process \eqref{eq:mctv} with $\rho = 0$, for a selection of parameter settings (note differing time and GC scales). Parameters are set to the reference parameters \eqref{eq:refparms} unless otherwise labelled in the plots (see main text for discussion).} \label{fig:gc_ds_x}
\end{figure}
In the top figure, the causal feedback delay $\tau$ is set to a similar time scale as the node decay times $1/a, 1/b$. Now subsampling has a comparatively mild distortional effect. We see that oscillations are weak, the plateau below $\Dt = \tau$ has virtually disappeared, GCs peak between $\tau$ and $2\tau$ and decay of GCs for $\Dt > \tau$ is more gradual. The inflexional points for $\Dt \le \tau$ no longer lie near $\tau,\tau/2,\tau/3,\ldots$. In the middle figure, the decay time $1/a$ for $X$ is small (fast decay), while the decay time $1/b$ for $Y$ is large (slow decay). Compared to the reference parameter settings, distortion is comparatively mild (although oscillations are sharply peaked) and we see that for small $\Dt$ the subsampled GC no longer lies completely below the continuous-time GC (\cf~ the discussion at the end of \secref{sec:gcouc}). In the bottom figure the situation is reversed: $X$ decay is slow, while $Y$ decay is fast. Distortion is now strong; subsampled GC is markedly smaller than continuous-time GC (implying poor detectability under subsampling) and oscillations are pronounced. We note that the $c$ parameter (causal feedback strength) has a comparatively small qualitative effect on Granger causalities vs. subsample increment.

\subsection{Statistical inference: detection of causalities} \label{sec:ouminlag:statinf}

Next, we analyse the effects of subsampling frequency on the ability to detect non-zero continuous-time GC. We quantify detectability---\ie, statistical power---via the \emph{Type II error (false negative) rate}: that is, the probability of failure to reject the null hypothesis that $\gc\bY\bX = 0$, at a given significance level $\alpha$. This is given by (\apxref{sec:statinf:gcdetect})
\begin{equation}
	P_{II}(x;\alpha) = F_x\bracr{F_0^{-1}(1-\alpha)} \label{eq:PII1}
\end{equation}
where $F_x$ denotes the cumulative distribution function (CDF) for the estimator $\egc\bY\bX$, given that the actual causality $\gc\bY\bX$ is equal to $x$\;\footnote{Here $\gc\bY\bX$, $\egc\bY\bX$ and the distribution $F_x$ refer to \emph{discrete-time, subsampled} GC, since this is what is estimated in an empirical setting.}. Note that the distribution of $\egc\bY\bX$ depends on the estimation method. Notwithstanding the potential advantages of the state-space approach (\secref{sec:gc}), all Granger causality estimates in this and the following Section are based on VAR modelling\footnote{VAR models were, however, converted to state-space form as a computational device, as explained in \apxref{sec:statinf:ssgc}; this does \emph{not} affect $\egc\bY\bX$, which is still distributed as for VAR-based causality estimation.}, for reasons outlined in \apxref{sec:statinf:ssgc} - principally, the current lack of an analytical expression for the (asymptotic) distribution of $\egc\bY\bX$ in the state-space case. We did in fact repeat, as far as possible, all experiments in this and the following Section via state-space modelling, using surrogate data techniques where appropriate to estimate distributions for $\gc\bY\bX$. Results (not shown) did not differ significantly from the VAR case for our bivariate models with uni-directional causality; see \secref{sec:disc:genrem} for further discussion.

\begin{figure}
\begin{center}
\includegraphics{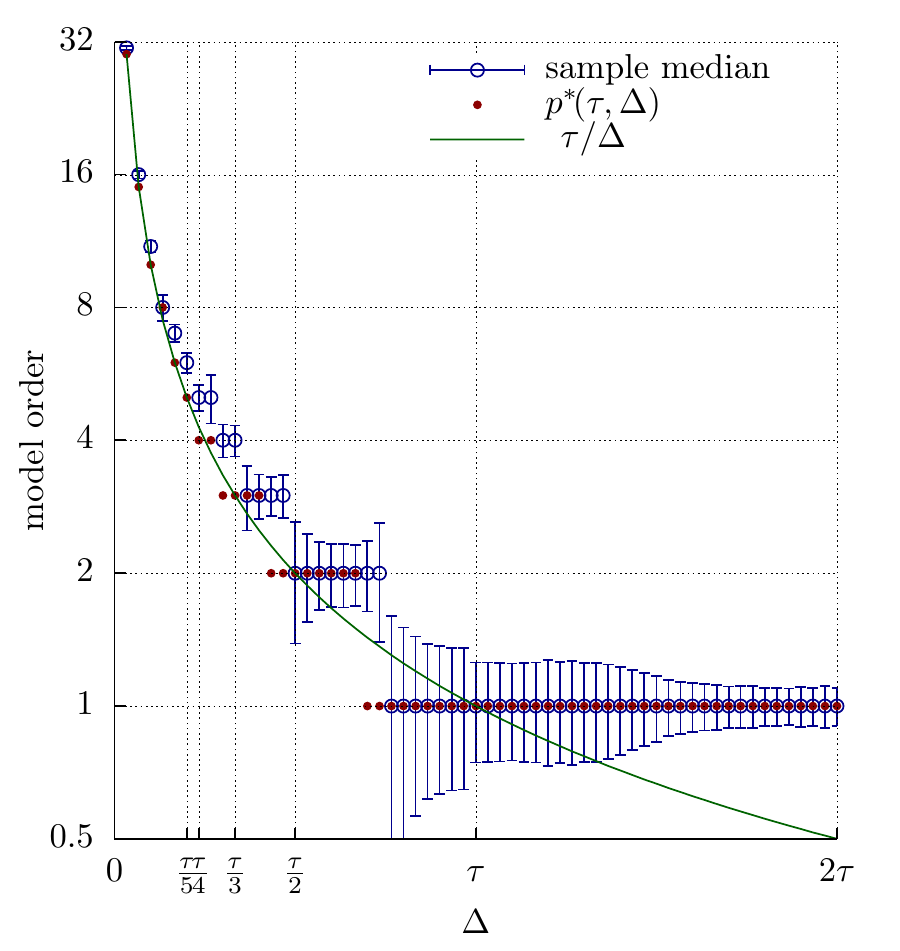}
\end{center}
\caption{Model orders (sample median) as estimated by the AIC from $1000 \times8000\,\text{ms}$ realisations of the minimal CTVAR \eqref{eq:mctv} with reference parameters \eqref{eq:refparms} and $\rho = 0$, plotted alongside the approximation $p^*(\tau,\Dt)$ of \eqref{eq:qDt}. Error bars indicate $\pm 1$ median absolute deviation. (Note the log scale.)} \label{fig:mo_ds}
\end{figure}

For $\egc\bY\bX$ based on a maximum-likelihood VAR model estimate of order $p$, for large sample size (\ie, number of observations) $m$, we have, asymptotically, $mF_x \sim\chi^2(d;mx)$ with $d = pn_xn_y$ degrees of freedom, where $n_x,n_y$ are the dimensions of the variables $\bX,\bY$ respectively  (\apxref{sec:statinf}). Since our subsampled minimal CTVAR is VARMA, rather than (finite-order) VAR (\secref{sec:ouminlag:subsamp}), in order to calculate $P_{II}(x;\alpha)$ \eqref{eq:PII1} we need to select an  empirical VAR model order $p$ for the subsampled process model appropriate to the sample size $m$. Here we use the standard Akaike Information Criterion (AIC) \citep{McQuarrie:1998} for model order selection (\apxref{sec:statinf}). Note that, unlike our results so far, which have been purely analytical, this requires simulation. For a wide range of parameter values we simulated the CTVAR process (see \apxref{sec:voudlsim} for details of our simulation method) and estimated the optimal model order according to the AIC. Results revealed that for a wide range of parameter values the optimal model order at sample interval $\Dt$ is well approximated by
\begin{equation}
	p^*(\tau,\Dt) \equiv \max\left(\left[\tau/\Dt\right],1\right) \label{eq:qDt}
\end{equation}
(where $[\cdot]$ denotes rounding to the nearest integer) independently of the number $m$ of observations (\figref{fig:mo_ds}).
This is intuitively reasonable\footnote{For models with more causal delays and more complex interactions between variables, we might expect a more intricate dependence of empirical VAR model order on model parameter values. Our intuition is, however, that our broad conclusions on detectability which follow from the simple form of \eqref{eq:qDt} would not change drastically.}: although the joint process is VARMA($2,q+1$), and thus in theory may only be represented as an infinite-order VAR, we might expect that with a sample interval of $\Dt$, at least $\tau/\Dt$ autoregression lags will be required to capture feedback at the causal delay $\tau$.

Next we must consider the relationship between sample size $m$ and sample interval $\Dt$. We consider a scenario where we have a CTVAR time series of fixed duration $T$ (measured in ms). The number of subsampled observations is then $m = \lfloor T/\Dt \rfloor$. This scenario corresponds to a realistic use case, where the experimenter has available an electrophysiological recording (at some base sampling frequency) of a given duration, and then has the option of downsampling the (discrete-time) recorded data.

We calculated $P_{II}(x;\alpha)$ from \eqref{eq:PII1} at significance level\footnote{As noted in \apxref{sec:statinf}, a multiple hypothesis correction ought to be made for joint significance testing of causalities in both the $Y\to X$ and $X \to Y$ directions. For simplicity we don't apply any correction here, but note that \eg, for a Bonferroni correction \citep{Hochberg:1987}, this would be equivalent to halving the significance level $\alpha$.} $\alpha = 0.05$. For a range of sample intervals $\Dt$, we used the $\chi^2$ asymptotic distribution of $\egc{Y(\Dt)}{X(\Dt)}$ given the known actual causality $x = \gc{Y(\Dt)}{X(\Dt)}$ as calculated from \eqref{eq:minmod:gcyx}. Model order was specified by the approximation $p^*(\tau,\Dt)$ of \eqref{eq:qDt}; note that while \eqref{eq:qDt} was empirically derived, the calculation of $P_{II}(x;\alpha)$ is purely analytic.

\begin{figure*}
\begin{center}
\includegraphics{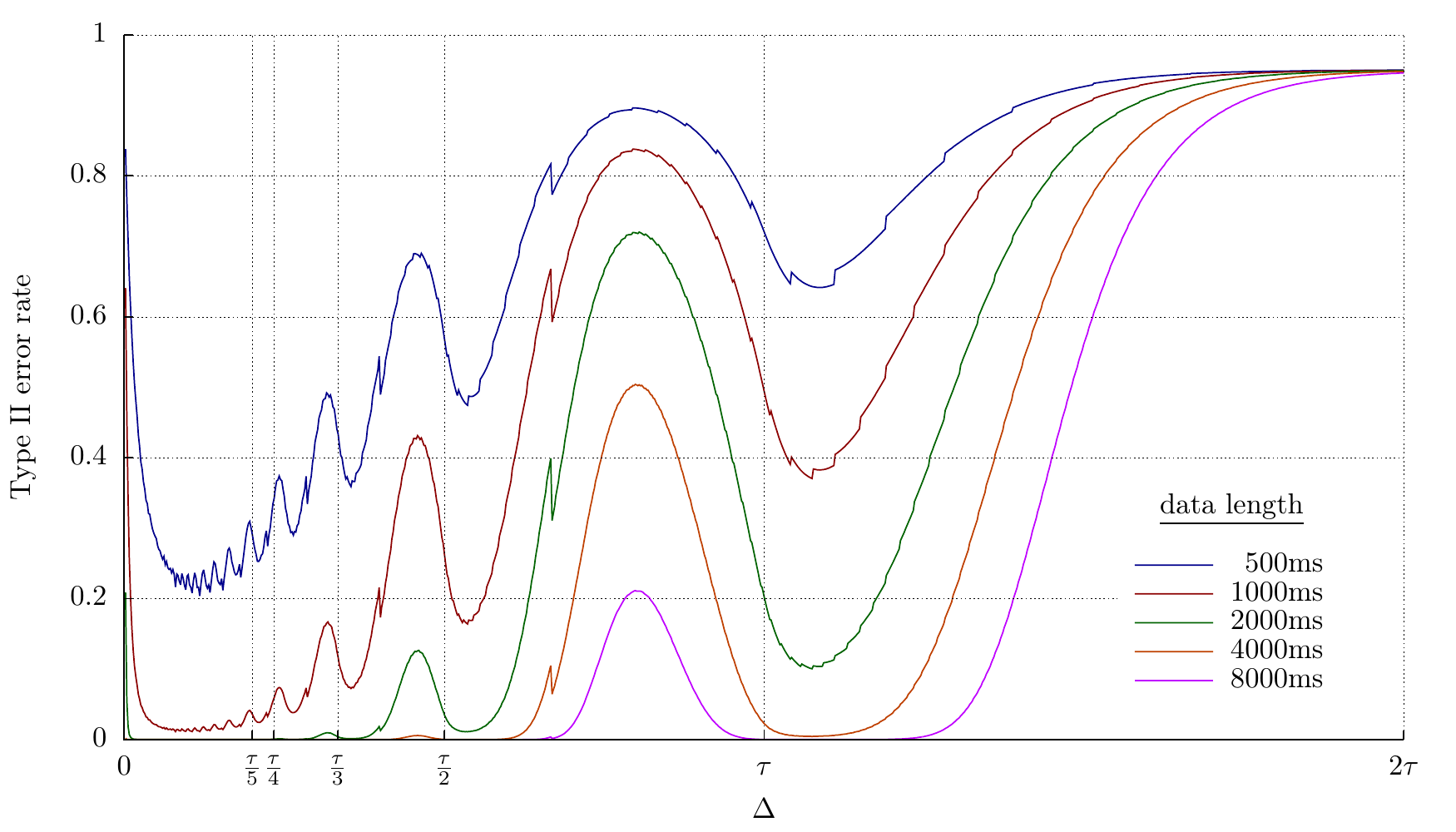}
\end{center}
\caption{Theoretical Type II error rate $P_{II}(x;\alpha)$ \eqref{eq:PII} at significance level $\alpha = 0.05$ for the CTVAR \eqref{eq:mctv} with reference parameters \eqref{eq:refparms} and $\rho = 0$, plotted against sample interval for a range of data sequence lengths. Calculations are based on model order $p^*(\tau,\Dt)$ of \eqref{eq:qDt} (\cf~\figref{fig:mo_ds}).} \label{fig:gc_ds_PII}
\end{figure*}

Results for the reference parameters and a range of data lengths from $T = 500\text{ms} - 8000\text{ms}$ are illustrated in \figref{fig:gc_ds_PII}.
There are three competing effects at play here: firstly, the dependence of ($\Dt$-subsampled) causal magnitude $\gc{Y(\Dt)}{X(\Dt)}$ on $\Dt$ (\figref{fig:gc_ref_ds}), secondly the decrease in sample size $m$ as $\Dt$ increases and thirdly the decrease in model order \eqref{eq:qDt} as $\Dt$ increases. Note that the last two factors, which affect the dispersion of the sample statistic, pull in opposite directions. In the large $\Dt$ limit, we see that, as a result of exponential decay of the actual causality $\gc{Y(\Dt)}{X(\Dt)}$ (\figref{fig:gc_ds}), the ability to detect causality in the $Y \to X$ direction degrades abruptly and rapidly at a point beyond the causal delay $\tau$. At the other extreme, if the subsampling frequency is too high ($\Dt \to 0$) detectability also becomes impossible as $\gc{Y(\Dt)}{X(\Dt)} \to 0$ (\figref{fig:gc_ref_ds}), exacerbated by the associated increase in model order and concommitant high variance of the sample statistic.

\begin{figure}
\begin{center}
\includegraphics{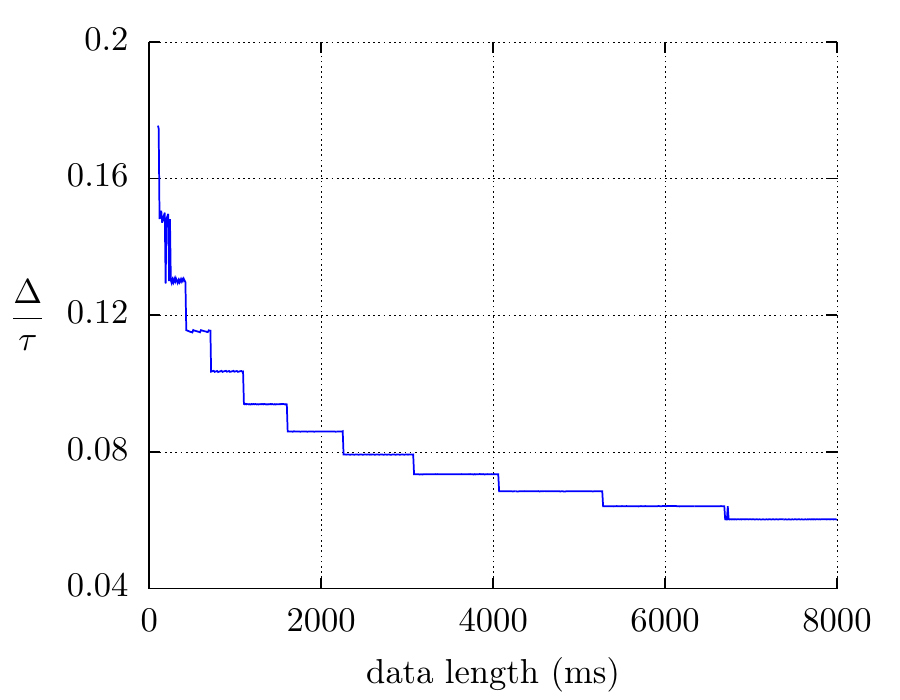}
\end{center}
\caption{Optimal detectability sweet spot $\Dt$ as a fraction of causal delay $\tau$, plotted against data length.} \label{fig:gc_ds_PII_optimal}
\end{figure}

In between these extremes, we see a sequence of detectability ``sweet spots'' and ``black'' spots - values of $\Dt$ which locally minimise (resp. maximise) the Type II error rate. There is an optimal sweet spot (\ie, a value of $\Dt$ which \emph{globally} minimises the Type II error rate) for detectability at a small sampling interval well below the causal delay $\tau$ (\figref{fig:gc_ds_PII_optimal}; for large data lengths it is hard to see these in \figref{fig:gc_ds_PII}, since they become indistinguishable from zero)\footnote{We remark that if, in contrast to the considered usage scenario, the number of \emph{observations} $m$ is held fixed and the duration of the signal $T$ allowed to vary, then numerical computation (results not shown) reveals that the optimal sample interval does not change with the number of observation, and lies slightly \emph{above} the causal delay at $\tau$. We have not been able to establish this analytically. This scenario might, however, be considered less typical.}.
This is followed by a series of oscillations in detectability as $\Dt$ increases. We see black spots---particularly just below the causal delay---where detectability becomes difficult or unfeasible, then recovers, before finally degrading beyond $\tau$. We also see that if the data length is too small, detectability becomes unreliable (or virtually impossible) at \emph{any} subsample frequency.

\begin{figure*}
\begin{center}
\includegraphics{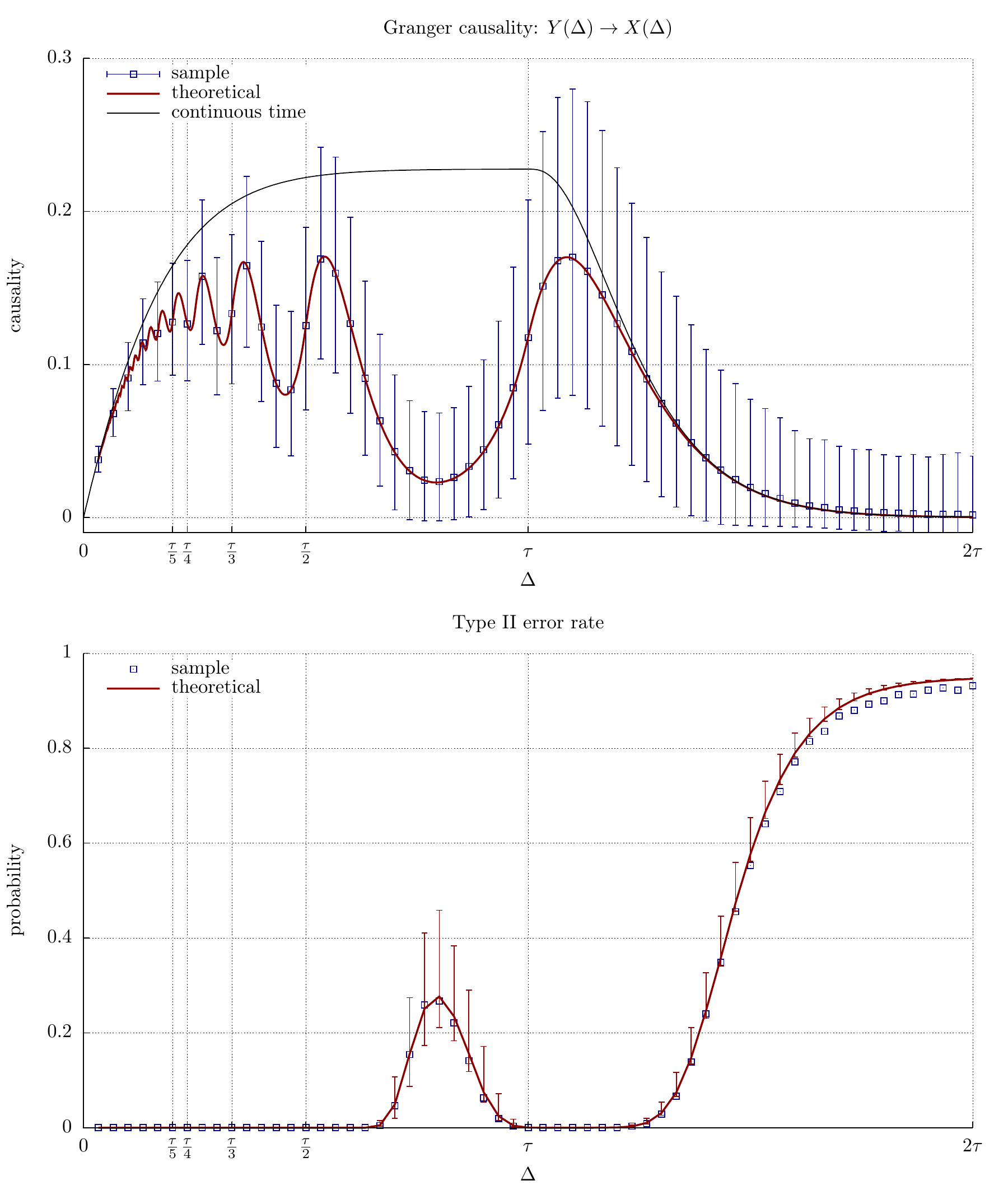}
\end{center}
\caption{Top figure: distribution of the de-biased (significant) sample Granger causality $\egc{Y(\Dt)}{X(\Dt)}$ for the minimal CTVAR \eqref{eq:mctv} with reference parameters \eqref{eq:refparms} and $\rho = 0$, based on $10,000$ realisations of length $8000\,\text{ms}$, plotted against sample interval $\Dt$. Points denote the mean, while error bars depict $95\%$ confidence intervals. The solid red line plots theoretical values \eqref{eq:minmod:gcyx}, while the solid black line plots the continuous-time GC $\gc YX(\Dt)$ at horizon $\Dt$. Bottom figure: Type II error rates at significance level $\alpha = 0.05$ for the same simulations (points). The solid red line plots the mean theoretical values \eqref{eq:PII} based on $\chi^2$ distributions, with error bars at $95\%$ confidence intervals (note that the dispersion in theoretical values is due to variance of the AIC-estimated model orders - see text for details).} \label{fig:sim_oulag_gc12_ds}
\end{figure*}

We then tested our analytical predictions for subsampled GC $\gc{Y(\Dt)}{X(\Dt)}$ \eqref{eq:minmod:gcyx} and Type II error rate $P_{II}(x;\alpha)$ \eqref{eq:PII1} against large-sample simulations, under a more realistic methodology where, rather than using the approximate model order $p^*(\tau,\Dt)$ of \eqref{eq:qDt}, model orders were estimated using the AIC per sample realisation of the CTVAR process\footnote{If the AIC yielded a model order of zero (\ie, it ``sees'' the process as pure white noise), the model order was set to $1$.}. For each sampling interval in the range  $\Dt = 1\,\text{ms} $ to $\Dt = 2\tau = 60\,\text{ms}$, $10,000$ stationary realisations of length $8000\,\text{ms}$ were generated. Causalities were calculated in sample using VAR model estimation and a state-space computational method (\apxref{sec:statinf:ssgc}), as described previously. \figref{fig:sim_oulag_gc12_ds} (top figure) plots the mean $\egc{Y(\Dt)}{X(\Dt)}$, de-biased according to \eqref{eq:gcbias} (note that due to sample fluctuations values can become slightly negative), for data sequences of length $8000\,\text{ms}$ (points, with error bars at $95\%$ empirical confidence intervals\footnote{Empirical confidence intervals were constructed at level $\alpha$ so that a fraction $1-\alpha/2$ of the data points lie below the upper bound and the same fraction lie above the lower bound.}), against the theoretical value $\gc{Y(\Dt)}{X(\Dt)}$ calculated according to \eqref{eq:minmod:gcyx} (red line). The continuous-time GC $\gc YX(\Dt)$ at horizon $\Dt$ is also displayed (black line).
We see excellent agreement of the sample estimates with theory. In the bottom figure, Type II error rates at significance level $\alpha = 0.05$ (calculated as the fraction of sample causality values for which the corresponding p-value is $> \alpha$) are plotted against values calculated as before according to the theoretical $\chi^2$ distributions of $\egc{Y(\Dt)}{X(\Dt)}$. Note that in this figure the error bars around the mean (again at $95\%$ confidence intervals) apply to the \emph{theoretical} values of $P_{II}$, since (for fixed $\Dt$) these vary with the varying model orders as estimated per-sample by the AIC. Again, agreement with theory is excellent (since error rates are calculated on the basis of \emph{asymptotic} statistics, we expect to see some deviation from theoretical predictions for larger values of $\Dt$, where the number of observations is smaller), and we see clear evidence of a detectability black spot at the peak between $\Dt = \tau/2$ and $\Dt = \tau$.

For completeness, we repeated the above experiment, this time in the non-significant $Y \to X$ direction, along with estimation of the \emph{Type I (false positive) error rate}.
\begin{figure*}
\begin{center}
\includegraphics{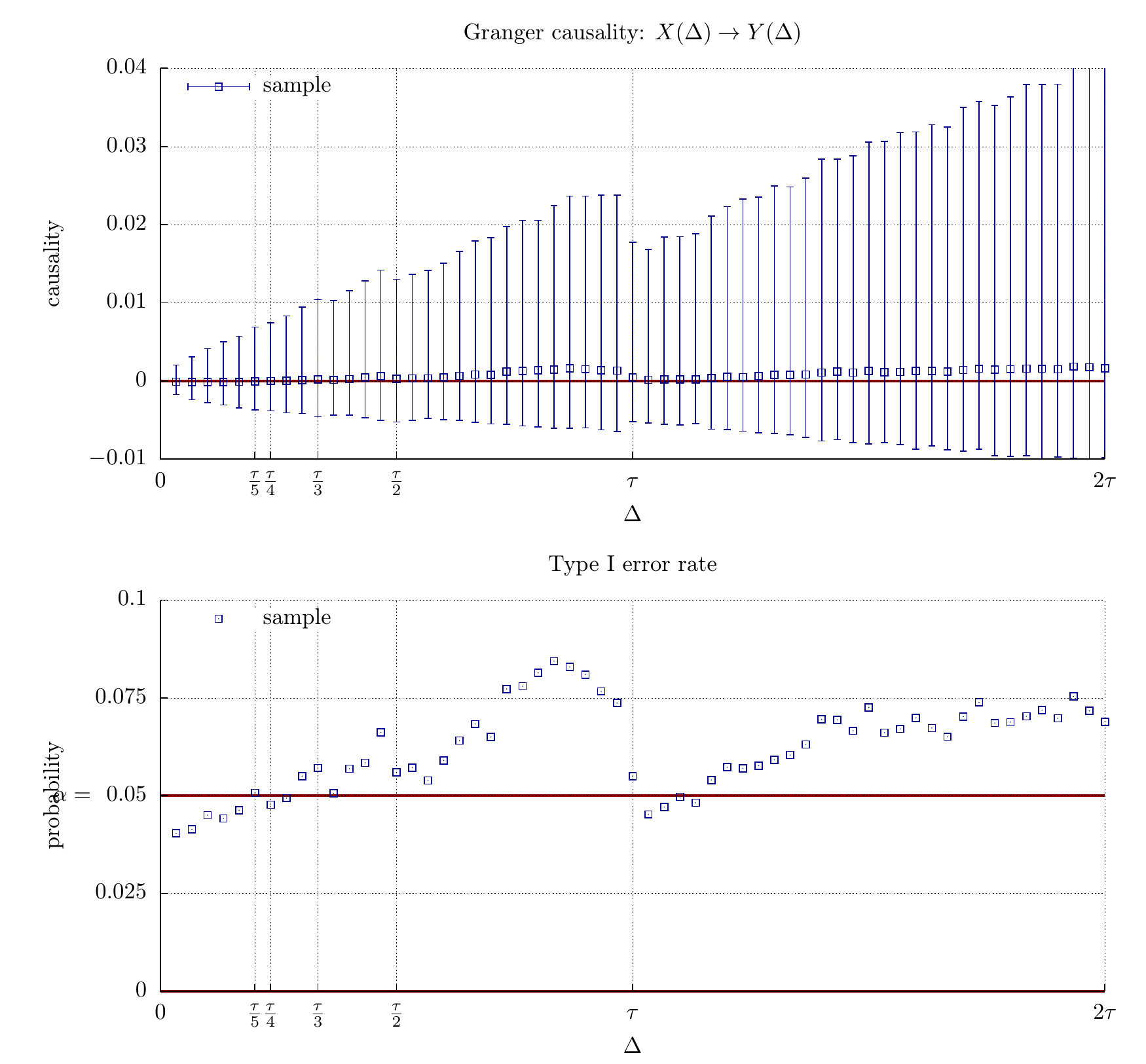}
\end{center}
\caption{Top figure: distribution of the de-biased (non-significant) sample Granger causality $\egc{X(\Dt)}{Y(\Dt)}$ for the CTVAR \eqref{eq:mctv}; parameters as in \figref{fig:sim_oulag_gc12_ds}. Bottom figure: Type I error rates at significance level $\alpha = 0.05$ for the same simulations (points). Note that here the theoretical causality is zero, while the theoretical Type I error rate is just $\alpha$ (solid line).} \label{fig:sim_oulag_gc21_ds}
\end{figure*}
Results are displayed in \figref{fig:sim_oulag_gc21_ds}. The top figure plots the mean de-biased $\egc{X(\Dt)}{Y(\Dt)}$ for data sequences of length $8000\,\text{ms}$, with error bars at $95\%$ confidence intervals. Some structure in the variation of the sample statistic with $\Dt$ is apparent, with an expected increase in variance with increasing $\Dt$. In the lower figure, Type I error rates at a significance level of $\alpha = 0.05$ are plotted. There is again some apparent structure in the variation of the error rate around the theoretical value of $\alpha$, in particular a peak just before $\Dt = \tau$. For larger values of $\Dt$ the Type I error rate is somewhat higher than the theoretical value of $\alpha = 0.05$; this may be explained by deviation of the sampling distribution of $\egc{X(\Dt)}{Y(\Dt)}$ from the theoretical (null) distribution---which again is asymptotic rather than exact---and also the failure of finite-order VAR modelling to capture the VARMA characteristics of the subsampled process.

We note that for the minimal CTVAR with reference parameters \eqref{eq:refparms} and $\rho = 0$, we have $\igc{X(\Dt)}{Y(\Dt)} \ll \gc{Y(\Dt)}{X(\Dt)}$, so that Solo's strong causality measure $\ggc{Y(\Dt)}{X(\Dt)}$ of \eqref{eq:ggc} \citep{Solo:2007} is virtually indistinguishable from $\gc{Y(\Dt)}{X(\Dt)}$ (this was also verified in simulation).

\section{Discussion} \label{sec:disc}

A key question when applying Granger causality to empirically sampled data, is how the relationship between the sampling rate and the time scale(s) of the underlying neurophysiological process affects Granger causal inference.  To address this, we introduce CTVAR processes---a generalisation of the standard vector Ornstein-Uhlenbeck process featuring finite-time, distributed lags---which capture the analogue, continuous-time nature of the underlying signal sampled by neurophysiological recording technologies. We develop a comprehensive theoretical basis, in both time and frequency domain, for CTVAR processes. We then analyse the process derived from a CTVAR process by subsampling at fixed time increments $\Dt$. We show how key quantities such as the autocovariance, CPSD, transfer function and residual noise covariance associated with the VAR model of the subsampled process may be derived from the parameters of a CTVAR model by spectral factorisation, and demonstrate that in the limit $\Dt \to 0$ these quantities approach their continuous-time counterparts.

Then, under the premise that Granger causality in continuous time is best considered at \emph{finite} prediction horizons, we develop a principled theoretical foundation for Granger causality in continuous time, based on the CTVAR formalism. For subsampled CTVAR processes, continuous-time GC appears as a natural limit of multistep discrete-time GC in the limit of a small subsampling increment. The existence of a zero-horizon Granger causality \emph{rate} is demonstrated, and shown to satisfy a continuous-time Geweke frequency decomposition. The properties of subsampled CTVAR processes are discussed, including the possibility of spurious causalities, and the failure of finite-horizon GC invariance under causal, invertible filtering.

Armed with this theoretical background, we consider the effects of subsampling on the \emph{detectability} of Granger causality. We proceed to an exact analytical solution of the subsampling problem for a minimal bivariate CTVAR process with finite causal delay, which we believe to be the first non-trivial full analytic solution in the literature for Granger causality in continuous time and subsampled continuous time. Analytic expressions are calculated for the autocovariance sequence, transfer function and CPSD of the subsampled process. We are able to factorise both the continuous-time and (in the case of uncorrelated residual noise) subsampled CPSD explicitly, leading to analytic expressions for Granger causalities for both the continuous-time and subsampled processes. This facilitates a detailed analysis of the effects of subsampling on detectability. Theoretical predictions are confirmed by large-sample simulations.

\subsection{Relationship to previous work} \label{sec:disc:prework}

Distributed-lag continuous-time processes have previously been considered in the econometrics literature, going back to \cite{Sims:1971,Sims:1972,Geweke:1978}, who also considered subsampling. Distortional effects of subsampling on Granger causality, in particular spurious causality, have been variously addressed, in both continuous time \citep{FlorensFougere:1996,ComteRenault:1996,RenaultEtal:1998,McCrorieChambers:2006} and discrete time \citep{Wei:1981,Marcellino:1999,BreitungSwanson:2002,Solo:2007,Solo:2016,Zhou:2014}.

Closest in spirit to our work is the ``CIMA'' (continuous-time invertible moving-average) model introduced by \cite{ComteRenault:1996}. There, however, only statistical test criteria for (non)causality are considered; the authors stop short of introducing, as we do here, statistical Granger-Geweke measures which quantify \emph{magnitude}, rather than just (non)existence, of Granger-causal effects. Our measures, furthermore, are defined in terms of limits of the corresponding discrete-time subsampled quantities under progressively finer subsampling.

\cite{Solo:2007} distinguishes between ``strong'' and the conventional ``weak'' Granger causality in discrete time (\secref{sec:gc}), noting that only the former is strictly preserved under subsampling. We argue, however, that strong GC is unsatisfactory as a \emph{directional} causality measure, as it inextricably combines the effects of time-directed and contemporaneous feedback. Using a state-space approach, \cite{Solo:2016} expands on the distortion induced by subsampling, including the possibilty of spurious causality [we note (\secref{sec:ouminlag:subsamp}) that for our minimal CTVAR spurious GC does not in fact arise].

\cite{Zhou:2014} consider Granger causality for discrete-time VAR models and a continuous-time integrate-and-fire neural simulation. They report oscillations at a near-constant frequency in estimated causalities plotted against sample interval, with causalities almost vanishing in the troughs. As we have noted, such oscillations are lacking in our minimal CTVAR. This is consistent with the explanation put forward by Zhou \etal\ (demonstrated also for discrete-time VAR models) that these oscillations are associated with periodic behaviour in the pre-subsampled signal, which manifest as peaks in the power spectra. For our minimal CTVAR we note that there are no peaks away from $\lambda = 0$ at $\Dt = 0$ (\figref{fig:cpsd_ds}) - underlining that the simplicity of the model helps isolate the effects of subsampling on GC inference.

\subsection{Original contributions} \label{sec:disc:contrib}

The original contributions of this study may be briefly summarised as:
\begin{enumerate}
	\item Theoretical analysis of CTVAR continuous-time, distributed-lag vector autoregressive processes as an appropriate model for Granger-causal analysis of neural systems.
	\item Natural and intuitive definitions of quantitative finite- and zero-horizon continuous-time Granger-Geweke measures, as limits of corresponding discrete-time quantities; these measures are proposed as ``ground truth'' targets for functional inference of neural systems based on discretely sampled neurophysiological recordings.
	\item Analysis of the relationship between continuous-time GC for a CTVAR process and discrete-time GC for a subsampling of the process.
	\item Demonstration of the non-invariance of finite-horizon GC under causal, invertible filtering.
	\item Complete analytical solution of a minimal, but non-trivial bivariate finite-lag CTVAR, both in continuous time and subsampled, revealing:
	\begin{enumerate}
		\item Exponential decay of subsampled GC for sampling intervals beyond causal feedback time scales.
		\item Resonance between causal and sampling frequencies, resulting in oscillations in subsampled GC for sampling intervals below causal feedback time scales.
	\end{enumerate}
	\item Analysis of the \emph{detectability} of Granger causalities under subsampling. This analysis reveals, in particular, (a) exponential decay of detectability beyond causal feedback time scales, and (b) the existence of detectability black spots and sweet spots in the sample rate.
\end{enumerate}

\subsection{Implications for causal inference from neurophysiological recordings} \label{sec:disc:neuro}

Analysis of the minimal CTVAR indicates that Granger causality in continuous time decays exponentially with increasing prediction horizon beyond causal feedback time scales, leading to exponential decay of subsampled GC with increasing sampling interval and a concomitant sharp drop in detectability. We conjecture that a similar effect will hold under more general conditions. Our analysis also reveals that, insofar as a choice of sample rates is available,  faster is not necessarily better: results for the minimal CTVAR indicate that, for a data segment of fixed length, detectability approaches zero as the sample interval $\Dt \to 0$. This may be viewed in the context of a more general phenomenon: that for some class of systems there is a finite optimal sampling rate for \emph{system identification} of a continuous-time model from subsampled data. \cite{Astrom:1969}, for example, proves that there is a finite optimal sampling rate for identification of the parameters of a (univariate) Ornstein-Uhlenbeck process, above which the variance of parameter estimates increases without bounds.

For our minimal CTVAR, then, there is an optimal sweet spot for sampling frequency that maximises detectability. Furthermore, as $\Dt$ decreases below the causal delay, Granger causality oscillates as the sampling frequency $1/\Dt$ resonates with the causal delay frequency $1/\tau$. These oscillations are distinct from the fixed-period oscillations noted by \cite{Zhou:2014} and have not, as far as we are aware, been reported previously. They may manifest in detectability black spots below the causal delay. Again, we conjecture that these phenomena generalise, although in the case of distributed causal delays sub-causal oscillations are likely to be ``smeared''. An important implication for Granger-causal inference from electrophysiological recordings such as EEG, MEG, LFP, \etc, where the sample rate may be high ($> 1000\,\text{Hz}$) compared with neural time scales, is that downsampling may well be advantageous (anecdotally, it is common in these contexts to downsample to $100-500\,\text{Hz}$ or slower). But this result also suggests, further, that to maximise detectability a range of sampling frequencies should be tested, so as to locate sweet spots and avoid black spots. In addition, we remark (\cf~\secref{sec:voudlds}) that too high a sampling rate could potentially lead to a failure of the minimum-phase condition for the subsampled signal \citep{AstromEtal:1984}.

Exponential decay of detectability has particularly relevance for Granger-causal inference from fMRI recordings \citep{Seth:gcfmri:2013}, since typical fMRI sample rates (currently $0.5-3\,\text{secs}$) are substantially slower than typical synaptic delays ($10-50\,\text{ms}$) in neural systems. fMRI/GC is already controversial for other reasons. \cite{Seth:gcfmri:2013} argue that confounds due to regional variation of the hemodynamic response function (HRF), which mediates the generation of the BOLD (Blood-Oxygen-Level Dependent) signal from the underlying neural activity, are mitigated by the filter-invariance property of discrete-time ($1$-step) GC \citep{Barnett:gcfilt:2011}. \cite{Solo:2016} claims that, while arguably causal, the HRF is unlikely to be minimum phase, so that filter invariance fails. Recent work\footnote{Personal communication.}, however, casts some doubt on this contention, on the grounds that the HRF model analysed by \cite{Solo:2016} may be overly simplistic and thereby misleading. Notwithstanding, our finding that (non-zero) \emph{finite-horizon} GC in continuous time is \emph{not} in general invariant under filtering, implies that the HRF---\emph{even if causal and invertible}---may still distort magnitudes and impact detectability of non-zero causalities at the neural level; that is, even if we could capture the BOLD signal in continuous time, GC analysis might still fail to reflect ``ground truth'' GC at the neural level. Importantly, filter-induced distortion will not give rise to spurious causalities. Note that this potential confound is distinct from subsampling-induced distortions identified for the various fMRI/GC scenarios discussed in \cite{Solo:2016}, since only $1$-step discrete-time GC is considered there. It is, however, far from clear what form finite-horizon GC filter-induced distortion might take for HRF-type filtering; more research is required.

\subsection{General remarks, limitations and future research directions} \label{sec:disc:genrem}

Our research into continuous-time models and subsampling has thrown up a number of conjectures, caveats, technical issues and potential extensions, for which further research is required. These include:
\begin{enumerate}
	\item Further investigation is required into the precise conditions under which subsampling a CTVAR process may induce spurious causalities.
	\item While our analysis of the CTVAR model and continuous-time GC extends for the most part to the important case of \emph{conditional} Granger causality \citep{Geweke:1984}, the situation with subsampling is inevitably more complex. In particular, for conditional GC, eq.~\ref{eq:gcvanish} in discrete time \citep{Lutkepohl:1993} and eq.~\ref{eq:ctgcequiv} in continuous time \citep{DufourRenault:1998} no longer hold; that is, noncausality at the immediate prediction horizon will not generally imply noncausality at larger prediction horizons, since causal effects may propagate through an auxiliary variable. Since neurophysiological time series are in general highly multivariate, further research is necessary to extend our work beyond the bivariate scenario addressed in this study. Given the complexity already apparent in our minimal CTVAR model, this is likely to be highly challenging analytically, and empirical studies will probably be required.
	\item Our minimal CTVAR restricts detailed analysis to the case of unidirectional causality and a single causal feedback delay. In the case of multiple (possibly bidirectional) feedback at varying delays, the situation will be far more complex. We would expect that multiple resonances bewteen sample frequency and causal feedback frequencies will result in more complex oscillations that, we suspect, may manifest in distortion and compromised inference of \emph{relative} GC magnitudes by subsampling. Further research, both analytic and empirical, is required.
	\item We have also, in this study, omitted analysis of the effects of measurement noise on Granger-causal inference \citep{Solo:2007}, and its interaction with subsampling and filtering.
	\item Our definition of CTVAR processes is not mathematically complete in terms of the (spectral) conditions under which our requirements that (i) a subprocess of a CTVAR be a CTVAR and (ii) a subsampled CTVAR be a VAR will be satisfied. In addition, we have not identified precise conditions on a continuous-time CPSD which will guarantee a unique spectral factorisation.
	\item It seems likely that a principled definition for continuous-time finite-horizon spectral Granger causality exists. Such a measure should satisfy the Geweke frequency decomposition $\gc\bY\bX(h) = \int_{-\infty}^\infty \sgc\bY\bX(\lambda; h) \,d\lambda$\,.
	\item It would be of interest to investigate the relationship between subsampled GC resonance oscillations (and/or post-delay maxima) as observed in the minimal CTVAR and the general problem of inference of feedback time scales in linear systems \citep{Bjorklund:2003,WibralEtal:2013}.
	\item As mentioned in \secref{sec:ouminlag}, there is a substantial existing literature on Stochastic Delay-Differential Equations (SDDEs) \citep{Longtin:2010}, which, although in general not focused on Granger-causal/information-theoretic analysis, may be useful in that direction. A related promising research direction, particularly for non-stationary and/or nonlinear systems in continuous time, is the inference of causal/driving mechanisms via Fokker-Planck equations \citep{PrusseitLehnertz:2008,WahlEtal:2016}.
\end{enumerate}

In a more general vein, There is a growing consensus that state-space modelling should be the preferred method---at least in discrete time---for performing Granger-causal inference \citep{Solo:2007,ValdesSosaEtal:2011,Seth:gcfmri:2013,FristonEtal:2014,Seth:gcneuro:2015,Barnett:ssgc:2015,Solo:2016}. This consensus is based on observations that state-space processes---unlike VAR processes---are closed under subsampling, the addition of additive noise, linear digital filtering, and subprocess extraction). Consequently, powerful and efficient new methods for Granger-causal analysis of state-space systems have now been developed \citep{Barnett:ssgc:2015, Solo:2016}. It is not clear, however, how we might extend the state-space/GC paradigm to continuous time \emph{with distributed lags}, since a na\"ive approach would seem to require an infinite-dimensional state space. We have also noted a current sticking point for state-space Granger-causal inference: that the sampling distribution for Granger causality statistics based on (maximum-likelihood) state-space model estimation remains unclear, necessitating computationally costly surrogate methods for statistical inference; more research, both theoretical and empirical, is required in this area. A further promising avenue of research is the deployment of state-space methods to reconstruct models at finer time scales than the sampling frequency

\subsection{Summary} \label{sec:disc:summary}

Granger causal analysis of subsampled time-series data will inevitably be susceptible to distortion, and detectability in particular will degrade as the sampling interval increases beyond the natural time scale(s) of the underlying process. This is to be expected, since subsampling results in the loss of the predictive information which underpins Granger causality. In this study we have characterised how subsampling affects detectability through an exact analytic solution of the subsampling problem for a continuous-time processes with finite causal delay. Our analysis reveals a rapid decay of detectability for large subsampling intervals, but also the existence of detectability `black'' and ``sweet'' spots as the sampling frequency interacts with the underlying generative time scale(s). The theoretical basis for these findings provides a very general framework for further investigations of statistical inference on sampled continuous-time processes with causal interactions over multiple time scales. This encompasses a very wide range of possible scenarios. Overall, our results indicate that Granger causality analysis will be most successful when data are sampled fast enough to capture the relevant causal time scales, but not so fast as to impair detectability. Thus Granger causality analysis will be most effective when informed by sensible priors about domain-specific time scales. Further research using state-space approaches may shed light on these issues.

\subsection*{Acknowledgements}

We thank Adam Barrett for helpful discussion, and an anonymous reviewer for insightful comments. We are grateful to the Dr. Mortimer and Theresa Sackler Foundation, which supports the Sackler Centre for Consciousness Science.

\appendix

\section{Fourier transforms in discrete and continuous time} \label{sec:ftran}

For a discrete-time sequence $x = \{x_k \,|\, k \in \integers\}$ (the $x_k$ may be random or deterministic, real or complex, scalar, vector, matrix, \etc)  with time step $\Dt$, we define the (two-sided) Fourier transform $\hat{x} \equiv \bracc{\rcond{\hat{x}(\lambda)}{-\infty < \lambda < \infty}}$ as
\begin{equation}
	\hat{x}(\lambda) \equiv \Dt \sum_{k = -\infty}^\infty x_k e^{-2\pi i \Dt\lambda k} \label{eq:dft}
\end{equation}
which is periodic in $\lambda$ with period $f_s = 1/\Dt$, the sampling frequency. We shall often restrict $\hat{x}(\lambda)$ to the interval $-1/(2\Dt) \le \lambda < 1/(2\Dt)$, where $1/(2\Dt) = f_s/2$ is the Nyqvist frequency. Note that we scale $\hat{x}(\lambda)$ by the sample interval $\Dt$; this ensures that the transform has the same dimensions for discrete and continuous-time transforms (see below). The original sequence may be recovered via the inverse transform
\begin{equation}
	x_k = \int_{-\frac1{2\Dt}}^{\frac1{2\Dt}} \hat{x}(\lambda) e^{2\pi i \Dt\lambda k} \,d\lambda \label{eq:idft}
\end{equation}

For a continuous-time sequence $x = \{x(t) \,|\, t \in \reals\}$, we define the Fourier transform $\hat{x} \equiv \bracc{\rcond{\hat{x}(\lambda)}{-\infty < \lambda < \infty}}$ as
\begin{equation}
	\hat{x}(\lambda) \equiv \int_{-\infty}^\infty x(t) e^{-2\pi i \lambda t} \,dt \label{eq:cft}
\end{equation}
and the original sequence may be recovered via the inverse transform
\begin{equation}
	x(t) = \int_{-\infty}^\infty \hat{x}(\lambda) e^{2\pi i \lambda t} \,d\lambda \label{eq:icft}
\end{equation}
We note, in particular, that for a discrete-time sequence $x(\Dt) \equiv \{x(k\Dt) \,|\, k \in \integers\}$ obtained by subsampling the continuous-time sequence $x = \{x(t) \,|\, t \in \reals\}$ at regular intervals $\Dt$, we have
\begin{equation}
	\lim_{\Dt \to 0} \widehat{x(\Dt)} = \hat{x} \label{eq:ftlim}
\end{equation}
pointwise; \ie, for fixed $\lambda$, $\widehat{x(\Dt)}(\lambda) \to \hat{x}(\lambda)$ as $\Dt \to 0$, at least insofar as the sum in \eqref{eq:dft} converges to the (Rieman) integral in \eqref{eq:cft}.

\section{Invariance of discrete-time 1-step Granger causality under causal, invertible filtering} \label{sec:finv}

Let $\cG(z) = \sum_{k = 0}^\infty G_k z^k$ be a causal invertible filter (without loss of generality we assume $G_0 = I$). Stability and minimum-phase require that both $|\cG(z)|$ and $|\cG(z)|^{-1}$ are non-zero on the unit disc $|z| \le 0$. If $\bX_k = \Psi(z) \cdot \beps_k$ is the MA representation of a VAR process, then clearly (we denote quantities relating to the filtered process with a tilde) $\tilde\bX_k \equiv \cG(z) \cdot \bX_k = \tilde\Psi(z) \cdot \beps_k$, where $\tilde\Psi(z) = \cG(z)\Psi(z)$, will also be a (stable, minimum-phase) VAR with residual noise intensity $\tilde\Sigma = \Sigma$.

Suppose now that $\trans{[\trans\bX \trans\bY]}$ is a joint VAR process, and that $\cG_{xy}(z) \equiv 0$; \ie, $\cG(z)$ is lower block-triangular. Let $\bX_k = \Psi'_{xx}(z) \cdot \beps'_{x,k}$ be the reduced MA representation of $\bX_k$ alone. Then $\tilde\bX_k = \cG_{xx}(z) \cdot \bX_k = \tilde\Psi'_{xx}(z) \cdot \beps'_{x,k}$ where $\tilde\Psi'_{xx}(z) = \cG_{xx}(z) \Psi'_{xx}(z)$ so that the reduced residuals intensity $\tilde\Sigma'_{xx}$ of the filtered process $\tilde\bX$ alone is $\tilde\Sigma'_{xx} = \Sigma'_{xx}$. But $\tilde\Sigma_{xx} = \Sigma_{xx}$, so from \eqref{eq:gc} we have $\gc{\tilde\bY}{\tilde\bX} = \gc\bY\bX$. It follows straightforwardly from \eqref{eq:trfun}, \eqref{eq:specfac} and \eqref{eq:sgc} that $\sgc{\tilde\bY}{\tilde\bX}(\lambda) = \sgc\bY\bX(\lambda)$ for all $\lambda$.

\section{Non-invariance of discrete-time multistep Granger causality under causal, invertible filtering} \label{sec:nofinv}

With reference to the argument in \apxref{sec:finv}, consider now the case $m > 1$: \eqref{eq:varpredmse} gives $\tilde\MSE_m  = \Dt \sum_{\ell=0}^{m-1} \tilde B_\ell \Sigma \trans{\tilde B_\ell}$, where the filtered MA coefficients are $\tilde B_\ell = \sum_{k = 0}^\ell G_k B_{\ell-k}$. A moment's consideration reveals that, even with $\cG_{xy}(z) \equiv 0$, $\tilde\MSE_m$ will not in general block-decompose conveniently, unless $\Psi_{xy}(z) \equiv 0$.

This is perhaps best illustrated by example. Consider the VAR(1)
\begin{subequations}
\begin{align}
	X_k &= a X_{k-1} + c Y_{k-1} + \eps_{x,k} \label{eq:ni2:varx} \\
	Y_k &= b Y_{k-1} \hspace{33.5pt} + \eps_{y,k} \label{eq:ni2:vary}
\end{align} \label{eq:ni2:var}%
\end{subequations}
with $|a|,|b| < 1$ (for stability) and $\Sigma = I$. We have $\Phi(z) = I-Az$ with $A = \begin{bmatrix} a & c \\ 0 & b \end{bmatrix}$, so that
\begin{equation}
	\Psi(z) = \begin{bmatrix}
		\displaystyle \frac1{1-az} & \displaystyle \frac{cz}{(1-az)(1-bz)} \\[1em]
		0 & \displaystyle \frac1{1-bz}
	\end{bmatrix} \label{eq:ni2:ma}
\end{equation}
We have $B_1 = A$, so that from \eqref{eq:varpredmse}
\begin{equation}
	\MSE_{2,xx} = \bracs{\Sigma + B_1 \Sigma \trans{B_1}}_{xx} = \bracs{I + A \trans A}_{xx} = 1+a^2+c^2 \label{eq:ni2:mse2}
\end{equation}
Spectral factorisation yields
\begin{equation}
	S_{xx}(z) = \bracs{\Psi(z) \Psi(z)^*}_{xx} = \frac{|1-bz|^2 + c^2}{|1-az|^2 |1-bz|^2}, \quad |z| = 1 \label{eq:ni2:Sxx}
\end{equation}
We seek a reduced spectral factorization of the form $S_{xx}(z) = \Psi'_{xx}(z) \Sigma'_{xx} \Psi'_{xx}(z)^*$ on $|z| = 1$ with
\begin{equation}
	\Psi'_{xx}(z) = \frac{1-hz}{(1-az)(1-bz)}, \qquad \Sigma'_{xx} = \upsilon \label{eq:ni2:sf}
\end{equation}
Comparing with \eqref{eq:ni2:Sxx}, we obtain
\begin{subequations}
\begin{align}
	\upsilon h &= b \label{eq:ni2:hv1} \\
	\upsilon\bracr{1+h^2} &= 1+b^2+c^2 \label{eq:ni2:hv2}
\end{align} \label{eq:ni2:hv}%
\end{subequations}
so that $\upsilon$ satisfies the quadratic equation $\upsilon^2 - 2D\upsilon + b^2$, $D \equiv \frac12\bracr{1+b^2+c^2}$, and \eqref{eq:ni2:hv} has the solution\footnote{It may be confirmed that the positive square root should be taken in the expression for $\upsilon$; see \eg, \cite{Barnett:gcfilt:2011}.}
\begin{equation}
	\upsilon = D + \sqrt{D^2-b^2}, \qquad h = \frac1b \bracr{D - \sqrt{D^2-b^2}} \label{eq:ni2:vh}
\end{equation}
From \eqref{eq:gc}, the $1$-step GC $\gc YX$ is then just $\log\upsilon$ \citep{Barnett:gcfilt:2011}. From \eqref{eq:ni2:sf}, collecting the $z^1$ terms, we have $B'_{1,xx} = a+b-h$, so that from \eqref{eq:varpredmse} we find
\begin{equation}
	\MSE'_{2,xx} = \upsilon\bracs{1+(a+b-h)^2} \label{eq:ni2:mse2r}
\end{equation}
and from \eqref{eq:hgc} with \eqref{eq:ni2:mse2} we have
\begin{equation}
	\gc Y{X,2} = \log\frac{\upsilon\bracs{1+(a+b-h)^2}}{1+a^2+c^2} \label{eq:ni2:gc2}
\end{equation}
Note that $\gc YX = 0$ iff $c = 0$, in which case we may check that $\gc Y{X,2} = 0$ as expected. Also, if $a+b = 0$ but $c \ne 0$, then from \eqref{eq:ni2:hv2} we see that $\gc Y{X,2} = 0$ while $\gc YX > 0$, confirming our observation in \secref{sec:gc} that, for $m > 1$, vanishing $\gc Y{X,m}$ does not imply vanishing $\gc YX$.

Now let us define the causal filter
\begin{equation}
	\cG(z) \equiv \begin{bmatrix} \displaystyle \frac{1-gz}{1-az} & 0 \\[1em] 0 & 1 \end{bmatrix}
\end{equation}
$\cG_{xy}(z) \equiv 0$ and invertibility requires $|g| < 1$. Applying $\cG_{xy}(z)$ to the MA operator  \eqref{eq:ni2:ma} we see immediately that the filtered process is identical to the original process \eqref{eq:ni2:var}, but with $a$ replaced by $g$. Then, since $a$ appears explicitly in the expression \eqref{eq:ni2:gc2} for $\gc Y{X,2}$, we see that filter-invariance fails for $2$-step GC (note that $\gc YX$ does not depend on $a$, so that the $1$-step GC is invariant under $\cG$).

\section{Estimation and statistical inference for discrete-time Granger causality} \label{sec:statinf}

Further motivation for the definition \eqref{eq:gc} of (time domain) Granger causality stems from a \emph{maximum likelihood} (ML) perspective. For simplicity we consider only $1$-step GC, although much of our exposition translates to $m$-step prediction via the relations \eqref{eq:varpredmse} and \eqref{eq:hgc}. Given a finite set of observations of the joint process $\trans{[\trans\bX \trans\bY]}$, the full and reduced $1$-step predictions correspond to the respective linear regression models
\begin{subequations}
\begin{align}
	\bX_k &= \sum_{\ell = 1}^\infty A_{xx,\ell} \bX_{k-\ell} + \sum_{\ell = 1}^\infty A_{xy,\ell} \bY_{k-\ell} + \beps_{x,k} \label{eq:est:full} \\
	\bX_k &= \sum_{\ell = 1}^\infty A'_{xx,\ell} \bX_{k-\ell} \hspace{62pt} + \beps'_{x,k} \label{eq:est:red}
\end{align} \label{eq:est}%
\end{subequations}
We may then ask whether the full model \eqref{eq:est:full} furnishes a more likely model (in the ML sense) for the data than the (nested) reduced model \eqref{eq:est:red}. Truncating the regressions to an appropriate finite model order $p$, which in an empirical setting may be estimated by standard model selection techniques such as the Akaike or Bayesian error criterion, cross-validation, \etc\ \citep{McQuarrie:1998}, the Neyman-Pearson lemma \citep{NeymanPearson:1928,NeymanPearson:1933} tells us that the uniformly most powerful (UMP) test for the null hypothesis
\begin{equation}
	H_0 : A_{xy,1} =  A_{xy,2} = \ldots = A_{xy,p} = 0  \label{eq:gcH0}
\end{equation}
against the alternative hypothesis that at least one of the $A_{xy,k}$ is non-zero, is the \emph{log-likelihood ratio} statistic, which by standard theory is just
\begin{equation}
	\egc\bY\bX \equiv \log\frac{\dett{\hSigma'_{xx}}}{\dett{\hSigma_{xx}}} \label{eq:egc}
\end{equation}
where $\hSigma_{xx}, \hSigma'_{xx}$ are ML estimators for the residuals covariance matrices of the respective models. Thus, in principle, time-domain Granger causality \eqref{eq:gc} [and instantaneous causality \eqref{eq:igc}] may be estimated in sample by replacing the respective residuals covariance matrices by ML estimates (but see below). For Gaussian processes we note that the standard ordinary least squares \citep[OLS:][] {Hamilton:1994} or Levinson-Wiggins-Robinson \citep[LWR:][]{Levinson:1947,Whittle:1963,WigginsRobinson:1965,MorfEtal:1978} estimators for a residuals covariance matrix are ML estimators, while in the non-Gaussian case they are asymptotically equivalent \citep{Lutkepohl:2005}. Standard large-sample theory \citep{Wilks:1938,Wald:1943} then yields an asymptotic sampling distribution for the Granger causality estimator \eqref{eq:egc}: specifically, if the statistic is based on $m$ observations and the number of degrees of freedom is $d = pn_xn_y$, where $n_x,n_y$ are the dimensionalities of $\bX,\bY$ respectively, then $m\!\egc\bY\bX \sim \chi^2(d,m\!\gc\bY\bX)$ (non-central $\chi^2$ distribution if $\gc\bY\bX > 0$ or central if $\gc\bY\bX = 0$)\footnote{In the case of a univariate causal target (\ie\ $n_x = 1$) an alternative asymptotic sampling distribution is available for the $R^2$-like statistic $\exp(\egc\bY\bX)-1$, scaled by sample size, as a central $F$-distribution under the null and a non-central $F$-distribution under the alternative hypothesis. According to \cite{Hamilton:1994}, for small samples in particular, the $F$-distribution may be preferable (it has a fatter tail than the corresponding $\chi^2$ distribution).}.  Note that since Granger causality statistics are all non-negative, their estimators will be positively biased - but the expected bias may easily be calculated \citep{Zhou:2014} as
\begin{equation}
	\expect{\egc\bY\bX-\gc\bY\bX} = \frac dm \label{eq:gcbias}
\end{equation}
regardless of the actual causality $\gc\bY\bX$; \eqref{eq:gcbias} may be used to obtain asymptotically unbiased estimates of Granger causalities.

To estimate $\gc\bY\bX$ for discrete time-series data, a na\"ive implementation where full and reduced models \eqref{eq:est:full}, \eqref{eq:est:red} are estimated \emph{separately} in sample will not suffice - this leads to inaccurate (even potentially negative) causality estimates. Rather---after selecting a suitable model order---the full model must be estimated, and the reduced model estimate calculated from the full model parameters. This step essentially involves spectral factorisation \eqref{eq:specfac} of the reduced model CPSD, which may be effected computationally \eg, via Wilson's frequency-domain algorithm \citep{Wilson:1972,Dhamala:2008b,Dhamala:2008a}, Whittle's time-domain algorithm \citep{Whittle:1963,Barnett:mvgc:2014}, or via \emph{state-space} methods (see below). Alternatively, the full model CPSD may be estimated directly from the data by standard methods (so-called ``nonparametric'' estimation), and factored separately for the full and reduced models \citep{Dhamala:2008b,Dhamala:2008a}. To estimate $\sgc\bY\bX(\lambda)$, from \eqref{eq:sgc} we see that spectral factorisation is unnecessary; only the full model parameters are required\footnote{This is not true for the important case of \emph{conditional} Granger causality \citep{Geweke:1984}, where spectral factorisation is still required \citep{Barnett:mvgc:2014}. \cite{Chen:2006}, while recognising the issue, propose an invalid computational method which attempts to avoid spectral factorisation - see \cite{Solo:2016} for further commentary.}. As regards statistical inference, in contrast to the time-domain case no sampling distribution for $\esgc\bY\bX(\lambda)$ (asymptotic or exact) is known [see \cite{Geweke:1984} for a fuller discussion on this issue] and nonparametric subsampling or surrogate data techniques are best deployed for significance testing and derivation of confidence intervals.

\subsection{State-space methods} \label{sec:statinf:ssgc}

Recently, efficient and practical (linear, discrete-time) state-space methods have been proposed for Granger-causal estimation \citep{Barnett:ssgc:2015,Solo:2016}. Unlike VAR models, state-space models accommodate a moving-average component in the data parsimoniously. It is known that sub-model extraction, invertible filtering, additive noise and---crucially---subsampling all induce a moving-average component \citep{NsiriRoy:1993,Solo:2007,Barnett:gcfilt:2011}. The significant advantage of state-space over VAR-based estimation is that the class of state-space models---equivalently VARMA models---is closed under all these operations. This, together with the availability of efficient (non-iterative) \emph{state-space subspace} system identification algorithms \citep{VOandDM:1996}, makes state-space methods an attractive approach to Granger causality estimation.

Pertinently to this study, we remark that a critical difference between state-space and VAR estimation is that the (minimal) state-space model order \emph{is not increased by subsampling} \citep{Solo:2016}. An issue with state-space Granger causality estimation, however, is that the distribution of the sample statistic remains unclear. While eq.~(17) in \cite{Barnett:ssgc:2015} suggests that the number of degrees of freedom for a Granger-causal sample statistic based on a maximum-likelihood state-space estimate of order $p$ should again be $d = pn_xn_y$, this has not been established rigorously (or indeed empirically)\footnote{\cite{Solo:2016} states without proof that the degrees of freedom is $d = 2pn_xn_y$; again, we have been unable to verify this.}.

VAR-based Granger causality estimation may also be enhanced by state-space methods: an (estimated)  VAR model may easily be converted to a state-space model in ``innovations form'' \citep{HandD:2012}. The method of \cite{Barnett:ssgc:2015} and \cite{Solo:2016} may then be applied to yield Granger causality estimates. This procedure exploits an additional, \emph{computational}, advantage of the state-space approach, that calculation of a reduced innovations-form state-space model from the full model (the spectral factorisation step) is achieved by solution of a single discrete algebraic Riccati equation (DARE), for which stable and efficient algorithms exist \citep{ArnoldLaub:1984,LancasterRodman:1995}. Note that in this case, since it is a VAR model which is actually estimated, causality estimates follow a $\chi^2$ sampling distribution with $d = pn_xn_y$ degrees of freedom, where $p$ is the VAR model order. This technique was used for computation of all empirical VAR-based Granger causality estimates in this study.

\subsection{Detecting Granger causality} \label{sec:statinf:gcdetect}

Suppose that the Granger causality sample statistic $\EGC$ (for some given number of variables, model order and data sample size) has a cumulative distribution function $F_x(u)$, given that the \emph{actual} Granger causality of the underlying stochastic process is $x$. We know that in the large-sample limit, $F_x(u)$ approaches asymptotically a non-central $\chi^2$ (scaled by sample size) with non-centrality parameter $x$, or a central $\chi^2$ if $x = 0$. The number of degrees of freedom is $d = n_1 n_2 p$ where $n_1$ (resp. $n_2$) is the number of target (resp. source) variables and $p$ the VAR model order. To test for significance at level $\alpha$ of a sample Granger causality value of $u$, the null hypothesis of zero Granger causality is $x = 0$, with null distribution $F_0(u)$. We thus \emph{accept} the null hypothesis---\ie\ we take $u$ as \emph{non}significant---if $u \le F_0^{-1}(1-\alpha)$ [equivalently, the p-value $1-F_0(u)$ is $\ge \alpha$]. However, if $x$ (the true Granger causality) is in fact $ > 0$ then this is a Type II error; \ie\ a false negative. Thus, given an actual causality $x > 0$, the Type II error rate (\ie\ probability of a Type II error) is
\begin{equation}
	P_{II}(x;\alpha) \equiv \prob{\EGC \le F_0^{-1}(1-\alpha)} = F_x\bracr{F_0^{-1}(1-\alpha)} \label{eq:PII}
\end{equation}
Note that (for fixed sample size) as $x \to \infty$, $P_{II}(x;\alpha) \to 0$ and as $x \to 0$, $P_{II}(x;\alpha) \to 1$. We consider the Type II error rate  $P_{II}(x;\alpha)$ as a measure of \emph{detectability}\footnote{We prefer to talk about detectability rather than statistical \emph{power}; the power of the statistical test is, of course, just $1-P_{II}(x;\alpha)$.}  in finite sample of a Granger causality known to be equal to some $x > 0$: if $P_{II}(x;\alpha)$ is too large then we will infer false negatives---that is, fail to detect a significant causality---unacceptably frequently. We might, then, describe a G-causal value $x$ as ``undetectable at significance level $\alpha$'' if $P_{II}(x;\alpha) > \alpha'$, where $\alpha'$ represents an ``acceptable'' incidence of Type II errors (we could take $\alpha' = \alpha$).

Note that the Type I error rate at significance level $\alpha$ (that is, the probability of falsely rejecting the null hypothesis of zero causality when the true causality actually \emph{is} zero) is, trivially, just $\alpha$. Note too, that for joint significance testing of \emph{multiple} Granger causality statistics, a multiple hypothesis, family-wise error rate or false discovery rate correction should be applied \citep{Hochberg:1987}.

\section{Inversion of the MA operator for a CTVAR process} \label{sec:ctgenfun}

Integrating \eqref{eq:cvmagen} by parts we find
\begin{align*}
	\zeta\Psi(\zeta)
	&= \zeta \int_{u = 0}^\infty B(u) \,e^{-\zeta u} \,du \\
	&= - \int_{u = 0}^\infty B(u) \,d\bracr{e^{-\zeta u}} \\
	&= \int_{u = 0}^\infty \dot B(u) \,e^{-\zeta u} \,du - \Big[{B(u) \,e^{-\zeta u}}\Big]_{u = 0}^\infty \\
	&= \int_{u = 0}^\infty \dot B(u) \,e^{-\zeta u} \,du + I \quad\text{by } \eqref{eq:makern0} \text{ and } \re(\zeta) \ge 0 \\
	&= \int_{u = 0}^\infty \int_{s = 0}^u A(s) B(u-s) \,e^{-\zeta u} \,ds \,du + I \quad\text{by } \eqref{eq:makern1} \\
	&= \int_{s = 0}^\infty \int_{u = s}^\infty A(s) B(u-s) \,e^{-\zeta u} \,du \,ds + I  \\
	&= \int_{s = 0}^\infty \int_{v = 0}^\infty A(s) B(v) \,e^{-\zeta(s+v)} \,dv \,ds + I \quad v = u-s \\
	&= \int_{s = 0}^\infty  A(s) \,e^{-\zeta s} \,ds \int_{v = 0}^\infty B(v) \,e^{-\zeta v} \,dv + I \\
	&= \bracs{\zeta I-\Phi(\zeta)} \Psi(\zeta) + I
\end{align*}
so that $\Psi(\zeta) = \Phi(\zeta)^{-1}$ on $\re(\zeta) \ge 0$ as required.

\section{ODE for the MA kernel of a CTVAR process} \label{sec:ctma}

Writing \eqref{eq:ctvarma} as $\bX(s) = \int_{u = -\infty}^s B(s-u) \,d\bW(u)$ we have, working to first order in $ds$:
\begin{align*}
	& d\bX(s) \\
	&= \int_{u = -\infty}^{s+ds} B(s+ds-u) \,d\bW(u) - \int_{u = -\infty}^s B(s-u) \,d\bW(u) \\
	&= \int_{u = -\infty}^s B(s+ds-u) \,d\bW(u) - \int_{u = -\infty}^s B(s-u) \,d\bW(u) \\
	&\phantom= + \int_{u = s}^{s+ds} B(s+ds-u) \,d\bW(u) \\
	&= \int_{u = -\infty}^s \dot B(s-u) \,d\bW(u) \,ds + \int_{u = s}^{s+ds} B(s+ds-u) \,d\bW(u) \\
	&= \int_{u = -\infty}^s \dot B(s-u) \,d\bW(u) \,ds + B(0) \,d\bW(s)
\end{align*}
so setting $t \equiv s-u$
\begin{equation}
	 d\bX(s) = \int_{t = 0}^\infty \dot B(t) \,d\bW(s-t) \,ds + B(0) \,d\bW(s) \label{eq:ctma:1}
\end{equation}
Now
\begin{align*}
	& \int_{u = 0}^\infty A(u) \bX(s-u) \,du \\
	&= \int_{u = 0}^\infty A(u) \int_{v = 0}^\infty B(v) \,d\bW(s-u-v) \,du \\
	&= \int_{u = 0}^\infty A(u) \int_{t = u}^\infty B(t-u) \,d\bW(s-t) \,du  \quad t \equiv u+v \\
	&= \int_{t = 0}^\infty \bracs{\int_{u = 0}^t A(u) B(t-u) \,du} d\bW(s-t)
\end{align*}
So from \eqref{eq:ctvar}
\begin{multline}
	 d\bX(s) = \int_{t = 0}^\infty \bracs{\int_{u = 0}^t A(u) B(t-u) \,du} d\bW(s-t) \,ds \\ + d\bW(s) \label{eq:ctma:2}
\end{multline}
Equating the right-hand sides of \eqref{eq:ctma:1} and \eqref{eq:ctma:2}, which must hold for all $s$, we obtain eqs. \eqref{eq:makern}.

\section{Limiting approximations for a subsampled CTVAR process} \label{sec:doulim}

Firstly we derive an asymptotic expansion for the scaled CPSD $\Dt S(\Dt\lambda;\Dt)$ of a subsampled CTVAR process $\bX$. We assume that the continuous-time autocovariance function $\Gamma$ has an inverse Laplace transform\footnote{This will be the case if $\Gamma(t)$ is bounded continuous or in $L^\infty(0,\infty)$.}; \ie\ $\Gamma(t)$ may be represented as a Laplace transform
\begin{equation}
	\Gamma(t) = \int_0^\infty \Lambda(s) e^{-ts} \,ds \qquad\text{for } t \ge 0 \label{eq:gamlap}
\end{equation}
The Wiener-Kintshine theorem  \eqref{eq:ccpsd} applied to \eqref{eq:gamlap} then yields
\begin{equation}
	S(\lambda) = \int_0^\infty\bracs{\frac{\Lambda(s)}{s+2\pi i\lambda} + \frac{\trans{\Lambda(s)}}{s-2\pi i\lambda}} ds \label{eq:lapx:Slambda}
\end{equation}
We now calculate the CPSD $S(\lambda;\Dt)$ of the subsampled process as follows (we don't assume that $\Dt$ is small):
\begin{align*}
	&S(\lambda;\Dt) \\
	&= \Dt \sum_{k = -\infty}^\infty \Gamma(k\Dt) e^{-2\pi i\Dt\lambda k} \qquad\text{from eq.~\eqref{eq:cpsd}} \\
	&= \Dt \sum_{k = 0}^\infty \Gamma(k\Dt) e^{-2\pi i\Dt\lambda k} + [*] - \Gamma(0) \\
	&= \Dt \sum_{k = 0}^\infty \bracs{\int_0^\infty \Lambda(s) e^{-\Dt sk} \,ds} e^{-2\pi i\Dt\lambda k} + [*] - \Gamma(0) \ \ \text{from \eqref{eq:gamlap}} \\
	&= \Dt \int_0^\infty \Lambda(s) \bracs{\sum_{k = 0}^\infty e^{-\Dt(s+2\pi i\lambda)k}} ds + [*] - \Gamma(0) \quad\text{rearranging} \\
	&= \Dt \int_0^\infty \Lambda(s) \bracs{1-e^{-\Dt(s+2\pi i\lambda)}}^{-1} ds + [*] - \Gamma(0)
\end{align*}
where $[*]$ indicates the Hermitian transpose of the preceding term, so that
\begin{equation}
	S(\lambda;\Dt) = \Dt \int_0^\infty \bracs{\frac{\Lambda(s)}{\displaystyle 1-e^{-\Dt(s+2\pi i\lambda)}} + \frac{\trans{\Lambda(s)}}{\displaystyle 1-e^{-\Dt(s-2\pi i\lambda)}}} ds - \Gamma(0) \label{eq:lapx:SLam}
\end{equation}
Now let us define the coefficients $C_n$ by
\begin{equation}
	\frac x{1-e^{-x}} = \sum_{n=0}^\infty C_n x^n \label{eq:lapx:Cn}
\end{equation}
so $C_0 = 1, C_1 = \shalf, C_2 = \tfrac1{12}, C_3 = 0, C_4 = -\tfrac1{720},$ \etc\ We then have (again $[*]$ indicates the Hermitian transpose of the preceding term)
\begin{align*}
	& S(\lambda;\Dt) \\
	&= \Dt \int_0^\infty \sum_{n=0}^\infty C_n [\Dt(s+2\pi i\lambda)]^{n-1} \Lambda(s) \,ds + [*] - \Dt\Gamma(0) \\
	& \hspace{4.5cm} \text{from \eqref{eq:lapx:SLam} and \eqref{eq:lapx:Cn}} \\
	&= \sum_{n=0}^\infty \Dt^n C_n \int_0^\infty (s+2\pi i\lambda)^{n-1} \Lambda(s) \,ds + [*] - \Dt\Gamma(0)
\end{align*}
From \eqref{eq:lapx:Slambda} we see that the $n = 0$ terms are just $S(\lambda)$ and from \eqref{eq:gamlap} $\Gamma(0) = \int_0^\infty \Lambda(s) \,ds$, so that the $n = 1$ terms cancel with the trailing $-\Dt\Gamma(0)$. We thus have have
\begin{equation}
	S(\lambda;\Dt) = S(\lambda) + \sum_{n=2}^\infty \Dt^n C_n \int_0^\infty (s+2\pi i\lambda)^{n-1} \Lambda(s) \,ds + [*] \label{eq:lapx:Slamb}
\end{equation}
From \eqref{eq:gamlap} we have
\begin{equation}
	\overset{(k)}\Gamma(0) = (-1)^k \int_0^\infty s^k \Lambda(s) \,ds \label{eq:lapx:Gamderivs}
\end{equation}
From the CTVAR Yule-Walker equations \eqref{eq:cyw}, $\overset{(k)}\Gamma(0)$ is symmetric for even $k$ and $\dot\Gamma(0) + \trans{\dot\Gamma(0)} = -\Sigma$. Expanding \eqref{eq:lapx:Slamb} in powers of $\Dt$ and using \eqref{eq:lapx:Gamderivs}, we find to $\bigO{\Dt^4}$:
\begin{equation}
	S(\lambda;\Dt) = S(\lambda) + \tfrac1{12}\Dt^2 \Sigma + \tfrac1{720}\Dt^4 (\Omega+12\pi^2 \lambda^2\Sigma) + \bigO{\Dt^5}
\end{equation}
where we have set $\Omega \equiv \dddot\Gamma(0) + \trans{\dddot\Gamma(0)}$. We note that the lowest-order approximation $S(\lambda;\Dt) = S(\lambda) + \bigO\Dt$ may be derived more simply by replacing the sum in the expression $S(\lambda;\Dt) = \sum_{k = -\infty}^\infty \Dt\Gamma(k\Dt) e^{-2\pi i\lambda k\Dt}$ [\cf\ eq.~\eqref{eq:cpsd}] by an integral over $t = k\Dt$ in the limit $\Dt \to 0$ \citep{Zhou:2014}.

Next, we note with \cite{Zhou:2014} that, since the analytic extension $\Psi(z;\Dt)$ of $\frac1\Dt H\Big(\frac\omega{2\pi\Dt};\Dt\Big)$ is holomorphic on the interior of the unit disc $|z| \le 1$ (\secref{sec:varGC}), by the Mean-Value Property for holomorphic functions \citep{Gamelin:2001}, we have
\begin{equation}
	\int_{-\frac1{2\Dt}}^{\frac1{2\Dt}} H(\lambda;\Dt) \,d\lambda = I
\end{equation}
from which we may conclude that $H(\lambda;\Dt) = \bigO1$  in $\Dt$. Now, since
\begin{align}
	H(\lambda;\Dt) \Sigma(\Dt) H(\lambda;\Dt)^* &= S(\lambda;\Dt) \notag \\
	&\to S(\lambda) = H(\lambda) \Sigma H(\lambda)^*
\end{align}
as $\Dt \to 0$, it follows that $H(\lambda;\Dt) = H(\lambda) + \bigO\Dt$ and $\Sigma(\Dt) = \Sigma + \bigO\Dt$ as required.

\section{Simulating CTVAR processes} \label{sec:voudlsim}

We simulate CTVAR processes via a straightforward generalisation of Newton's method to stochastic integro-differential equations. To this end, we develop a VAR approximation for the subsampled process\footnote{For our minimal CTVAR, since we can explicitly solve the spectral factorisation problem for the subsampled process with a rational transfer function (see \apxref{sec:minoulagdsgc} below), we could in principle generate realisations of the subsampled process \emph{exactly} as a VARMA at any desired time resolution. However, the factorisation is unwieldy, so we prefer to use an approximation which, furthermore, is also applicable to problems for which a spectral factorisation cannot easily be obtained analytically.}. Firstly, we define the coefficients\footnote{We use the notation $A_k[\Dt]$ to distinguish these quantities from the $A_k(\Dt)$ which, under our notational convention, denote the VAR coefficients of the subsampled process; they will not generally coincide exactly.}
\begin{equation}
	A_k[\Dt] = \int_{(k-1)\Dt}^{k\Dt} A(u) \,du \qquad\text{for }  k = 1,2,\ldots \label{eq:coeffx}
\end{equation}
A useful result here, is that if $\varphi(t)$ is \emph{Lipschitz continuous}\footnote{This condition might be relaxed, depending on the form of the kernel $A(u)$.} on $[0,\infty)$---in particular if $\varphi(t)$ is everywhere differentiable and has \emph{bounded derivatives} on $[0,\infty)$---then it is straightforward to show that
\begin{equation}
	\sum_{k = 1}^\infty A_k[\Dt] \varphi(k\Dt) = \int_0^\infty A(u) \varphi(u) \,du + \bigO\Dt \label{eq:phisum}
\end{equation}
Now consider the autoregression
\begin{equation}
	\bX\big(k\Dt) = \bX\big((k-1)\Dt\big) + \Dt \sum_{l=1}^\infty A_\ell[\Dt] \cdot \bX\big((k-\ell)\Dt\big) + \boeta_k \label{eq:ctvarx}
\end{equation}
Assuming that both $\Gamma(t)$ and $\dot\Gamma(t)$ satisfy the condition for \eqref{eq:phisum} to hold, then from \eqref{eq:cyw} we may verify after some algebra that
\begin{equation}
	\Cov{\boeta_\ell}{\boeta_{\ell-k}} = \delta_{k0} \Dt\Sigma + \bigO{\Dt^3} \label{eq:epscov}
\end{equation}
so that the residuals are ``almost'' white and in this sense \eqref{eq:ctvarx} ``almost'' specifies a VAR\footnote{See \cite{Sims:1971,Geweke:1978} for detailed analysis of the \emph{exact} VAR satisfied by the subsampled CTVAR. Roughly, the approximation \eqref{eq:ctvarx} can be expected to be reasonable provided $\bX(t)$ does not fluctuate too wildly at the time scale of the sample interval $\Dt$. In particular, our minimal CTVAR satisfies this condition for $\Dt \ll \min(1/a, 1/b)$. See also \cite{Bergstrom:1966,Sargan:1974} for different approaches to discrete approximation of SDEs.}.

To generate an approximate realisation of the subsampled process, then, we generate realisations $\bx_k \approx \bX(k\dt): k = 1,2,\ldots$ for very small time increments $\dt \ll 1$, by the recursion
\begin{equation}
	\bx_k = \bx_{k-1} + \dt \sum_{l=1}^L A_\ell[\Dt] \cdot \bx_{k-\ell} + \boeta_k \label{eq:ctvarxx}
\end{equation}
where $L$ is large enough that $A_\ell[\Dt] \ll 1$ for $\ell > L$, with iid normal residuals $\boeta_k \sim \normal(\boldsymbol 0; \dt\Sigma)$. An integration time increment $\dt = 0.01\,\text{ms}$ was used for all simulations in this study.

\section{Calculation of the cross-power spectral term for the subsampled minimal CTVAR process with finite causal delay} \label{sec:Sxy}

We have
\begin{align*}
	S_{xy}(z;\Dt)
	&= \sum_{k = 1}^\infty \Gamma_{xy}(k\Dt) z^k + \sum_{k = 0}^\infty \Gamma_{yx}(k\Dt) z^{-k} \\
	&= \sum_{k = 1}^{q-1} \Gamma_{xy}(k\Dt) z^k + \sum_{k = q}^\infty \Gamma_{xy}(k\Dt) z^k + \sum_{k = 0}^\infty \Gamma_{yx}(k\Dt) z^{-k} \\
	&= \sum_{k = q}^\infty \theta \bracr{\frac 1{a+b} e^{a\tau} \alpha^k - \frac 1{2b} e^{b\tau} \beta^k} z^k \\
	&\hspace{1cm}+ \sum_{k = 1}^{q-1} \frac 1{2b} \eta e^{-b\tau} \beta^{-k} z^k
	+ \sum_{k = 0}^\infty \frac 1{2b} \eta e^{-b\tau} \beta^k z^{-k} \\
	&= \sum_{k = q}^\infty \theta \bracr{\frac 1{a+b} e^{a\tau} \alpha^k - \frac 1{2b} e^{b\tau} \beta^k} z^k \\
	&\hspace{1cm}+ \frac 1{2b} \eta e^{-b\tau} \bracs{\sum_{k = 1}^{q-1} \beta^{-k} z^k + \sum_{k = 0}^\infty \beta^k z^{-k}} \\
	&= \frac 1 c \theta\eta e^{a\tau} \frac{(\alpha z)^q}\mfa - \frac 1{2b} \theta e^{b\tau} \frac{(\beta z)^q}\mfb - \frac 1{2b} \eta e^{-b\tau} \frac{(\gamma z)^q}{1-\gamma z}
\end{align*}

\section{Spectral factorisation of the subsampled minimal CTVAR: zero residuals correlation} \label{sec:minoulagdsgc}

Here we perform the spectral factorisation of the subsampled CPSD \eqref{eq:oudscpsd}---as required for discrete-time GC calculations in the time \eqref{eq:gc} and frequency \eqref{eq:sgc} domains---for the full and reduced regressions. Since spectral factorisation are \emph{unique} \citep{Wilson:1972}, we achieve this via an \textit{ad hoc} approach suggested by the structure of the process and the form of the CPSDs. Throughout this section $z = e^{-i\omega}$, where $\omega = 2\pi\Dt\lambda$ is the angular frequency. For compactness, we generally drop the $\Delta$ in $S(z;\Dt)$, \etc

For the full regression, we attempt a rational VARMA factorisation of the CPSD \eqref{eq:oudscpsd} of the form
\begin{equation}
	A(z) S(z) A^*(z) = B(z) \Sigma B^*(z)
\end{equation}
so that $A(z)$ represents the VAR factor, $B(z)$ the VMA factor and $\Sigma = \begin{bmatrix} \sigma_{xx} & \sigma_{xy} \\ \sigma_{xy} & \sigma_{yy} \end{bmatrix}$ the residuals covariance matrix. $A(z),B(z)$ are matrix polynomials in $z$ with $A(0) = B(0) \equiv I$ and $\Sigma$ is symmetric positive-definite. Since $Y$ [and hence $Y(\Dt)$] is autonomous, there is clearly no causality in the $X(\Dt) \to Y(\Dt)$ direction. We also note that any \emph{instantaneous} GC must arise from the moving-average component of the joint subsampled process (see below), so we attempt a factorisation of the form
\begin{multline}
	\begin{bmatrix}
		A_{xx}(z) & 0 \\ 0 & A_{yy}(z)
	\end{bmatrix}
	\begin{bmatrix}
		S_{xx}(z) & S_{xy}(z) \\ S_{xy}(\cz) & S_{yy}(z)
	\end{bmatrix}
	\begin{bmatrix}
		A_{xx}(\cz) & 0 \\ 0 & A_{yy}(\cz)
	\end{bmatrix}
	= \\
	\begin{bmatrix}
		B_{xx}(z) & B_{xy}(z) \\ 0 & B_{yy}(z)
	\end{bmatrix}
	\begin{bmatrix}
		\sigma_{xx} & \sigma_{xy} \\ \sigma_{xy} & \sigma_{yy}
	\end{bmatrix}
	\begin{bmatrix}
		B_{xx}(\cz) & 0 \\ B_{xy}(\cz) & B_{yy}(\cz)
	\end{bmatrix} \label{eq:mctvarsf}
\end{multline}
The component equations are then
\begin{subequations}
\begin{align}
	xx: && |A_{xx}|^2 S_{xx} &= \sigma_{xx} |B_{xx}|^2 + 2\sigma_{xy} \re\bracc{B_{xx} B^*_{xy}} + \sigma_{yy} |B_{xy}|^2 \\
	xy: && A^*_{yy} A_{xx} S_{xy} &= B^*_{yy} \bracs{\sigma_{xy} B_{xx} + \sigma_{yy} B_{xy}} \\
	yy: && |A_{yy}|^2 S_{yy} &= \sigma_{yy} |B_{yy}|^2
\end{align}%
\end{subequations}
Examination of the CPSD suggests we try
\begin{subequations}
\begin{align}
	A_{xx}(z) &= (1-\alpha z) (1-\beta z) \label{eq:Axx} \\
	A_{yy}(z) &= 1-\beta z
\end{align}%
\end{subequations}
The $yy$ equation yields $B_{yy}(z) \equiv 1$, so that
\begin{equation}
	\sigma_{yy} = w
\end{equation}
Defining
\begin{subequations}
\begin{align}
	|M(z)|^2 &\equiv w \bracr{u_1 |1-\beta z|^2 + u_2 |1-\alpha z|^2} \label{eq:Mofz1} \\
	L(z) &\equiv \beta\left[v_1 (1-\beta z) (1-\gamma z) + v_2 (1-\gamma z) (1-\alpha z)\right. \notag \\
	&\hspace{1cm} \left. + v_3 (1-\alpha z) (1-\beta z) \right]
\end{align}%
\end{subequations}
(note that $|M(z)|^2$ may be factorised), the $xy$ equation then yields
\begin{equation}
	\sigma_{xy} B_{xx}(z) + \sigma_{yy} B_{xy}(z) = -L(z) z^{q-1} \label{eq:Bxy}
\end{equation}
on $|z| = 1$. But since $B_{xx}(z), B_{xy}(z)$ are polynomials (by assumption), \eqref{eq:Bxy} holds for all $z$ in the complex plane, so that in particular setting $z = 0$ we obtain
\begin{equation}
	\sigma_{xy} = \left\{
		\begin{array}{rcl}
			 -\beta P && q = 1 \\
			0 && q > 1
		\end{array}
	\right.
\end{equation}
with
\begin{equation}
	P \equiv v_1+v_2+v_3 \label{eq:Pdef}
\end{equation}
Next, the $xx$ equation yields (after some algebra)
\begin{equation}
	D |B_{xx}(z)|^2 = |M(z)|^2 - |L(z)|^2 \label{eq:Bxx}
\end{equation}
on $|z| = 1$, where
\begin{equation}
	D \equiv \dett\Sigma = \sigma_{xx} \sigma_{yy} - \sigma^2_{xy}
\end{equation}
Thus we need to solve the factorisation problem \eqref{eq:Bxx} for $D,B_{xx}(z)$ (see below) and we then have the residuals variance $\sigma_{xx}$ for the full regression.

Note that $|M(z)|^2$ for $z = e^{-i\omega}$ is linear in $\cos\omega$ while  $|L(z)|^2$ is linear in $\cos\omega, \cos2\omega$. Therefore \eqref{eq:Bxx} may be factored for $B_{xx}(z)$ a $2$nd order polynomial and it follows from \eqref{eq:Bxy} that $B_{xy}(z)$ is of order $q+1$. In summary, we have
\begin{subequations}
\begin{align}
	A_{xx}(z) \text{ is of order } & 2   \\
	A_{yy}(z) \text{ is of order } & 1   \\
	B_{xx}(z) \text{ is of order } & 2   \\
	B_{xy}(z) \text{ is of order } & q+1 \\
	B_{yy}(z) \text{ is of order } & 0
\end{align}%
\end{subequations}
while the remaining coefficients vanish. We see that the joint subsampled process $(X(\Dt),Y(\Dt))$ is thus VARMA($2,q+1$). Note that $Y(\Dt)$ is just VAR($1$).

For the full regression, we write
\begin{equation}
	L(z) = \beta(P-Qz+Rz^2) \label{eq:Lofz}
\end{equation}
where
\begin{subequations}
\begin{align}
	Q  &= v_1(\beta+\gamma) + v_2(\gamma+\alpha) + v_3(\alpha+\beta) \\
	R  &= v_1\beta\gamma + v_2\gamma\alpha + v_3\alpha\beta
\end{align}%
\end{subequations}
We have
\begin{equation}
	|M(z)|^2 = 2w(\varphi - \psi \cos\omega) \label{eq:Mofz2}
\end{equation}
where
\begin{subequations}
\begin{align}
	\varphi &\equiv \shalf\bracs{u_2(1+\alpha^2) + u_1(1+\beta^2)} \\
	\psi &\equiv u_2\alpha + u_1\beta
\end{align} \label{eq:phipsi}%
\end{subequations}
and setting
\begin{equation}
	B_{xx}(z) = 1-Uz+Vz^2 \label{eq:Bxx1}
\end{equation}
the factorisation problem \eqref{eq:Bxx} becomes
\begin{multline}
	D\bracs{1+U^2 + V^2 -2U(1+V) \cos\omega + 2V \cos2\omega} = \\ 2w(\varphi - \psi \cos\omega) - \beta^2 \left[P^2+Q^2 + R^2 \right. \\ \left.- 2Q(P+R) \cos\omega+ 2PR \cos2\omega\right]
\end{multline}
Since this must hold for all $\omega$, we have
\begin{subequations}
\begin{align}
	D(1+U^2 + V^2) &= 2\fa \\
	DU(1+V)        &= \fb \\
	DV             &= \fc
\end{align}%
\end{subequations}
where we have set
\begin{subequations}
\begin{align}
	\fa &\equiv w\varphi - \shalf\beta^2 (P^2 +Q^2 + R^2) \\
	\fb &\equiv w\psi - \beta^2 Q(P+R) \\
	\fc &\equiv - \beta^2 PR
\end{align}%
\end{subequations}
so that
\begin{subequations}
\begin{align}
	U &=  \frac\fb{D+\fc} \\
	V &=  \frac\fc D\\
\end{align}%
\end{subequations}
with $D$ the largest root of the $4$th degree polynomial equation
\begin{equation}
	(D+\fc)^2 \bracs{(D-\fa)^2 - \bracr{\fa^2 - \fc^2}} + \fb^2 D^2 = 0 \label{eq:4deg}
\end{equation}
Setting
\begin{equation}
	D = \fd + \sqrt{\fd^2-\fc^2} \label{eq:4degDsol}
\end{equation}
we find that \eqref{eq:4deg} is satisfied if $4(\fd-\fa)(\fd+\fc) +\fb^2 = 0$, so that \eqref{eq:4degDsol} is the required solution with
\begin{equation}
	\fd \equiv \shalf\bracr{\fa-\fc + \sqrt{(\fa+\fc)^2-\fb^2}} \label{eq:4degKsol}
\end{equation}
Note that if $\Dt$ divides $\tau$ exactly (for $\Dt \le \tau$), then $P \equiv 0$, so $\fc \equiv 0$ and largest root is
\begin{equation}
	D = \fa + \sqrt{\fa^2-\fb^2}
\end{equation}

For the reduced regression we need to factorise $S_{xx}(z)$, the power spectrum of $X(\Dt)$ alone \eqref{eq:oudscpsd:xx}. The VARMA factorisation takes the form
\begin{equation}
	|a(z)|^2 S_{xx}(z) = \sigma'_{xx} |h(z)|^2
\end{equation}
where $\sigma'_{xx}$ is the residuals variance of the reduced regression, which is easily seen to be solvable by $a(z) = (1-\alpha z) (1-\beta z)$, $h(z) = 1-pz$, so that  $X(\Dt)$ is VARMA($2,1$). We find:
\begin{subequations}
\begin{align}
	\sigma'_{xx}(1+p^2) &= 2\varphi \\
	\sigma'_{xx} p  &= \psi
\end{align} \label{eq:sigrxxs}%
\end{subequations}
so that, eliminating $p$, the solution for $\sigma'_{xx}$ is (again we need the largest root)
\begin{equation}
	\sigma'_{xx} = \varphi + \sqrt{\varphi^2-\psi^2}
\end{equation}

Finally, to calculate $\sgc Y X(z)$, \eqref{eq:sgc} yields
\begin{equation}
	\sgc Y X(z) = -\log\bracs{1-S_{xx}(z)^{-1} \Sigma_{yy|x} |H_{xy}(z)|^2}
\end{equation}
We thus require $S_{xx}(z)$, $\Sigma_{yy|x}$ and $H_{xy}(z)$. Again, we already have $S_{xx}(z)$ from \eqref{eq:oudscpsd}, while $\Sigma_{yy|x} = \sigma_{yy} - \sigma_{xx}^{-1}\sigma_{xy}^2 = \sigma_{xx}^{-1}D$. From \eqref{eq:mctvarsf}, we have $H_{xy}(z) = A_{xx}(z)^{-1} B_{xy}(z)$. We already know $A_{xx}(z)$ from \eqref{eq:Axx} and from \eqref{eq:Bxy} we have $B_{xy}(z) = -\sigma_{yy}^{-1} \bracs{\sigma_{xy} B_{xx}(z) + L(z) z^{q-1}}$, with $B_{xx}(z)$ given by \eqref{eq:Bxx1} and $L(z)$ by \eqref{eq:Lofz}. The expression for $\sgc Y X(z)$ may be simplified somewhat by noting that $S_{xx}(z) |A_{xx}(z)|^2 = \sigma_{yy}^{-1} |M(z)|^2$, where, from \eqref{eq:Bxx}, $|M(z)|^2 =	D |B_{xx}(z)|^2 + |L(z)|^2$.

From \eqref{eq:mctvarsf} and $B_{yy}(z) \equiv 1$, we see that for the $\Dt$-subsampled joint process to be minimum-phase it is necessary and sufficient that all roots of the equation $B_{xx}(z) = 1-Uz+Vz^2 = 0$ \eqref{eq:Bxx1} lie strictly outside the unit disc $|z| \le 1$ in the complex plane. It is easily checked that the process $X(\Dt)$ is always minimum phase: from \eqref{eq:sigrxxs} we have $p = \Big(\varphi - \sqrt{\varphi^2-\psi^2}\Big)\big/\psi$, so that $|p| < 1$ is always satisfied and the (single) root of $h(z) = 0$ thus always lies outside the unit disc. The process $Y(\Dt)$ is VAR($1$) and thus trivially minimum-phase.

\ifdefined\ARXIV
\else
\newpage\noindent\textbf{\large References}
\fi

\bibliographystyle{elsarticle-harv}

\bibliography{downsample}



\end{document}